%% file: arxiv3.tex
\documentclass[prx,aps,amsfonts,showpacs,nofootinbib,longbibliography,notitlepage,twocolumn,groupedaddress]{revtex4-2}

\include{preamble}

\usepackage[normalem]{ulem} 
\usepackage[dvipsnames]{xcolor}
\usepackage{comment}
\usepackage{subfigure}
\begin{document}

\title{Measurement-induced phases of matter require feedback}

\author{Aaron J. Friedman}
\affiliation{Department of Physics and Center for Theory of Quantum Matter, University of Colorado, Boulder CO 80309, USA}

\author{Oliver Hart}
\affiliation{Department of Physics and Center for Theory of Quantum Matter, University of Colorado, Boulder CO 80309, USA}

\author{Rahul Nandkishore}
\affiliation{Department of Physics and Center for Theory of Quantum Matter, University of Colorado, Boulder CO 80309, USA}

\begin{abstract}
    We explore universality and phases of matter in hybrid quantum dynamics combining chaotic time evolution and projective measurements. We develop a unitary representation of measurements based on the Stinespring Theorem, which we crucially identify with the time evolution of the system and measurement apparatus, affording significant technical advantages and conceptual insight into hybrid dynamics. We diagnose spectral properties in the presence of measurements for the first time, along with standard, experimentally tractable probes of phase structure, finding no nontrivial effects due to measurements in the absence of feedback. We also establish that nonlinearity in the density matrix is neither sufficient nor necessary to see a transition, and instead identify utilization of the measurement outcomes (i.e., ``feedback'') as the crucial ingredient. After reviewing the definition of a phase of matter, we identify nontrivial orders in \emph{adaptive} hybrid dynamics---in which measurement outcomes determine future unitary gates---finding a genuine measurement-induced absorbing-state phase transition in an adaptive quantum East model. In general, we find that only deterministic and constrained Haar-random dynamics with active feedback and without continuous symmetries can realize genuine, measurement-induced phases of matter.
\end{abstract}

\date{\today}
\maketitle
\tableofcontents

\section{Introduction} 
\label{sec:intro}
Understanding the possible phases of nonequilibrium quantum matter is a key frontier in quantum science. While most work has focused on well-isolated systems undergoing unitary time evolution \cite{ETH1, ETH2, ETH3, mblarcmp, mblrmp, NahumRUC1, NahumOperator, RUCNCTibor, CDLC1, RUCconTibor, RUCconVedika, U1FRUC, ConstrainedRUC}, the most general quantum dynamics also includes \emph{measurements}. It was recently observed that increasing the frequency 
of measurements can drive an \emph{entanglement transition} \cite{og-MIPT, FisherMIPT1, chan, FisherMIPT2, ChoiQiAltman, RomainHolographic2} between area- and volume-law scaling of the entanglement entropy of the system's state. Interest in the landscape of quantum dynamics---and transitions in various physicals quantities---that depend on the rate $\measrate$ of projective measurements has since exploded (see \cite{RUCreview} for a review). 

Here we investigate whether the transitions 
induced by projective measurements on top of chaotic local time evolution  \cite{og-MIPT, FisherMIPT1, chan, FisherMIPT2, ChoiQiAltman, RomainHolographic2} amount to transitions between distinct \emph{phases of matter} in any reasonable understanding of the term. Disappointingly, we find that \emph{none} of the measurement-induced transitions reported in the literature \cite{og-MIPT, FisherMIPT1, chan, FisherMIPT2, ChoiQiAltman, RomainHolographic2, RUCreview, ZabaloMIPT, GullansHuseOP, GullansHusePRX, MIPT-ATA, UtkarshChargeSharp, hsieh, barkeshli, MIPT-exp} correspond to a transition between distinct phases of matter in any physically meaningful or historically consistent sense. We also analyze spectral properties in the presence of measurements for the first time, finding that the measurement-induced entanglement transition (MIET) is distinct from thermalization transitions \cite{ETH1, ETH2, ETH3, mblarcmp, mblrmp}. 

We further establish that nonlinearity in the density matrix is neither necessary nor sufficient for a quantity to detect a measurement-induced transition. Instead, we establish utilization of the measurement outcomes as the crucial ingredient to realizing measurement-induced phenomena. We then identify absorbing-state transitions \cite{Absorbing, GarrahanAbsorb, HayeAbsorbing} in nonsymmetric adaptive circuits---in which measurement outcomes determine subsequent unitary gates---as the only presently viable route to genuine measurement-induced phases of matter. In particular, we numerically simulate an adaptive 1D quantum East model \cite{QuantumEast2020} in which the combination of measurements and outcome-dependent feedback realize an absorbing-state phase transition in the directed percolation universality class \cite{HayeAbsorbing}. 

Key to our analysis of hybrid dynamics is the \emph{unitary} representation of projective measurements that we develop via the  Stinespring Dilation Theorem \cite{Stinespring} (see also \cite{SpeedLimit, AaronDiegoFuture, AaronYifanFuture}). The unitary measurement channel acts on a dilated Hilbert space that includes the state of the measurement apparatus, and is unique up to the identification of a ``default'' outcome, corresponding to the initial state of the apparatus \cite{AaronDiegoFuture}. While unitary descriptions of measurements as a bookkeeping tool have been known \cite{ChoiQiAltman}, the utility of our formalism follows from the identification of that unitary with the physical \emph{time evolution} of the system and measurement apparatus during the measurement process \cite{AaronDiegoFuture}. As a result, though formally equivalent to the Kraus representation \cite{ChoisThm}, 
our unitary representation \cite{Stinespring, SpeedLimit, AaronDiegoFuture, ChoisThm, AaronYifanFuture} has significant technical and conceptual advantages. In Sec.~\ref{sec:Stinespring} we provide a ``textbook'' exposition of our unitary Stinespring measurement formalism.

The remainder of this work considers whether and when projective measurements can give rise to new phases of matter or universal dynamics, compared to chaotic time evolution alone. We consider standard diagnostics of phase structure and universality in ``generic'' hybrid quantum circuits \cite{RUCNCTibor, NahumOperator, NahumRUC1, CDLC1}, as well as those ``enriched'' by conservation laws \cite{RUCconTibor, RUCconVedika, U1FRUC, ConstrainedRUC, UtkarshChargeSharp, hsieh, barkeshli} and/or kinetic constraints \cite{RitortKCM, garrahan2010kinetically, KCMRydberg, Valado_2016, GarrahanLectures, QuantumEast2020, ConstrainedRUC}. In Sec.~\ref{sec:hybridcirc} we detail the  hybrid circuits of interest using the Stinespring representation, focusing on canonical models \cite{og-MIPT, FisherMIPT1, chan, FisherMIPT2, ChoiQiAltman, RomainHolographic2, RUCreview, ZabaloMIPT, GullansHuseOP, GullansHusePRX, MIPT-ATA, UtkarshChargeSharp, hsieh, barkeshli, MIPT-exp} in which the protocol at time $t$ is not conditioned on prior measurement outcomes.

In Sec.~\ref{sec:OneCopy} we consider generic, experimentally tractable probes common to condensed matter and atomic, molecular, and optical (AMO) physics, corresponding to expectation values, correlation functions, and response functions. We show that \emph{none} of these probes see a transition due to projective measurements, distinguishing the MIET from all historical examples of genuine phase transitions, including thermalization transitions \cite{ETH1, ETH2, ETH3, mblarcmp, mblrmp}, which also manifests in the scaling of the system's entanglement entropy. 

In Sec.~\ref{sec:SFF} we consider the spectral form factor (SFF) \cite{BohigasChaos, RMT_SFF, YoshidaSFF, ShenkerRMT, ProsenRMTChaosPRX, CDLC1, bertini2018exact, CDLC2, U1FRUC, SubirSFF, ConstrainedRUC, SamJohnFeynman}. Though not experimentally measurable, the SFF is a standard diagnostic for quantum dynamics that is sensitive to thermalization transitions \cite{ETH1, ETH2, ETH3, mblarcmp, mblrmp}. As with the standard probes of Sec.~\ref{sec:OneCopy}, we find no effect due to measurements in most hybrid protocols, establishing that the MIET \cite{og-MIPT, FisherMIPT1, chan, FisherMIPT2, ChoiQiAltman, RomainHolographic2} is not a thermalization transition. The blindness of the SFF persists even when defined to be quadratic in the density matrix $\DensMat (t)$ and postselected, establishing that nonlinearity of a quantity in $\DensMat (t)$ is not \emph{sufficient} to observe a transition. Additionally, in Sec.~\ref{sec:adaptive} we find that adaptive hybrid circuits can realize transitions captured by standard expectation values, establishing that nonlinearity of a quantity in $\DensMat (t)$ is not \emph{necessary} to observe a transition. Thus, the intuition in the literature \cite{og-MIPT,UtkarshChargeSharp} that measurement-induced transitions are associated with (or generally require) nonlinear functions of the density matrix $\DensMat (t)$ is incorrect.

Instead, we find that measurement-induced transitions in any quantity are only possible when the measurement outcomes are \emph{utilized}. In Sec.~\ref{sec:can't work}, we compare the findings of Secs.~\ref{sec:OneCopy} and \ref{sec:SFF} with the MIET literature  \cite{og-MIPT, FisherMIPT1, chan, FisherMIPT2, ChoiQiAltman, RomainHolographic2, RUCreview, ZabaloMIPT, GullansHuseOP, GullansHusePRX, MIPT-ATA, UtkarshChargeSharp, hsieh, barkeshli, MIPT-exp} to argue that (\emph{i}) it is the utilization of outcomes---rather than nonlinearity in $\DensMat (t)$---that is \emph{sine qua non} for a measurement-induced transition and (\emph{ii}) the measurement-induced \emph{entanglement} transition is not a transition between distinct phases of matter in any physically meaningful or historically consistent sense of the term.

As we discuss in Sec.~\ref{sec:can't work}, one means by which to ``use'' the measurement outcomes---at least on paper---corresponds to \emph{postselection}. Generally, postselected probes can be grouped into two categories: measures of entanglement \cite{og-MIPT, chan, FisherMIPT1, FisherMIPT2, RomainHolographic2, ChoiQiAltman, ZabaloMIPT, GullansHusePRX, GullansHuseOP, MIPT-ATA, UtkarshChargeSharp, hsieh, barkeshli, MIPT-exp, MPAF_cross} and the disconnected parts of $n$-point functions \cite{UtkarshChargeSharp, hsieh, barkeshli, learnability}. Such quantities have been reported 
to show a transition as a function of measurement rate $\measrate$. However, we do not consider such probes herein, and instead refer the reader to the foregoing references for further details. Rather, in Sec.~\ref{subsec:what is a phase}, we review the standard, historical criteria that phases of matter must fulfill. 

In Sec.~\ref{subsec:no postselect}, we discuss the well-known practical issue with postselection \cite{RUCreview}, along with a \emph{conceptual} issue that does not appear to have been widely recognized and cannot be sidestepped by any  protocol (as far as we are aware). The practical issue is that postselection is exponentially costly in spacetime volume---while \cite{og-MIPT} refers to postselection as a ``severe statistical challenge,'' it is fully \emph{impossible} in the thermodynamic limit---the only limit in which phases of matter are defined. The conceptual issue is that postselected probes are incompatible with the very idea of a phase of matter---they are either not robust, experimentally impossible, or require infinite resources to evaluate in the limit where phases are defined. Most importantly, knowledge of a sample's phase must afford quantitative predictions about that sample. Yet Sec.~\ref{sec:OneCopy} establishes that the experimentally observable properties of a hybrid protocol are unrelated to the MIET.

In the remainder of Sec.~\ref{sec:can't work}, we explore alternatives to postselected probes of measurement-induced transitions. In Sec.~\ref{subsec:no postselect} we consider various attempts in the literature to ameliorate the postselection problem(s), finding that none succeed. In Sec.~\ref{subsec:classifiers}, we further preclude the use of neural networks, classifiers, and similar methods from diagnosing phase structure. Essentially, these methods do not probe the system itself, but instead ``guess'' the phase with high probability. Finally, in Sec.~\ref{subsec:Need Feedback}, we argue that only circuits with feedback---i.e., classical decoding \cite{Sam-meas, JongYeonDecode} or adaptive dynamics \cite{SpeedLimit, SthitadhiSteer, GornyiMeas, TomMIPT, preselect, MIPT_wormhole, ODea}---can realize genuine, measurement-induced phases of matter. However, it is unclear whether, in practice, classical postprocessing can be scaled to detect the MIET \cite{Sam-meas, JongYeonDecode}.

This leaves adaptive hybrid protocols \cite{SthitadhiSteer, TomMIPT, preselect, MIPT_wormhole, GornyiMeas, ODea, JongYeonDecode}---in which the composition of the circuit at time $t$ may be conditioned on the outcomes of prior measurements---as the most promising route toward realizing genuine, measurement-induced phases of matter. We consider such protocols in Sec.~\ref{sec:adaptive}, ruling out large swaths of adaptive hybrid circuits as being incompatible with measurement-induced \emph{phase} transitions (MIPTs), which separate physically distinct phases of matter, while also identifying which classes of models \emph{can} realize MIPTs. Importantly, we note that the adaptive MIPT is unrelated to the MIET \cite{ODea}.

We consider maximally chaotic adaptive hybrid protocols both with and without block structure. We first observe that Haar-random time evolution without structure is incompatible with a genuine phase: Because no operators survive even a single layer of time evolution, a robust order parameter (whether local or nonlocal) is impossible. We then consider models with discrete symmetries---because only a handful of system-spanning operators are robust to time evolution, robust order remains impossible. We consider the example of a $\Ints^{\,}_2$-symmetric Ising circuit acting on a system of qubits to illustrate this result.

We then investigate adaptive dynamics in the presence of continuous Abelian symmetries; while such models can realize robust order, they cannot realize phase transitions. Chaotic dynamics with continuous symmetries conserve an extensive number of local ``charge'' operators, which are thereby robust to chaotic time evolution. Hence, it is always possible to devise adaptive protocols that ``steer'' the system toward an absorbing state \cite{Absorbing, GarrahanAbsorb} with a particular expectation value of the charge operators, which acts as a local order parameter. We illustrate this result via numerical simulation of qubits on a 1D chain, where the chaotic dynamics conserve a $\U{1}$ symmetry corresponding to $\sum_j \, \PZ{j}$. We randomly intersperse these dynamics with measurements of $\PZ{j}$ (at rate $\measrate$), followed by the operation $\PX{j}$ if the outcome is $1$. The unitary dynamics lead to a symmetric simple exclusion process (SSEP) \cite{Schutz_SSEP} of the charge operators, while the adaptive measurements convert $\PZ{j}$ to $\ident$. As a result, for \emph{any} $\measrate>0$, one always finds $\expval{\PZ{j}(t)}=\Order{1}$ for sufficiently late times $t = \text{poly}(\Nsite)$. In other words, for any initial state, this protocol steers the system into the state $\ket{\bvec{0}}$ (all spins up) up to a vanishing density of defects $\ket{1}$. However, because the charge operators $\PZ{j}$ are conserved under time evolution, there is no competition between the chaotic dynamics and adaptive gates, precluding a sharp transition as a function of measurement rate $\measrate$.

Finally, we consider adaptive dynamics in models with kinetic constraints, which admit robust transitions between distinct phases of matter. It is crucial that the underlying unitary evolution competes with the adaptive measurements. The latter generically removes the measured operators in the Heisenberg picture, so that measuring operators of the form of the local order parameter leads to nonzero expectation values in generic initial states. However, for a phase transition to exist, the unitary dynamics must compete with the adaptive measurements by spawning the measured operators. We note that this is generically the case in nonconserving deterministic models (including Hamiltonian models), but in the context of Haar-random circuits requires kinetic constraints without continuous symmetries. 

We explore this MIPT via numerical simulation of a 1D quantum East model \cite{QuantumEast2020} on $\size$ qubits. The unitary dynamics apply a Haar-random gate to qubit $j$ if the ``East'' neighbor is in the state $\ket{1}$, and does nothing otherwise. The adaptive part of the protocol steers the system toward the dynamically stationary absorbing state $\ket{\bvec{0}}$ \cite{Absorbing, GarrahanAbsorb, HayeAbsorbing} by measuring $\PZ{j}$ and applying $\PX{j}$ if the outcome is 1. We find a critical measurement rate $\measrate^{\,}_{\rm c} \approx 0.038$ and identify universal exponents consistent with directed percolation \cite{HayeAbsorbing} via high-quality scaling collapse. For $\measrate > \measrate^{\,}_{\rm c} $, the system reaches the absorbing state $\ket{\bvec{0}}$  in time $t = \Order{\size}$, while for $\measrate < \measrate^{\,}_{\rm c} $, the absorbing state requires time $t = \exp (\size)$, or the absorbing state is not reached. These results are consistent with concurrent works on adaptive dynamics \cite{TomMIPT, preselect, MIPT_wormhole, JongYeonDecode, ODea}, and generalize to arbitrary adaptive hybrid protocols. At the time of this writing, the absorbing-state transitions realized in these models are the \emph{only} physically meaningful examples of measurement-induced phase transitions.

\section{The Stinespring formalism} 
\label{sec:Stinespring}
The Stinespring Dilation Theorem \cite{Stinespring} provides an alternative representation of measurement channels that is technically better suited to the scrutiny of hybrid circuits and conceptually more revealing. The crucial insight herein is that the resulting unitary representation is not merely a bookkeeping tool, but reflects the time evolution of the system and measurement apparatus \cite{AaronDiegoFuture}. This allows for, e.g., the evolution of operators in the Heisenberg picture in the presence of projective measurements.

\subsection{Stinespring dilation and isometric measurement}
\label{subsec:SSIsometric}
The Stinespring Dilation Theorem \cite{Stinespring} states that all quantum operations (or ``quantum channels'') can be represented via isometries and [partial] traces acting on the density matrix. By Choi's theorem \cite{ChoisThm}, Stinespring's isometric channels are equivalent to both the Kraus and Completely Positive Trace-Preserving (CPTP) map formulations of quantum channels. However, we find that the Stinespring formalism is  more convenient and intuitive in the context of measurements.
 
An isometry is a norm-preserving map from a given Hilbert space $\Hilbert^{\,}_A$ to a ``dilated'' Hilbert space $\Hilbert^{\,}_B$. The dimensions $\HilDim^{\,}_{A,B} = \operatorname{dim} (\Hilbert^{\,}_{A,B})$ give the number of many-body states in each space, and must obey  $\HilDim^{\,}_B \geq \HilDim^{\,}_A$. The isometry $\isometry \, : \, \Hilbert^{\,}_A \mapsto \Hilbert^{\,}_B$ satisfies $\norm{ \isometry \ket{u}} = \norm{ \ket{u} }$ for all vectors $\ket{u} \in \HilDim^{\,}_A$; equivalently, we have that $\isometry^{\dagger} \isometry = \ident^{\,}_A$, while $\isometry \isometry^{\dagger}$ projects onto the subspace  $\Hilbert^{\,}_A \subset \Hilbert^{\,}_B$. When $\HilDim^{\,}_A = \HilDim^{\,}_B$, $\isometry$ is then \emph{unitary} (i.e., $\isometry \isometry^{\dagger} = \ident$). Hence, unitary channels---e.g., corresponding to time evolution---are a proper subset of isometric channels.

\begin{figure}[t!]
    \centering
    \includegraphics[width=.85\columnwidth]{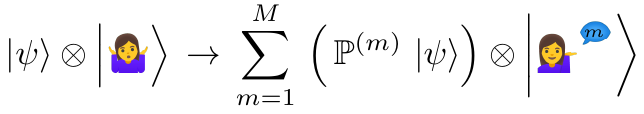}
    \caption{Cartoon depiction of measurement of $\mobserv$ \eqref{eq:ASpectralDecomp} in the Stinespring formulation. The wave function only \emph{appears} to collapse upon measurement---in reality, the observer and system become entangled, and the post-measurement state is a superposition of states where outcome $m$ obtains and the observer records outcome $m$. The dilated Hilbert space encodes this entanglement using an ``outcome register.''}
    \label{fig:manyworlds}
\end{figure}

The rationale for dilating the Hilbert space in the context of measurement is that the act of measurement entangles the state of the physical system with that of the apparatus or observer (which reflects the \emph{outcome}), as depicted in Fig.~\ref{fig:manyworlds}. The dilated Hilbert space,
\begin{equation}
    \label{eq:Dilated Hilbert}
    \Hilbert^{\,}_{\rm dil} \, = \, \Hilbert^{\,}_{\rm ph} \otimes \Hilbert^{\,}_{\rm ss} \, , ~~
\end{equation}
includes an ``outcome'' (or ``Stinespring'') register that records the measurement outcome quantumly. 

An observable $\mobserv$ has the spectral decomposition
\begin{equation}
\label{eq:ASpectralDecomp}
    \mobserv \, = \, \sum\limits_{m=1}^{\Noutcome} \, \Eig{m} \, \Proj{\,}{(m)} \, ,~~
\end{equation}
where $\set{ \eig{m} }$ are the ${\Noutcome}$ \emph{unique} eigenvalues of $\mobserv$ (which are also the possible measurement outcomes), and $\Proj{\,}{(m)}$ is a projector onto the eigenstate (or set of eigenstates) of $\mobserv$ with eigenvalue $\eig{m}$. That is, $\mobserv \, \Proj{\,}{(m)} = \eig{m} \Proj{\,}{(m)}$; when the spectrum of $\mobserv$ is nondegenerate, $\Proj{\,}{(m)} = \BKop{m}{m}$ projects onto a single eigenstate of $\mobserv$. The projectors are self adjoint and idempotent,
\begin{equation}
\label{eq:ProjectorHermitianIdempotent}
    \Proj{}{(m)} \, = \, \left( \Proj{}{(m)} \right)^{\dagger} \, = \,  \left( \Proj{}{(m)} \right)^2  \, , ~~
\end{equation}
as well as orthogonal and complete,
\begin{equation}
\label{eq:ProjectorOrthoComplete}
    \Proj{}{(m)} \Proj{}{(n)} \, = \, \kron{m,n}\Proj{}{(n)} \, ~~~\text{and}~~~~ \sum\limits_m \,  \Proj{}{(m)} \, = \, \ident \, .~~
\end{equation}
The isometry corresponding to measurement of $\mobserv$ \eqref{eq:ASpectralDecomp} is captured pictorially in Fig.~\ref{fig:manyworlds}. The initial state of the system is $\ket{\psi}$, while the initial state of the observer is trivial (i.e., prior to measurement, the observer's state is one dimensional). Measuring $\mobserv$ \eqref{eq:ASpectralDecomp} leads to
\begin{equation}
    \label{eq:IsoMeas}
    \ket{\psi} \, \to \, \ket{\psi'} \, = \, \sum\limits_{m=1}^{\Noutcome} \, \left( \Proj{\,}{(m)} \ket{\psi} \right) \otimes \ket{m}^{\,}_{\rm ss} \, , ~~
\end{equation}
where the ``ss'' subscript labels the ${\Noutcome}$-dimensional ``Stinespring'' or ``outcome'' register, where $\Noutcome$ is the number of \emph{unique} eigenvalues of $\mobserv$ \eqref{eq:ASpectralDecomp}. The Stinespring states are orthonormal and complete,
\begin{align}
\label{eq:SS complete}
    \inprod{m}{n}^{\vpp}_{\rm ss} \, = \, \kron{m,n} ~~~1 \leq m,n \leq \Noutcome \, ,~~
\end{align}
and the Stinespring state $\ket{m}^{\vpp}_{\rm ss}$ encodes the fact that outcome $\eig{m}$ was observed upon measurement. 

The isometry corresponding to the channel \eqref{eq:IsoMeas} is
\begin{equation}
\label{eq:Meas Iso}
    \isometry^{\vpd}_{[\mobserv]} \, = \, \sum\limits_{m=1}^{\Noutcome} \,  \Proj{\rm ph}{(m)} \otimes \ket{m}^{\,}_{\rm ss}
\end{equation}
which maps the physical state $\ket{\psi} \in \Hilbert^{\,}_{\rm ph}$ to the \emph{dilated} state $\ket{\psi'}\in \Hilbert^{\rm dil} =\Hilbert^{\rm ph} \otimes \Hilbert^{\rm ss} $ upon measuring $\mobserv$ \eqref{eq:ASpectralDecomp}.

Importantly, unlike the Kraus representation of measurement channels, the isometry captured by \eqref{eq:IsoMeas} is \emph{unique}, up to the choice of the precise state of the Stinespring register (``ss'') that reflects the measurement outcome $m$. The Stinespring states need only obey \eqref{eq:SS complete}. 

The standard axioms of quantum mechanics dictate that the post-measurement state (for a given outcome $m$) is simply $\Proj{\,}{(m)} \ket{\psi}$ up to normalization. The Dilation Theorem \cite{Stinespring} then implies that the ``full'' post-measurement state $\ket{\psi'}$ is generically a superposition of \emph{dilated} states wherein the ``collapsed'' physical wavefunction for the system $ \proj{(m)} \ket{\psi}$ is entangled with a bookkeeping (or ``Stinespring'') register that reflects the observation of outcome $m$ (reflected in the Stinespring register in \eqref{eq:IsoMeas} and Fig.~\ref{fig:manyworlds}). 

Because \eqref{eq:IsoMeas} is isometric, there is no further need to normalize the wavefunction. Explicitly, we have
\begin{align}
    \inprod{\psi'}{\psi'} \, &= \, \sum\limits_{m,n=1}^{\Noutcome} \, \matel*{\psi}{\Proj{\,}{(m)}\Proj{\,}{(n)}}{\psi} \otimes \inprod{m}{n} \notag \\ 
    &= \, \sum\limits_{m=1}^{\Noutcome} \matel*{\psi}{\Proj{\,}{(m)}}{\psi} \, = \, \inprod{\psi}{\psi} \, = \, 1 \, ,~~\label{eq:IsoPostMeasNorm}
\end{align}
where we used completeness of the projectors to resolve the identity. Note that the explicit renormalization of the wavefunction required in the Copenhagen description of measurement is cancelled by normalization of the Stinespring part of 
\eqref{eq:IsoMeas}.

\subsection{Unitary measurement} 
\label{subsec:SSUnitary}
We now move beyond the isometric representation \eqref{eq:IsoMeas} to a more powerful formulation. Note that the isometric channel $\isometry$ defined in \eqref{eq:Meas Iso} corresponding to the measurement of $\mobserv$ \eqref{eq:ASpectralDecomp} is unique (up to the choice of basis \eqref{eq:SS complete} for the Stinespring register), and the Stinespring Dilation Theorem \cite{Stinespring} guarantees that it accurately captures the measurement of $\mobserv$ \eqref{eq:ASpectralDecomp}. Choi's Theorem \cite{ChoisThm} guarantees that the isometric map \eqref{eq:IsoMeas} is equivalent to the Kraus representation of measurements. We stress that, like Kraus operators, the isometric channel $\isometry \, : \, \ket{\psi} \to \ket{\psi'}$ \eqref{eq:Meas Iso} is merely a bookkeeping tool for representing measurements.

However, the key advantage to our Stinespring formalism is \emph{not} the isometric representation \eqref{eq:Meas Iso}, but the \emph{unitary} formalism that it implies. Importantly, any isometry that maps between Hilbert spaces of the same dimension is unitary by definition. This also means that isometries can always be embedded in unitaries. Regarding \eqref{eq:IsoMeas}, this is accomplished by including the Stinespring (outcome) registers from the outset. Doing so requires that we identify a ``default'' Stinespring state; without loss of generality, we choose this to be $\ket{0}$, so that the unitary measurement channel $\measunitary$ maps 
\begin{equation}
   \label{eq:UniMeas}
   \Umeas{\mobserv} \, : \, \ket{\psi}^{\vpd}_{\rm ph} \otimes \ket{0}^{\vpd}_{\rm ss} \, \mapsto \, \sum\limits_{m=1}^{\Noutcome} \, \left( \Proj{\,}{(m)} \ket{\psi} \right)^{\vpd}_{\rm ph} \otimes \ket{m}^{\vpd}_{\rm ss} \, ,~~
\end{equation}
which is uniquely satisfied by the dilated unitary operator 
\begin{equation}
    \label{eq:Meas Uni}
    \Umeas{\mobserv} \, = \, \sum\limits_{m=1}^{\Noutcome} \,  \Proj{\rm ph}{(m)} \otimes \SShift{m}{\rm ss} \,, ~~
\end{equation}
where $\SShift{\,}{\rm ss}$ acts on the Stinespring register as the $\Noutcome^{\,}_i$-state Weyl $\shift{}$ operator \eqref{eq:ShiftOpDef}. The Weyl operators $\sshift{}$ and $\sweight{}$ are defined in App.~\ref{app:OperatorGym}, and unitarily extend the Pauli $\PX{}$ and $\PZ{}$ operators to allow for $\Noutcome > 2$ unique measurement outcomes. Specifically, $\SShift{m}{\rm ss}$ shifts the state of the Stinespring register by $m$ modulo $\Noutcome$, taking the default Stinespring state $\ket{0}$ to the state $\ket{m}$ in \eqref{eq:IsoMeas}. One can easily check that the operator $\measunitary$ \eqref{eq:Meas Uni} is unitary by using the orthogonality of the projectors \eqref{eq:ProjectorOrthoComplete}. 

Advantages of the unitary representation \eqref{eq:Meas Uni} include (\emph{i}) not ``spawning'' new Stinespring registers and (\emph{ii}) not needing to renormalize the density matrix with each measurement. However, a far greater advantage results from recognizing that the Stinespring register (labelled ``ss'') \emph{physically} corresponds to the state of the \emph{measurement apparatus} \cite{AaronDiegoFuture}. In this sense, the unitary $\measunitary$ \eqref{eq:Meas Uni} is not merely a formal bookkeeping tool, but represents the actual time evolution of the combined system and measurement apparatus during the projective measurement. A key achievement of our formalism is that $\measunitary$ \eqref{eq:Meas Uni} can be used to evolve \emph{operators} under a projective measurement.

For example, one measures $\PZ{}$ on an atomic qubit (where $\ket{0}$ denotes the atom's ground state and $\ket{1}$ corresponds to an excited states) using fluorescent measurement \cite{AaronDiegoFuture,SpeedLimit}. The apparatus is a photon detector, which is initialized in the state $\ket{0}^{\,}_{\rm ss}$, and transitions to the state $\ket{1}^{\,}_{\rm ss}$ upon detecting a photon (i.e., a ``click'' outcome in which, e.g., a photon excites an electron in a nickel atom in the detector). Fluorescent light is shone on the qubit atom, and if the atom is in the ground state $\ket{0}$, no photons are absorbed are emitted, while if the atom is in the excited state $\ket{1}$, a photon is absorbed and another emitted and detected by the apparatus. This is precisely what is represented by \eqref{eq:Meas Uni} in the $\LocDim=2$ limit, and generalizes to other projective measurements \cite{AaronDiegoFuture}.

For generality, consider a dilated (and possibly adaptive) hybrid protocol $\evo$ consisting of both unitary time evolution and projective measurements, where the choice and location of gates of both types may depend on the outcomes of prior measurements. Because the protocol $\evo$ is known, the set of all observables that \emph{might} be measured while applying $\evo$ is also known (where the specific sequence of measurements may be adaptively conditioned on the ``trajectory'' of outcomes). We define the set $\set{ \, \mobserv^{\,}_{i} \, | \, 1 \leq i \leq \Nmeas \,}$ containing the $\Nmeas$ ``protocol observables'' that may be measured in the application of $\evo$. Correspondingly, we create $\Nmeas$ Stinespring (or ``outcome'') registers, where the $i$th Stinespring register has  $\Noutcome^{\,}_i$ states corresponding to the $\Noutcome^{\,}_i$ unique eigenstates of the measured observable $\mobserv^{\,}_i$. These registers reflect the states of the $\Nmeas$ distinct measurement apparati used\footnote{In practice, the same apparatus can be used for multiple measurements (after recording the outcome and resetting the apparatus); theoretically, it is more convenient (and equivalent) to imagine assigning a different apparatus to each measurement.}.

We now work in the ``dilated Hilbert space'' \eqref{eq:Dilated Hilbert}, corresponding to all physical and Stinespring registers. By convention, we initialize all Stinespring registers in the ``default'' state $\ket{0}^{\,}_{{\rm ss},i}$; the initial \emph{dilated} density matrix is
\begin{equation}
    \label{eq:DilatedInitialState}
    \DensMatSS \, = \, \DensMat^{\,}_{\rm ph} (0) \otimes \left[ \bigotimes\limits_{i=1}^{\Nmeas} \,  \BKop{0}{0}^{\,}_{{\rm ss},i} \, \right] \, ,~~
\end{equation}
where $\DensMat^{\,}_{\rm ph} (0)$ is the initial density matrix for the \emph{physical} degrees of freedom at $t=0$.

For convenience, we define the shorthand ``vector'' notation for the states of all Stinespring registers,
\begin{align}
    \label{eq:OutcomeVector}
    \BKop{\bvec{n}}{\bvec{n}}^{\,}_{\rm ss} \, &\equiv \, \bigotimes\limits_{i=1}^{\Nmeas} \, \BKop{n^{\,}_{i}}{n^{\,}_{i}}^{\,}_{{\rm ss},i} \,, ~~
\end{align}
where the vector $\bvec{n}$ contains the $\Nmeas$ outcomes $\set{ n^{\,}_{i} }$.

With $\Nmeas>1$ measurements, the unitary measurement of the observable $\mobserv^{\,}_i$ \eqref{eq:ASpectralDecomp} is denoted by
\begin{equation}
    \label{eq:UnitaryMeas1}
    \measunitary^{\vpd}_{i} \, = \, \Umeas{\mobserv^{\,}_i}  \, = \, \sum\limits_{m=0}^{\Noutcome^{\,}_i-1} \, \Proj{\rm ph}{(m)} \otimes \SShift{m}{\Noutcome^{\,}_i;i} \, , ~~
\end{equation}
where $\SShift{\,}{\Noutcome^{\,}_i;i}$ is a $\Noutcome^{\,}_i$-state Weyl shift operator \eqref{eq:ShiftOpDef} that acts only on the Stinespring register $i$ with $\Noutcome^{\,}_i$ unique states, and $\Proj{\rm ph}{(m)}$ is defined in \eqref{eq:ASpectralDecomp}. 

The key insight, which furnishes numerous results in the remainder, is that the unitary $\measunitary$ \eqref{eq:UniMeas} is not merely a bookkeeping tool, but reflects the actual, physical time evolution of the system of interest and measurement apparatus \cite{AaronDiegoFuture}. The identification of the measurement unitary with time evolution allows for evolution of operators in the Heisenberg picture and the analysis of spectral properties in the presence of measurements. This holds for measurements of qubit systems, and appears to hold even for the measurement of unbounded observables (i.e., in photonic systems), and other systems \cite{AaronDiegoFuture}.

\subsection{Measurement outcomes}
\label{subsec:SSTrajectory}
The Stinespring formulation simplifies the recovery of various quantities involving the outcomes of one or more projective measurements. For example, consider the Stinespring description of the measurement of a single observable $\mobserv$ \eqref{eq:ASpectralDecomp}, where the state of the physical system is initially given by the density matrix $\DensMat^{\,}_{\rm ph}$ as in \eqref{eq:DilatedInitialState}.

Independent of the formalism used to describe measurement, considering only the physical Hilbert space, the expectation value of $\observ$ is given by
\begin{align}
    \expval{\mobserv}^{\,}_{\DensMat} \, &=  \, \tr{ \, \mobserv \, \DensMat \,} \, = \,   \sum\limits_{m=1}^{\Noutcome} \, \Eig{m} \, p^{\,}_m  \, , ~~    \label{eq:AExpVal0}
\end{align}
where the probability to obtain outcome $m$ depends on the density matrix $\DensMat^{\,}_{\rm ph}$ according to
\begin{align}
    \label{eq:ProbN}
    p^{\,}_m \, = \, \tr{ \, \proj{m} \, \DensMat^{\,}_{\rm ph} \,} \, , ~~
\end{align}
and the normalized post-measurement wavefunction after obtaining outcome $m$ is given by
\begin{equation}
    \label{eq:PostMeasNstate}
    \ket{\psi^{\,}_m} \, = \, \frac{\Proj{\,}{(m)} \ket{\psi}}{~~~~\matel{\psi}{\Proj{\,}{(m)}}{\psi}^{1/2}} \, ,~~
\end{equation}
with the corresponding density matrix given by
\begin{equation}
    \label{eq:PostMeasNDensMat}
    \DensMat^{\,}_m \, = \, \frac{\Proj{\,}{(m)} \DensMat \,  \Proj{\,}{(m)}}{\tr{\, \Proj{\,}{(m)} \, \DensMat^{\,}_{\rm ph} \,}} \, ,~~
\end{equation}
which is also correctly normalized, with $\DensMat^{\,}_m = \BKop{\psi^{\,}_m}{\psi^{\,}_m}$. 

In the \emph{dilated} Hilbert space \eqref{eq:Dilated Hilbert}, using the notation \eqref{eq:OutcomeVector}, the dilated density matrix $\DensMatSS$ is given by
\begin{equation*}
    \tag{\ref{eq:DilatedInitialState}}
    \DensMatSS \, = \, \DensMat^{\,}_{\rm ph} \otimes \BKop{0}{0}^{\,}_{\rm ss} \, , ~~
\end{equation*}
and we define expectation values in the dilated Hilbert space $\Hilbert^{\,}_{\rm dil}$ according to 
\begin{align}
    \expval{ \,  \mobserv^{\vpp}_{\rm ph} \otimes \mobserv^{\prime}_{\rm ss} \, }^{\,}_{\DensMatSS} \, &= \, \tr{ \, \big( \, \mobserv^{\vpp}_{\rm ph} \otimes \mobserv^{\prime}_{\rm ss} \, \big) \, \DensMatSS \,} 
    \, .~~ \label{eq:SSExpValDef}
\end{align}
For example, the probability $p^{\,}_m$ to obtain outcome $m$ is given in terms of a dilated expectation value \eqref{eq:SSExpValDef} as
\begin{align}
    p^{\,}_n \, &= \, \expval{ \, \ident^{\,}_{\rm ph} \otimes \BKop{n}{n}^{\,}_{\rm ss} \, }^{\,}_{\DensMatSS} \notag \\
    &= \,\tr{ \, \big( \, \ident^{\,}_{\rm ph} \otimes \BKop{n}{n}^{\,}_{\rm ss}\, \big) \, \DensMatSS \,} \,, ~~ \label{eq:DilatedProbN}
\end{align}
and the expectation value for a Stinespring-measured observable $\mobserv$ \eqref{eq:ASpectralDecomp} can similarly be written as
\begin{align}
    \expval{\mobserv}^{\,}_{\DensMatSS} \, &= \, \expval{ \, \ident^{\,}_{\rm ph} \otimes \sum\limits_{m=1}^{\Noutcome} \, \Eig{m} \, \BKop{m}{m}^{\,}_{\rm ss}~}^{\,}_{\DensMatSS} \, = \, \sum\limits_{m=1}^{\Noutcome} \, \Eig{m} p^{\,}_m \notag \\
    &= \, \tr{ \,  \big( \, \ident^{\,}_{\rm ph} \otimes \sum\limits_{m=1}^{\Noutcome} \, \Eig{m} \, \BKop{m}{m}^{\,}_{\rm ss}\, \big) \, \DensMatSS \,}\, ,~~\label{eq:DilatedExpA}
\end{align}
and we note that all information about the measurement outcome is stored in the Stinespring register.

The preceding formulae for $p^{\,}_n$ \eqref{eq:DilatedProbN} and $\expval{\mobserv}^{\,}_{\DensMatSS}$ \eqref{eq:DilatedExpA} generalize straightforwardly to the case of multiple measurements by  including multiple Stinespring registers. The $\Nmeas$ outcomes of measuring the observables $\set{\, \mobserv^{\,}_i \, }$ are stored in the outcome vector $\bvec{n} = \{ n^{\,}_1 , \dots , n^{\,}_{\Nmeas} \} $ \eqref{eq:OutcomeVector}. After any number of measurements, the dilated density matrix $\DensMatSS$ is a sum over terms of the form $\BKop{\bvec{n}}{\bvec{m}}$, which maps the outcome states $\set{\, m^{\,}_i \,}$ to the outcome states $\set{\, n^{\,}_i \,}$. In the foregoing formulae, one simply replaces $\BKop{n}{n}$ with $\BKop{\bvec{n}}{\bvec{n}}$ \eqref{eq:OutcomeVector}. For concreteness, the joint probability to obtain outcomes $\bvec{n}$ is given by
\begin{align}
    p^{\,}_{\bvec{n}} \, &= \, \expval{ ~ \ident^{\,}_{\rm ph} \otimes \BKop{\bvec{n}}{\bvec{n}}^{\,}_{\rm ss} ~}^{\,}_{\DensMatSS} \notag \\
    &= \,\tr{ \, \big( \, \ident^{\,}_{\rm ph} \otimes \BKop{\bvec{n}}{\bvec{n}}^{\,}_{\rm ss}\, \big) \, \DensMatSS (t)\,} \,, ~~ \label{eq:ProbNvec}
\end{align}
and all other joint and conditional expectation values follow straightforwardly from the various definitions above.

With multiple rounds of measurements, it may be useful to restrict various quantities to particular outcome \emph{trajectories} (or alternatively, average over all outcomes) of a \emph{subset} $\MSites$ of Stinespring registers, while leaving the rest of the density matrix (or some operator) intact. The density matrix (or any operator) is projected onto a particular outcome trajectory $\bvec{n}$ by the operator
\begin{equation}
    \label{eq:SSTrajectoryProjector}
    \SSProj{\bvec{n}} \, \equiv \, \ident^{\,}_{\rm ph} \otimes \BKop{\bvec{n}}{\bvec{n}}^{\vpp}_{\rm ss} \, , ~~
\end{equation}
which may correspond to a subset $\MSites$ of outcomes. The normalized density matrix following \eqref{eq:SSTrajectoryProjector} is then
\begin{equation}
    \DensMatSS^{\,}_{\bvec{n}} \, = \, \frac{\underset{\MSites}{\trace} \left[ \, \DensMatSS (t) \, \BKop{\bvec{n}}{\bvec{n}} \, \right]}{\underset{\rm dil}{\trace} \left[ \, \DensMatSS (t)\, \BKop{\bvec{n}}{\bvec{n}} \, \right]} \, , ~~
    \label{eq:SSDensMatTrajectory}
\end{equation}
where the denominator is simply $p^{\,}_{\bvec{n}}$ \eqref{eq:ProbNvec}. The trace in the numerator runs over 
$\MSites$ (where the corresponding measurements have already occurred). Note that \eqref{eq:SSDensMatTrajectory} reproduces \eqref{eq:PostMeasNDensMat} for a single measurement, while \eqref{eq:SSDensMatTrajectory} gives the post-measurement density matrix following a particular sequence of outcomes.

We can also compute the probability-weighted average over all  outcomes. This involves summing \eqref{eq:SSDensMatTrajectory} over all $\bvec{n}$, weighted by $p^{\,}_{\bvec{n}}$, which cancels the denominator in \eqref{eq:SSDensMatTrajectory}. Summing the numerator over $\bvec{n}$ leads to $\sum_{\bvec{n}} \BKop{\bvec{n}}{\bvec{n}} = \ident$ by completeness. The average over the subset of outcomes $\MSites$ is then
\begin{equation}
    \label{eq:DensMatAvgOutcomes}
    \DensMatSS^{\vpd}_{\MSites} (t) \, = \, \Emean{\DensMatSS (t)}^{\,}_{\MSites} \, = \, \sum\limits_{\vec{n}} \, p^{\,}_{\vec{n}} \, \DensMatSS^{\,}_{\vec{n}} \, = \, \underset{\MSites}{\trace} \left[ \, \DensMatSS (t) \, \right] \,,~~
\end{equation}
where the ``arrow'' vector symbol indicates that the set $\vec{n} \subset \bvec{n}$ is a subset of all the Stinespring registers. In general, \eqref{eq:DensMatAvgOutcomes} gives the reduced density matrix for the physical degrees of freedom and any remaining Stinespring registers over whose outcomes we do not [yet] wish to average. If $\MSites$ includes all outcome registers, then $\DensMatSS^{\,}_{\MSites} \to \DensMat^{\,}_{\rm ph}$ is the density matrix for the physical system averaged over all measurement outcomes.

\subsection{Comment on particular trajectories}
\label{subsec:NoTrajectories}
Considering calculations involving monitored quantum systems in the Stinespring formalism, the two options presented so far for treating the Stinespring (outcome) degrees of freedom are to (\emph{i}) project this object onto a particular outcome trajectory, via \eqref{eq:SSTrajectoryProjector} or (\emph{ii}) average over all possible outcomes via \eqref{eq:DensMatAvgOutcomes}. We now motivate our restriction to the latter case in the remainder.

Importantly, because the recorded outcome of generic quantum measurements performed in some state $\ket{\Psi}$ are always random (as seen in Fig.~\ref{fig:manyworlds}), models of quantum systems are generally unable to make deterministic predictions about the outcome of any particular experimental ``shot''. However, the probability $p^{\,}_m$ to obtain the $m$th outcome \eqref{eq:ProbN} can be determined from the initial state $\ket{\Psi}$; accordingly, one can make predictions about \emph{expectation values} of measurements, and statistics associated with higher moments (or cumulants). 

These statistics depend on the particular details of the many-body wavefunction $\ket{\Psi}$, which cannot be determined in a single experiment. Recovering statistics for measurement outcomes therefore generically requires many experimental ``shots'' to synthesize. Importantly, for the statistics to be meaningful, each shot should use the same many-body state $\ket{\Psi}$. However, measuring the state to extract statistics inherently disturbs (i.e., changes) it. Since the wavefunction cannot be accessed in its entirety (nor could that information be stored classically in the thermodynamic limit), and because the no-cloning theorem precludes copying the many-body state prior to a measurement, the same state $\ket{\Psi}$ must be prepared from scratch in each experimental shot. 

Additionally, in performing the experiment, one has no control over the outcome of a given measurement. Thus, realizing a particular measurement trajectory $\bvec{n}$ multiple times requires a number of experimental shots that scales as $\Order{\LocDim^{\measrate \, \Nsite \, \tfin}}$, where $\measrate$ is the measurement rate, $\Nsite \sim \size^{\SpaceDim}$ is the number of sites (in $\SpaceDim$ spatial dimensions with linear size $\size$), and $\Nsite \, \tfin$ is the total spacetime volume of the circuit, which becomes infinite in the thermodynamic limit of interest. This is also known as the ``postselection problem'' \cite{RUCreview}, which we discuss in Sec.~\ref{sec:can't work} when we discuss the salient features of genuine phases of matter.

However, more familiar phases of matter (e.g., in the context of condensed matter or AMO systems) do not seem to require preparing multiple identical versions of the same state across a divergent number of experimental shots. This is due in part to the fact that, in a stable phase of matter, while measuring the system disturbs its state, it does not take the state out of the phase. Hence, one can still extract universal features---which are by definition insensitive to microscopic variations in particular samples and derive from a coarse-grained description of any sample in the phase---without reconstructing the state from scratch. Importantly, the universal probes have low variance from sample-to-sample, providing a notion of \emph{typicality}, which allows one to extract statistics without performing infinitely many shots. 

The quantities we consider in Sec.~\ref{sec:OneCopy} (and later in Sec.~\ref{sec:adaptive}) have this property. One can quickly check that the variances of generic observables under Haar-random evolution is exponentially small in the system size. Thus, the observables measured in the application of the circuit---as well as the probe observables used to diagnose phase structure---enjoy a notion of typicality. Hence, the statistics for some quantity averaged over  the Haar ensemble and measurement outcomes is well approximated by the statistics of a finite number of independent shots spanning a vanishing fraction of all possible realizations. 

However, the same is not possible along a particular trajectory $\bvec{n}$, where even two shots may require infinite time, and typicality cannot be invoked. After all, there is no notion of ``approximate'' postselection. For example, low-temperature equilibrium states are well captured by ground-state properties, as the low-temperature state has an exponentially large probability of sampling ground-state properties. In contrast, the probability that a random shot successfully produces the desired trajectory $\bvec{n}$ is exponentially small. We conclude that quantities evaluated with respect to \eqref{eq:SSTrajectoryProjector} are not experimentally viable, while quantities evaluated with respect to \eqref{eq:DensMatAvgOutcomes} are.

\section{Hybrid circuit models} 
\label{sec:hybridcirc}
We now detail the treatment of random quantum circuit protocols including measurements in the Stinespring formalism, both with and without symmetries. In keeping with the spirit of standard quantum circuits, we restrict our measurements to \emph{local} observables---i.e., measurements should act on finitely many neighboring sites.

\subsection{The spacetime lattice}
\label{subsec:SpacetimeLattice}
We begin by constructing the ``spacetime lattice'', which defines the dilated Hilbert space $\Hilbert^{\,}_{\rm dil}$ \eqref{eq:Dilated Hilbert} for a given hybrid circuit. We consider systems with $\Nsite$ total $\LocDim$-state qudits (with on-site Hilbert space $\Hilbert^{\,}_j = \Comps^{\LocDim}$), so that
\begin{align}
    \label{eq:Hilbert Phys}
    \Hilbert^{\,}_{\rm ph} \, = \, \bigotimes\limits_{j=1}^{\Nsite} \, \Hilbert^{\,}_j \, = \, \left( \Comps^{\LocDim} \right)^{\otimes \, \Nsite} \, ,~~
\end{align}
where, in $\SpaceDim$ spatial dimensions, $\Nsite \sim \size^{\SpaceDim}$.

We now consider the Stinespring degrees of freedom. For the \emph{isometric} representation of measurement discussed in Sec.~\ref{subsec:SSIsometric}, the additional Stinespring registers are ``spawned'' in real time as needed, and there is no need to codify the spacetime lattice. However, the \emph{unitary} implementation of measurement discussed in Sec.~\ref{subsec:SSUnitary} requires that all outcome registers be specified from the outset, and prepared in the initial state $\BKop{\bvec{0}}{\bvec{0}}^{\,}_{\rm ss}$ \eqref{eq:DilatedInitialState}. The prescription below for forming the spacetime lattice in some cases generates more Stinespring sites than are actually ``used'' by the circuit, to allow for the possibility that the choice of measurements at a given time depend on prior measurement outcomes. However, we note that tracing over any unused outcome registers (still in the state $\BKop{0}{0}$) gives the same result as not having included the register to begin with. Thus, the spacetime lattice contains all the degrees of freedom that one \emph{might} need for a particular calculation involving measurements, providing a convenient means of labelling outcome registers and implementing unitary measurement.

We first suppose that the hybrid protocol involves $\tfin$ total time steps, where we may take $\tfin \to \infty$. We next require that each time step contain at most $\MeasRounds$ ``rounds'' of measurements, labeled $\sigma$, where a ``round'' corresponds to a layer of nonoverlapping measurement gates. The number $\MeasRounds^{\,}_t$ of measurement rounds may vary between time steps, and the total number of rounds is
\begin{equation}
    \label{eq:Total meas rounds}
    \MeasRounds^{\,}_{\rm tot} \, = \, \sum\limits_{t'=1}^{\tfin} \, \MeasRounds^{\,}_{t'}  \, , ~~
\end{equation}
where $\MeasRounds^{\,}_{t'}$ the number of layers in time step $t'$. 

The measurement layer $t,\sigma$ of the circuit involves the measurement of some set $\MSites^{\,}_{t,\sigma}$ of observables labeled $\mobserv^{\,}_{t,\sigma,r}$, which may have different numbers of unique eigenvalues ($\Noutcome^{\,}_{t,\sigma,r}$), and act nontrivially on clusters $r$ containing any finite number of neighboring sites. In general, the spacetime lattice is compatible with \emph{adaptive} protocols, in which the composition of the circuit at time $t$ may be conditioned on the outcomes of prior measurements.

The definition of the layers in terms of nonoverlapping measurements implies that each physical site $j$ encounters \emph{at most one} measurement per round (i.e. at most $\Nsite$ observables per round). If a protocol na\"ively requires two or more consecutive measurements involving site $j$, one simply splits these measurements into two separate rounds, without loss of generality. Consequently, it is possible to label the observables $\mobserv^{\,}_{t,\sigma,j}$ unambiguously by, e.g., the leftmost site on which $\mobserv^{\,}_{t,\sigma,r}$ acts \cite{NahumOperator, RUCconTibor, RUCconTibor, RUCconVedika}); however, we  also use the label $r$ when convenient.

We further assume that the observables $\mobserv^{\,}_{t,\sigma,j}$ have at most $\LocDim$ unique eigenvalues.  This assumption holds for all string-like observables that can be measured directly in experiment---in general, operators that have more than $\LocDim$ unique outcomes correspond to superpositions of observables that can be measured directly (e.g., a local Hamiltonian term in a spin chain). Such expectation values are formed by making numerous individual measurements, each with at most $\LocDim$ unique eigenvalues a piece. However, such a procedure is not possible midway through a hybrid circuit protocol. Thus, the assumption that $\Noutcome^{\,}_{t,\sigma,r} \leq \LocDim$ is fully compatible with the allowed circuit observables.

Hence, because there are at most $\LocDim$ outcomes per measured observable, at most $\Nsite$ observables per round, and at most $\MeasRounds$ rounds per time step, we can store the outcomes of measuring the circuit observables $\mobserv^{\,}_{t,\sigma,j}$ using a total of $\Nsite \cdot \MeasRounds \cdot \tfin$ Stinespring sites with $\LocDim$ internal states each. If $\mobserv^{\,}_{t,\sigma,j}$ has $\Noutcome^{\,}_{t,\sigma,j} < \LocDim$  unique outcomes, we use only the lowest $\Noutcome^{\,}_{t,\sigma,j}$ levels of the Stinespring register.  

The composition of the \emph{spacetime lattice} follows from the discrete-time nature of the circuit along with the foregoing considerations. The result is a $\SpaceDim+1$-dimensional lattice with $ \size^{\SpaceDim} \, \times \, \left( \MeasRounds^{\,}_{\rm tot}  + 1\right)$ vertices labeled ``$j,\tau$'', where $j$ runs over physical sites. The last axis of the
lattice correspond to time: The temporal ``slices'' $\tau$ contain $\Nsite$ $\LocDim$-state qudits. The $\SpaceDim$-dimensional slice $\tau=0$ corresponds to the $\LocDim$-state \emph{physical} degrees of freedom in $\Hilbert^{\,}_{\rm ph}$. The slice $\tau$ corresponds to measurement layer $\sigma$ of time step $t$:
\begin{equation}
    \label{eq:Combined SS label}
    \tau (t, \sigma ) \, = \, \sum\limits_{t'=1}^{t-1} \, \MeasRounds^{\,}_{t'}  + \sigma \, ,~~
\end{equation}
and contains $\Nsite$ qudits to encode all possible measurement outcomes. The Stinespring site $j,\tau$ stores the outcome of measuring  $\set{ \mobserv^{\,}_{t,\sigma,j}}$ in layer $\sigma$ of time step $t$, and  requires $\Noutcome^{\,}_{t,\sigma,j}$ internal states (the number of unique eigenvalues of $\mobserv^{\,}_{t,\sigma,j}$). In principle, the result is recorded either in the first $\Noutcome^{\,}_{\varsigma,j}$ states $0,1,\dots,\Noutcome^{\,}_{\varsigma,j}-1$ of an $\LocDim$-state qudit, or equivalently, in an $\Noutcome^{\,}_{\varsigma,j}$-state qudit at $\varsigma,j$. In practice, this distinction is unimportant.

The Stinespring Hilbert space is given in terms of the $\tau > 0$ slices of the spacetime lattice as
\begin{align}
    \label{eq:Hilbert ss general}
    \Hilbert^{\,}_{\rm ss} \, = \, \bigotimes\limits_{\tau=1}^{\tfin} \, \bigotimes\limits_{\sigma=1}^{\MeasRounds^{\vpp}_t} \, \bigotimes\limits_{j=1}^{\Nsite} \,  \Comps^{\Noutcome^{\vpp}_{\tau,\sigma,j}}  \, ,~~
\end{align}
where $\Noutcome^{\vpp}_{j,\tau}$ is the number of unique eigenvalues of the corresponding observable $\observ^{\,}_{j,\tau}$ (and defaults to unity if no measurement is performed---i.e., a trivial outcome register). For convenience, we embed \eqref{eq:Hilbert ss general} in
\begin{align}
    \label{eq:Hilbert ss q}
    \Hilbert^{(\LocDim)}_{\rm ss} \, \equiv \,   \left( \Comps^{\LocDim } \right)^{\otimes \Nsite \, \left( \MeasRounds^{\,}_{\rm tot} +1\right) }  \, ,~~
\end{align}
so that all degrees of freedom in $\Hilbert^{\,}_{\rm dil}$ are $\LocDim$-state qudits.

\subsection{Measurement unitaries}
\label{subsec:MeasCirc}
Hybrid circuits evolve a quantum system using a combination of 
time-evolution and projective measurements. The former are generated by local unitary gates acting on the physical degrees of freedom \eqref{eq:Hilbert Phys}; the latter are represented via unitary (or isometric) gates according to the Stinespring formulation outlined in Sec.~\ref{sec:Stinespring}. The measurement gates in layer $\sigma$ of time step $t$ act on---and most importantly, entangle---the physical qudits in the slice $\tau=0$ \eqref{eq:Hilbert Phys} and the Stinespring qudits in the slice $\tau=t,\sigma$ \eqref{eq:Hilbert ss q} of the spacetime lattice (discussed in Sec.~\ref{subsec:SpacetimeLattice}). Their Hilbert space is given by \eqref{eq:Hilbert ss general}, and may be embedded in the simplified spacetime lattice \eqref{eq:Hilbert ss q} for convenience.

Following the literature \cite{og-MIPT, FisherMIPT1, chan, FisherMIPT2, ChoiQiAltman, RomainHolographic2}, we first consider \emph{nonadaptive} circuits, where the outcomes of measurements do not affect the selection of future gates (we survey the landscape of adaptive hybrid circuits in Sec.~\ref{sec:adaptive}). The primary effect of projective measurements in nonadaptive hybrid circuits is to purify the state, destroying entanglement generated via unitary time evolution. Accordingly, the hybrid circuits of interest \cite{og-MIPT, FisherMIPT1, chan, FisherMIPT2, ChoiQiAltman, RomainHolographic2} typically alternate between time-evolution gates and single- and two-site measurements. The measurements are generally probabilistic: An observable $\mobserv^{\,}_{t,\sigma,j}$ is measured on site $j$ in layer $\sigma$ of time step $t$ with probability $\measrate$, while no measurement is made with probability $1-\measrate$, independent of all other sites and prior time steps. One then looks for a sharp feature in the expectation value of some quantity as the parameter $\measrate$ is tuned from zero to unity. 

In this work we consider more generic hybrid protocols, which may have multiple rounds of measurement per time step, with observables of various bases, ranges, and degeneracies in each round. We require that each physical site $j$ be measured at most once per round, reflecting the constraints of actual experiments. We restrict the circuit observables $\mobserv$ to those that can be measured in a single shot (e.g., precluding the measurement of sums of operators). We also assume that the number of unique outcomes $\Noutcome^{\,}_{t,\sigma,r}$ for each observable is at most $\LocDim$.

We now consider projective measurements in the context of the circuit. The measurement of  $\mobserv^{\,}_{t,\sigma,r}$ on cluster $r$ in layer $\sigma$ of time step $t$ is represented by the unitary,
\begin{align}
    \label{eq:MeasUnitaryGate}
    \measunitary^{\vpd}_{t,\sigma,r} \, &\equiv \, \sum\limits_{m=0}^{\Noutcome-1} \,  \Proj{r,0}{(m)} \, \otimes \, \SShift{m}{\Noutcome;r,t} \, ,~~
\end{align}
where the ``$t,\sigma,r$'' indices on $\Noutcome$ have been omitted for visual convenience and $\Proj{r,0}{(m)}$ projects the \emph{physical} cluster $r,0$ onto the $m$th eigenspace of $\mobserv^{\,}_{t,\sigma,r}$ \eqref{eq:ASpectralDecomp}. The shift operator $\SShift{\,}{\Noutcome;j,t}$ \eqref{eq:ShiftOpDef} acts on the first $\Noutcome^{\vpp}_{t,\sigma,r} \leq \LocDim$ states of the \emph{outcome} register at $t,\sigma,r$.

The measurement layer $t,\sigma$ of the circuit is given by
\begin{equation}
    \label{eq:MeasCircLayer}
    \measunitary^{\vpd}_{t,\sigma} \, = \, \hspace{-2mm} \bigotimes\limits_{~~r \in \MSites^{\vpp}_{t,\sigma}}  \hspace{-2mm}  \measunitary^{\vpd}_{t,\sigma,r} \, , ~~
\end{equation}
where $\MSites^{\vpp}_{t,\sigma}$ is the set of sites that participate in measurement layer $\sigma$ of time step $t$. The measurement layers may be placed at any point in the single-time-step circuit relative  the time-evolution layers.

\begin{figure}[t!]
    \centering
    \includegraphics[width=.9\columnwidth]{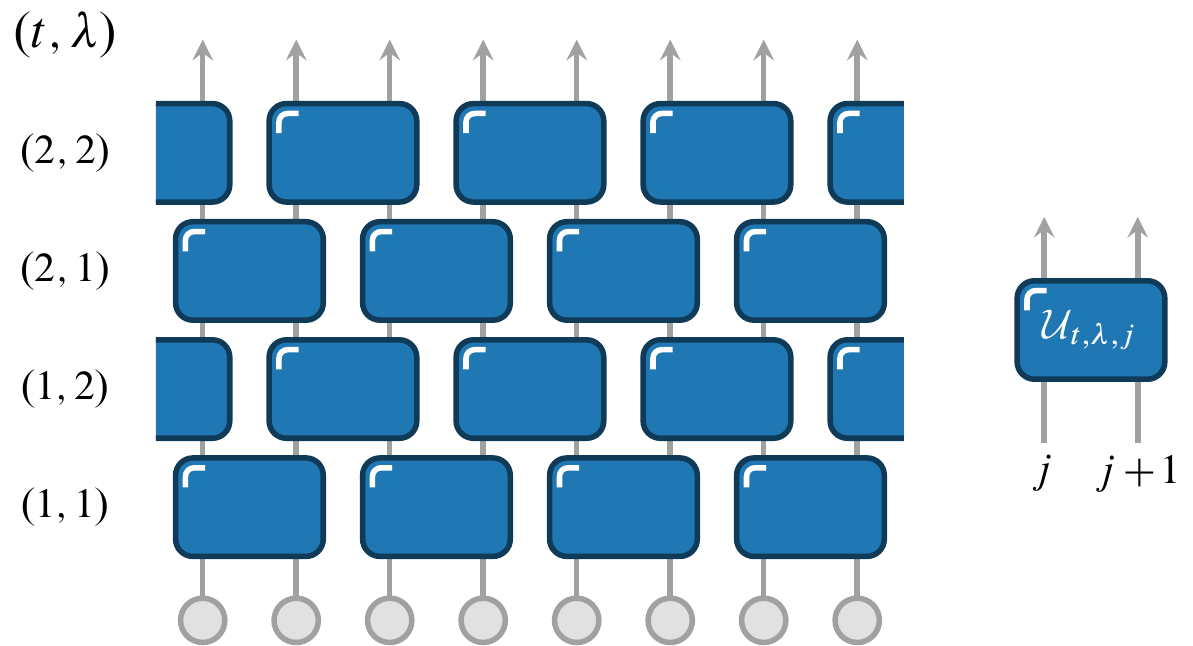}
    \caption{Depiction of a $1\SpaceDim$ brickwork quantum circuit comprising two-site gates (left) and a single two-site unitary gate corresponding to time step $t$, unitary layer $\lambda$, and site $j$ (right). Each time step $t$ consists of $\ell=2$ layers labeled $\lambda$, corresponding to even and odd bonds, as indicated by the labels $(t,\lambda)$. The two-site gates are drawn independently---in both space $j$ and time $t$---from the Haar ensemble \cite{YoshidaCBD}.}
    \label{fig:brickcirc}
\end{figure}

\subsection{Time evolution}
\label{subsec:Tevo}

Time evolution gates act solely on the physical Hilbert space (i.e., the $\tau=0$ slice of the spacetime lattice). Locality of the dynamics is captured by evolving the system using a ``circuit'' of $\ell$-site unitary ``gates''. A local  $1\SpaceDim$ circuit with the default $\ell=2$ is depicted in Fig.~\ref{fig:brickcirc}. 

The unitary gates are \emph{local} in that the gate $\gate^{\,}_{t,\lambda,j}$ (in layer $\lambda$ of time step $t$) acts on a cluster $r$ comprising a contiguous region of $\ell$ neighboring qudits. We then arrange these gates into $\ell$ layers per time step, so that each site $j$ realizes each of the $\ell$ positions in the various gates in each time step. In $1\SpaceDim$, gates are uniquely labeled by the leftmost qubit $j \in r$, and in layer $\lambda$, the label $j$ for the gate $\gate^{\,}_{t,\lambda,j}$  satisfies $j~{\rm mod}~\ell \, = \, \lambda$. In this way, the $\ell$ layers of time-evolution gates tile the system in a brickwork geometry in every time step of the circuit. Fig.~\ref{fig:brickcirc} depicts the $\ell=2$ case, where the two layers alternate between even and odd bonds. Note that extension to higher dimensions is relatively straightforward: The $2\SpaceDim$ analogue of the brickwork circuit involves four layers of four-site gates acting on plaquettes of the square lattice.

The state $\ket{\psi}$ of the physical system is evolved from time $t$ to time $t+1$ according to 
\begin{equation}
    \ket{\psi (t+1)} \, = \, \evo^{\vpd}_t \, \ket{\psi (t)} \, ,~~
    \label{eq:TevoStep}
\end{equation}
where $\evo^{\,}_t$ is a circuit given by the \emph{time-ordered} product
\begin{equation}
    \evo^{\vpd}_t \, = \, \prod\limits_{\lambda=1}^{\ell} \evo^{\vpd}_{t,\lambda} \, , ~~
    \label{eq:TevoLayer}
\end{equation}
where $\evo^{\,}_{t,\lambda}$ is a \emph{layer} of the circuit comprising mutually commuting $\ell$-site unitary gates,
\begin{equation}
    \evo^{\,}_{t,\lambda} \, \equiv \, \bigotimes\limits_{r \in \lambda} \, \gate^{\vpd}_{t,\lambda,r} \, ,~~
    \label{eq:TevoGate}
\end{equation}
where $\gate^{\vpd}_{t,\lambda,r}$ is the gate that acts on the physical sites ($\tau=0$) in cluster $r$. Absent any symmetries or constraints, we draw the gates $\gate^{\,}_{t,\lambda,r}$ independently for each cluster $r$ in each layer $\lambda$ of the circuit from the unitary group $\U{\LocDim^{\ell}}$ with uniform measure (i.e., the gates are Haar-random $\LocDim^\ell \times \LocDim^\ell$ unitaries). We also define the full evolution from the initial state to time $t$ via
\begin{equation}
    \ket{\psi (t)}  \, = \, \evo (t) \, \ket{\psi (0)} ~,~~~ \evo (t) \, \equiv \, \prod\limits_{s=1}^{t-1} \evo^{\vpd}_{s} \, , ~~
    \label{eq:Tevo to t}
\end{equation}
where the product is time ordered. 

Generally speaking, minimal brickwork circuits (e.g., as depicted in Fig.~\ref{fig:brickcirc} for the $1\SpaceDim$ case) are sufficient to reproduce the maximally chaotic physics realized by a $\LocDim^{\Nsite} \times \LocDim^{\Nsite}$ many-body random unitary \cite{YoshidaCBD,RUCNCTibor,CDLC1,RUCconTibor,RUCconVedika,U1FRUC,ConstrainedRUC}.

\subsection{Symmetries, constraints, and block structure}
\label{subsec:Blocktevo}
The circuits of Sec.~\ref{subsec:Tevo} are designed to be fully generic, and therefore lack many ingredients present in typical systems---namely,  \emph{symmetries}. We now extend these generic circuits \eqref{eq:TevoGate} to capture arbitrary Abelian symmetries \cite{RUCconTibor,RUCconVedika,U1FRUC,ConstrainedRUC} and/or kinetic constraints \cite{RitortKCM,garrahan2010kinetically,KCMRydberg,Valado_2016,GarrahanLectures,QuantumEast2020,ConstrainedRUC}. 

We note that the generators of Abelian symmetries can always be represented using a common basis. In general, we take this to be the eigenbasis of the Weyl operator $\weight{\,}$ \eqref{eq:WeightOpDef}, which reduces to the Pauli $Z$ matrix for $\LocDim=2$ (the Weyl operator basis is detailed in App.~\ref{app:OperatorGym}). The local eigenstates of $\weight{\,}$ form the \emph{computational basis}. 

In models with kinetic constraints (but no conservation laws), we express the constraints in the computational basis (the $\weight{\,}$ eigenbasis, without loss of generality) \cite{ConstrainedRUC}. With both symmetries and constraints present, we require that they share a common basis---i.e., that the ``charge'' basis is also the computational basis. The rationale is that, if the constraint were formulated in the $\shift{}$ basis, e.g., while charges correspond to the $\weight{}$ basis, then the projector onto configurations of fixed charge would mix between states that do and do not satisfy the constraint (and vice versa). However, configurations that violate the constraint by definition should be dynamically ``frozen.'' Hence, our results apply to the ``natural'' class of symmetry-compatible constraints---as well as to generic models with \emph{either} Abelian symmetries or constraints \cite{ConstrainedRUC}.

The unitary gate acting on some $\ell$-site cluster $r$ can be written in the block-diagonal form
\begin{align}
    \label{eq:GenProjGate} 
    \gate^{\vpd}_r \, &\equiv \, \sum_{\alpha} \, \Proj{r}{(\alpha)} \, \Haar^{\vpd}_{r,\alpha} \, \Proj{r}{(\alpha)} \, , ~~
\end{align}
where $\Proj{r}{(\alpha)}$ projects onto block $\alpha$ (which contains a set of $n^{\,}_{\alpha} \geq 1$ states of cluster $r$ that are allowed by the symmetries and/or constraints to mix under dynamics). The unitary $\Haar^{\,}_{r,\alpha}$ acts purely within block $\alpha$ \cite{RUCconTibor, RUCconVedika, U1FRUC, ConstrainedRUC}, and is  \emph{independently} drawn for each $\alpha$ from the unitary group $\U{n^{\,}_{\alpha}}$ with uniform (Haar) measure \cite{RUCconTibor, RUCconVedika, YoshidaCBD, U1FRUC, ConstrainedRUC}. 

\begin{figure}[t!]
    \centering
    \includegraphics[width=.85\columnwidth]{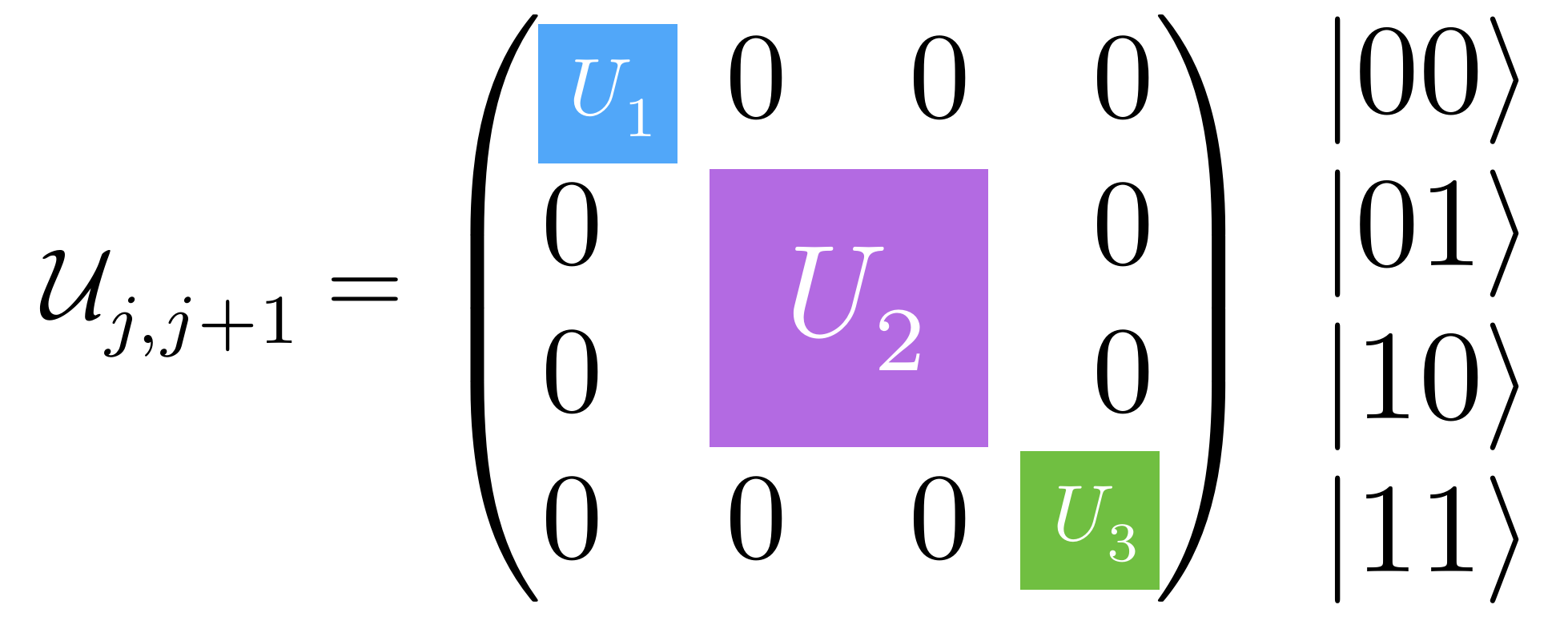}
    \caption{Block-diagonal gate for a $1\SpaceDim$ circuit with $\mathsf{U}(1)$ conservation of $Z$ (or ``charge''), for  $\LocDim = \ell = 2$ . Blocks with different charges cannot mix under dynamics; consequently, the unitary gate acts nontrivially only within blocks of fixed charge, and the unitaries in different blocks are independently drawn. 
    }
    \label{fig:U1gate}
\end{figure}

Unitarity of $\gate^{\,}_r$ \eqref{eq:GenProjGate} requires that the projectors be complete and idempotent,
\begin{align}
    \label{eq:BloccGateProjectorCompleteness}
    \sum_{\alpha} \, \Proj{r}{(\alpha)} \, = \, \ident \,,~\quad~ \Proj{r}{(\alpha)}  \Proj{r}{(\alpha')}\,=\, \kron{\alpha,\alpha'}  \Proj{r}{(\alpha)} \, ,~~
\end{align}
and each $\Proj{r}{(\alpha)}$ can be written as a sum over projectors  onto particular $\ell$-site computational-basis configurations of $r$ that belong to the block $\alpha$, 
\begin{align}
    \label{eq:blockprojectorsum}
    \Proj{r}{(\alpha)} \, = \, \sum\limits_{\vec{a} \in \alpha} \, \prod\limits_{j \in r} \, \BKop{a^{\,}_j}{a^{\,}_j}^{\vpd}_j \, , ~~
\end{align}
where $\vec{a} = (a^{\,}_{j}, a^{\,}_{j+1}, \dots , a^{\,}_{j+\ell-1})$ labels $\weight{}$-basis configurations of  $r$. Each block $\alpha$ contains at least one configuration $a$ and every configuration appears in some block. Fig.~\ref{fig:U1gate} depicts the block-diagonal gate for a $\U{1}$-symmetric $1 \SpaceDim$ circuit with $\LocDim=\ell=2$; further examples appear in \cite{ConstrainedRUC}. 

The blocks $\alpha$ may correspond to configurations of the $\ell$-site cluster $r$ with definite conserved charge $\totcharge^{\,}_r = \sum_{j \, \in \, r \,} \charge^{\,}_j$, configurations that satisfy some constraint(s), or a combination thereof. Note that configurations that do not satisfy the constraint(s) belong to their own trivial blocks $\alpha$ (with $n^{\,}_{\alpha}=1$). In the case of symmetries, even single-state blocks have $1 \times 1$ Haar unitaries (which capture generic interactions in these states); however, in the case of constraints, we may take $\Haar^{\,}_{r,\alpha} = \ident$ \cite{ConstrainedRUC}, meaning that interactions (if present) are conditioned on satisfaction of the constraint. In this case, \eqref{eq:GenProjGate} becomes
\begin{equation}
\label{eq:KinProjGate}
    \gate^{\vpd}_{t,\lambda,r} \, \equiv \, \sum\limits_{\alpha} \, \Proj{r}{(\alpha)} \, \Haar^{\vpd}_{r,\alpha} \, \Proj{r}{(\alpha)} +\sum\limits_{\beta}\, \Proj{r}{(\beta)} \, , ~
\end{equation}
and \eqref{eq:BloccGateProjectorCompleteness} is modified to include both $\alpha$ and $\beta$ blocks.

Note that \emph{any} combination of Abelian symmetries and (symmetry-compatible) constraints can be captured by gates of the form \eqref{eq:GenProjGate}. Moreover, both conservation laws and constraints may be imposed either within the $\LocDim$-dimensional Hilbert space \cite{RUCconTibor,ConstrainedRUC} or through the introduction of ancilla degrees of freedom \cite{RUCconVedika,U1FRUC}. Our only other assumption is that all blocks can be represented in a common basis---namely, the computational basis.

\subsection{Floquet circuits}
\label{subsec:FloqCirc}
\begin{figure}[t!]
    \centering
    \includegraphics[width=.4\columnwidth]{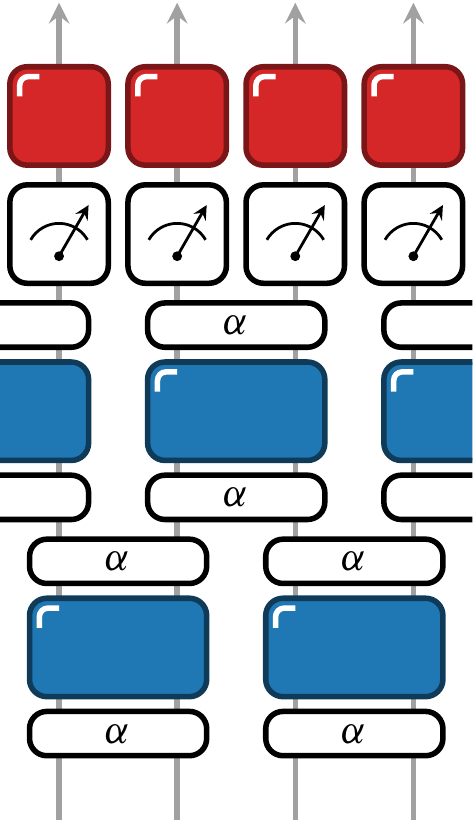}%
    \hspace{1cm}
    \includegraphics[width=.4\columnwidth]{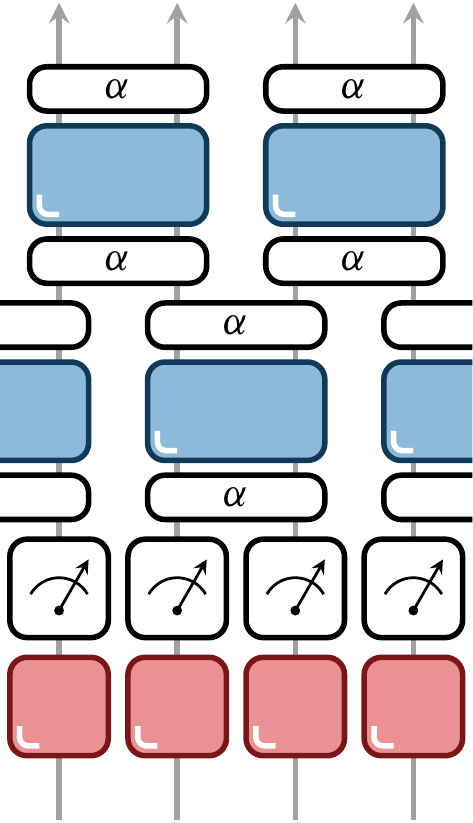}
    \caption{Circuit diagram of a four-site cut of a $1\SpaceDim$ Floquet unitary \eqref{eq:HybridFloqDef} for $\ell=2$, with operators multiplied from bottom to top.  \emph{Left}---Diagrammatic depiction of $\Floq$: The block-diagonal time-evolution gates \eqref{eq:GenProjGate} act first (where the blue objects are the Haar unitaries $\Haar^{\,}_{\alpha}$ and the unfilled objects labeled  $\alpha$ are projectors onto blocks $\alpha$), followed by measurement unitaries \eqref{eq:MeasUnitaryGate}, and finally a cyclic shift \eqref{eq:T shift} of all time slices $\tau \neq 0$ of the lattice. \emph{Right}---Diagram for $\Floq^{\dagger}$: The alternate coloring of the gates indicates complex conjugation, while their reversed ordering and inverted marker indicate transposition. The legs correspond to all dilated sites labeled $j$, and the circuit continues to the left and right. Note that only the two $\alpha$s sandwiching a given gate are related \eqref{eq:GenProjGate}.}
    \label{fig:FloqDiag}
\end{figure}

In calculations pertaining to dynamics, universal properties are well-captured by \emph{noisy} quantum circuits (i.e., those that are random in space and time). However, one can also use \emph{spectral properties} to diagnose universal quantum dynamics \cite{BohigasChaos, RMT_SFF, CDLC1, U1FRUC, ConstrainedRUC}, but a spectrum only exists if evolution to arbitrary times can be captured by a single operator. In static systems, that operator is the Hamiltonian; in Floquet systems---which are periodic in time---that operator is the ``Floquet unitary'' (or single-period evolution operator) $\Floq$. Evolution by $t$ time steps is realized by $\evo (t) = \Floq^{\, t}$ \eqref{eq:Tevo to t}.

It is straightforward to realize Floquet extensions of the noisy hybrid circuits described thus far. We emphasize that, while the measurement gates $\set{\measunitary^{\,}_{\sigma,r}}$ are periodic, their \emph{outcomes} need not be.  Because evolution is periodic, we need only determine the single-period evolution circuit $\Floq$---i.e., we simply specify the single-period circuit $\Floq$ for the first time step, and later times are reached via repeated application of the same Floquet operator $\Floq$. In other words, all Haar-random unitaries are independently drawn only in the first time step (to determine $\Floq$); the set of measurement observables $\set{\mobserv^{\,}_{\sigma,r}}$ and sites $\MSites$ to measure are similarly chosen only in the first time step, after which the evolution then repeats in time.

The time-independent analogue of the measurement unitary \eqref{eq:MeasUnitaryGate} is given straightforwardly by
\begin{align}
    \label{eq:MeasFloquetGate}
    \measunitary^{\vpd}_{\sigma,r} \, &= \, \sum\limits_{m=0}^{\Noutcome-1} \, \Proj{\sigma;r,0}{(m)} \, \otimes \, \SShift{m}{\Noutcome;1,r} \, ,~~
\end{align}
so that the measurement outcomes are always stored in layer $\tau=1$ (with $\tau=0$ the physical layer). 

Following a given round of measurements, one applies a cyclic time translation operator $\Tshift$ to the Stinespring slices of the spacetime lattice defined in Sec.~\ref{subsec:SpacetimeLattice}. That operator shifts the time slice $\tau > 0$ to position $\tau - 1~{\rm mod}~\MeasRounds^{\,}_{\rm tot}$, while leaving the physical slice $\tau=0$ untouched, i.e.,
\begin{align}
    \label{eq:T shift}
    \Tshift \, &\equiv \, \sum\limits_{\set{  n^{\vpp}_{\tau,j} }} \bigotimes\limits_{j=1}^{\Nsite} \bigotimes_{\tau=1}^{\MeasRounds^{\,}_{\rm tot}} \, \BKop{n^{\vpp}_{j,\tau+1}}{n^{\vpp}_{j,\tau}}  \, , ~~
\end{align}
which is only sensible if we include $\Nsite$ Stinespring registers with $\LocDim$ internal states in each slice $\tau$, as described in in Secs.~\ref{subsec:SpacetimeLattice}~and~\ref{subsec:MeasCirc}. Note that $\Floq$ may act trivially on any number of Stinespring registers, and that $\measunitary$ \eqref{eq:MeasFloquetGate} need not use all levels of a given Stinespring register. 

To summarize, consider a hybrid Floquet evolution with $\MeasRounds$ measurements per time step $s$. The operator $\Floq$ comprises layers $\lambda$ of time-evolution gates interspersed with layers $\sigma$ of projective measurements. For each measurement layer $\sigma$, we apply each of the the unitary measurement channels $\measunitary^{\,}_{\sigma,r}$ \eqref{eq:MeasFloquetGate}---corresponding to all measurements of clusters $r$ in measurement round $\sigma$----and the outcomes are recorded in the Stinespring layer $\tau=1$. Following the round $\sigma$, we apply the cyclic temporal shift operator $\Tshift$ \eqref{eq:T shift}, moving the most recent measurement results to the slice $\tau=\MeasRounds^{\,}_{\rm tot}$. After the final measurement round $\sigma=\MeasRounds$ in the final time step $s=T$,  the final application of $\Tshift$ moves all measurement outcomes to their expected locations---i.e., slice $\tau=t,\sigma$ \eqref{eq:Combined SS label} corresponds to round $\sigma$ of measurements in time step $t$ for all $t,\sigma$. A diagrammatic depiction of the Floquet operator \eqref{eq:HybridFloqDef} appears in Fig.~\ref{fig:FloqDiag}. 

In a simple example wherein all measurement layers follow all time-evolution layers in each time step, the Floquet operator is given by the time-ordered product
\begin{align}
    \label{eq:HybridFloqDef}
    \Floq \, &= \, \TimeOrder \,  \prod\limits_{\sigma=1}^{\MeasRounds} \, \Big[ \Tshift \,   \bigotimes\limits_{r \in \sigma}^{\Nsite} \, \measunitary^{\vpd}_{\sigma,r}\,  \Big] \,  \prod\limits_{\lambda=1}^{\ell} \bigotimes\limits_{r \in \lambda} \,  \gate^{\vpd}_{\lambda,r} \, ,~~
\end{align}
where $\TimeOrder$ is a time-ordering operator, which captures the general case in which the measurement layers and time-evolution layers are interspersed. Note that a cluster $r$ can always be uniquely assigned to a Stinespring site $j$. 

Once $\Floq$ is determined, we simply apply this operator repeatedly to reach later times. The measurement protocol (i.e., what, where, and when to measure) are the same in each time step; however, the outcomes need not be the same. This is transparent in the Stinespring formalism \eqref{eq:UnitaryMeas1}, as all outcomes occur. In the noisy case (with no time-translation symmetry), we simply absorb the temporal shift $\Tshift$ \eqref{eq:T shift} 
into the initial density matrix, take $\measunitary^{\vpd}_{t,j}$ as defined in \eqref{eq:MeasUnitaryGate}, and draw all time-evolution gates randomly and independently in both space and time.

\section{Observables and correlations}
\label{sec:OneCopy}
We now consider expectation values and correlations of generic observables, which can diagnose dynamical 
universality and phase structure in experiment without fine tuning. These quantities generally require multiple independent experimental ``shots'' to evaluate (i.e., to recover statistics), where only the initial state of the system $\DensMat^{\,}_0$ \eqref{eq:DilatedInitialState} need be the same in  each shot. That state is evolved under the hybrid circuit, and the outcomes of various ``probe'' measurements $\observ$ (as distinct from the ``circuit'' observables $\mobserv$ measured as part of the hybrid protocol $\evo$) are recorded and averaged without postprocessing. We further restrict to \emph{nonadaptive} circuits, where the outcomes of circuit measurements are not utilized. 

Such quantities are the most straightforward to measure in generic condensed matter and AMO experiments. Moreover, they do not involve postselection or outcome decoding, are insensitive to sample-to-sample variations (e.g., gate errors and different measurement outcomes), are generally not fine tuned, and have a well-defined notion of typicality (meaning that they can be sampled in practice), and commonly diagnose phase structure.  

Mathematically, these quantities involve only a \emph{single copy} of the dilated density matrix $\DensMatSS$ \eqref{eq:DilatedInitialState}---i.e., they are linear functions of $\DensMatSS$. While it has been claimed in the literature \cite{og-MIPT, UtkarshChargeSharp} that linear functions of the density matrix are necessarily blind to MIPTs, a key result of this work is that nonlinearity in the density matrix is neither necessary nor sufficient for detecting MIPTs. This conclusion is motivated by the analyses of Secs.~\ref{sec:SFF} and \ref{sec:adaptive}.

\subsection{Quantities of interest}
\label{subsec:Onefold Quantities}
\begin{figure}
    \centering
    \includegraphics{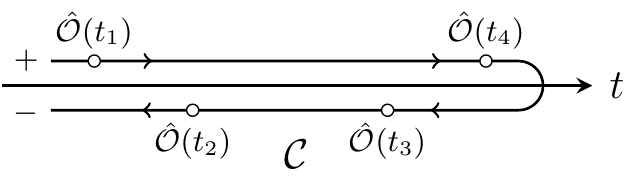}
    \caption{Schematic depiction of the Keldysh contour $\mathcal{C}$. While the time arguments do not satisfy $t^{\,}_n > t^{\,}_{n-1}$, they are nevertheless orderable on the Keldysh contour, which contains only one forward and one backward time evolution.}
    \label{fig:Keldysh}
\end{figure}

The ``standard'' probes of phase structure we consider correspond to  quantities of the general form 
\begin{align}
    \expval{ \, \observ^{\,}_1 (t^{\,}_1) \cdots \observ^{\,}_n (t^{\,}_n)\,}^{\,}_{\DensMatSS} \, , ~~
    \label{eq:n-point-expval}
\end{align}
where the time arguments $\{t_i\}$ can be ordered on a closed-time (Keldysh) contour, as shown schematically for $n=4$ in Fig.~\ref{fig:Keldysh} \cite{OllieNonLinear}. Such quantities correspond, e.g., to the time-dependent expectation value of a local observable ($n=1$), linear response correlation functions ($n=2)$, and time-ordered nonlinear response functions (or $n$-point correlators) \cite{OllieNonLinear}. We note that out-of-time-ordered correlators (OTOCs) \cite{NahumOperator,RUCNCTibor} cannot be evaluated in practice for $\measrate > 0$ since the measurement process is ``irreversible'' from an experimental standpoint. 

Importantly, quantities of the form \eqref{eq:n-point-expval} can be extracted using independent experimental ``shots'' performed on a single sample in finite time and without fine tuning. The outcomes of measurements performed as part of the hybrid evolution are not subsequently utilized. Importantly, these types of quantities are immune to shot-to-shot noise, do not include postselection or feedback, and the number of observables measured to construct the probe itself is finite (at most $\Order{\Nsite}$). In Sec.~\ref{sec:can't work}, we justify restricting to quantities of the form \eqref{eq:n-point-expval}  by arguing that \emph{all} phases of matter can be defined using quantities of this type (including nonequilibrium and topological phases).

In a given calculation, the probe observables $\{ \observ^{\,}_i  \}$ in \eqref{eq:n-point-expval} may be treated separately from the circuit observables $\{ \mobserv^{\,}_{t,\sigma,r}\}$, in either the Schr\"odinger evolution of density matrices or the Heisenberg evolution of operators (made possible by our Stinespring formalism of Sec.~\ref{sec:Stinespring}). 

The probe observables' evolution in the Heisenberg picture factorizes according to
\begin{align}
    \left( \evo^{\dagger}_1 \observ^{\,}_1 \evo^{\vpd}_1 \right)  &\left( \evo^{\dagger}_{2,1}   \observ^{\,}_2 \evo^{\vpd}_{2,1} \right) \cdots     \left(  \evo^{\dagger}_{n,\dots,2,1} \observ^{\,}_n \evo^{\vpd}_{n,\dots,2,1} \right) \notag \\
    = \, \evo^{\dagger}_1 \observ^{\,}_1   &\evo^{\dagger}_{2} \observ^{\,}_2  \cdots \evo^{\dagger}_{n} \observ^{\,}_n \evo^{\vpd}_n \cdots \evo^{\vpd}_2 \evo^{\vpd}_1 \, ,~~
    \label{eq:n-fold Heisenberg}
\end{align}
meaning that every unitary gate $\Haar$ and its conjugate $\Haar^{\dagger}$ appears exactly once in any such expression. Averaging the unitary gates in a given hybrid protocol over the Haar ensemble simplifies (perhaps within each charge or Krylov sector) to the onefold channel \eqref{eq:1FoldHaarAvg} described in App.~\ref{app:Haar Channel} \cite{YoshidaCBD,ConstrainedRUC}. Additionally, each measurement unitary gate $\measunitary$  (and its conjugate $\measunitary^\dagger$) appear exactly once.

\subsection{No signatures in ``generic'' chaotic dynamics}
\label{subsec:Observ Meas No Blocks}
We first consider the effect of measurements on ``generic'' circuits \cite{og-MIPT}---i.e., in which time-evolution is not enriched with block structure (i.e., neither conserved quantities nor kinetic constraints). In this case, the transition matrix for \emph{any} time-evolution layer $\lambda$ is given by \eqref{eq:Tevo Tmat no blocc}, with matrix elements given by \eqref{eq:Tgate Mat El No Sym}. Such Haar-random circuits model fully generic chaotic quantum systems, which generically equilibrate to the featureless, maximally mixed state $\DensMat^{\,}_{\rm therm} = \DensMat^{\,}_{\infty} \propto \ident$. Averaging over the Haar ensemble, each time-evolution transition matrix layer annihilates \emph{all} possible physical operators \emph{except} the identity. 

We now consider the combination of the transition matrices for time-evolution layer $\lambda$ \eqref{eq:Tgate Mat El No Sym} and the outcome-averaged measurement of $\mobserv^{\,}_{t,\sigma,r}$ on cluster $r$  \eqref{eq:Tmat meas elements nonadapt degen Z}. Noting that the time-evolution transition matrix factorizes over sites, the elements of the \emph{combined} transition matrix gate for cluster $r$---corresponding to time evolution and measurement in \emph{either order}---are given by
\begin{align}
    \widetilde{\tgate}^{(t,\sigma,r)}_{\vec{m},\vec{n};\vec{m}'',\vec{n}''} \, &\equiv \, \opmatel{\ubasisop{\vec{m},\vec{n}}}{\,  \tgate^{\vpd}_{t,\lambda,r} \, \tgate^{\vpd}_{t,\sigma,r} \, }{\ubasisop{\vec{m}'',\vec{n}''}} \notag \\
    &= \, \sum\limits_{\vec{m}',\vec{n}'} \, \tgate^{(t,\lambda,r)}_{\vec{m},\vec{n};\vec{m}',\vec{n}'} \tgate^{(t,\sigma,r)}_{\vec{m}',\vec{n}';\vec{m}'',\vec{n}''} \notag\\
    &= \, \kron{\vec{m},\vec{m}''} \, \kron{\vec{n},\vec{n}''} \, \kron{\vec{m}'',\vec{0}} \, \kron{\vec{n}'',\vec{0}} \notag \\
    &= \,\tgate^{(t,\lambda,r)}_{\vec{m},\vec{n};\vec{m}'',\vec{n}''}  , ~~
    \label{eq:generic t mat combined}
\end{align}
meaning that, on average, the combination of measurement layer $\sigma$ and time-evolution layer $\lambda$ (in either order) is identical to time evolution alone. In fact, \emph{any} number of measurement layers are trivialized by a \emph{single} layer of generic  time-evolution acting on all physical sites in the system (in either the Schr\"odinger or Heisenberg picture, for arbitrary circuit geometry and choice of $\set{\mobserv^{\,}_{t,\sigma,r}}$).

The blindness of \eqref{eq:n-point-expval} to projective measurements in generic hybrid circuits was first conjectured in \cite{og-MIPT}. However, it is not the case that projective measurements do nothing in such systems. Rather, the combination of generic, Haar-averaged time evolution and arbitrary measurement gates (applied in either order) annihilates the second term in the outcome-averaged measurement transition matrix gate \eqref{eq:Tmat meas elements nonadapt degen Z}, which would otherwise admit nontrivial dynamics with degenerate measurements. 

The first term in the transition matrix gate \eqref{eq:Tmat meas elements nonadapt degen Z} acts as $\sum_{n=1}^{\LocDim-1} \, \opBKop{\weight{n}}{\weight{n}}$ for $\weight{}$-basis measurements, on average. However, this differs from the generic time-evolution transition-matrix gate $\opBKop{\ident}{\ident} \, = \, \opBKop{\weight{0}}{\weight{0}}$ \eqref{eq:Tevo Tmat no blocc} only in the inclusion of the $n \neq 0 $ terms, which are annihilated upon contraction with \eqref{eq:Tevo Tmat no blocc} (in either order). Hence, the combined transition matrix \eqref{eq:generic t mat combined} for time-evolution layer $\lambda$ and any number of measurement layers (before or after $\lambda$) acts on the entire physical system as $ {\tmat} \vphantom{T}^{\,}_{t,\lambda} \, = \, \opBKop{\ident}{\ident}$ \eqref{eq:Tevo Tmat no blocc}, which is equivalent to $ {\tmat} \vphantom{T}^{\,}_{t,\lambda}$ alone. 

One might conclude from the conjecture of \cite{og-MIPT} as stated that quantities of the form \eqref{eq:n-point-expval} are blind to the competition between measurements and chaotic evolution. The analysis above suggests that such competition \emph{does not exist}---at least, not in experimentally observable quantities. Moreover, there is also no cooperation between these two features, nor do measurements trivialize chaotic dynamics. In fact, maximally chaotic dynamics trivialize the effects of measurements, and produce the infinite-temperature state $\DensMat^{\,}_{\infty}$ more quickly than measurements. In the absence of block structure, the most generally correct statement is that the combination of measurements (without feedback) and highly chaotic time evolution is indistinguishable from chaotic time evolution alone.

This appears to be a generic feature of measurement-based quantum protocols in which the measurement outcomes are not subsequently utilized \cite{SpeedLimit}. Because the outcomes are effectively discarded, their outcome-averaged effect is equivalent to allowing the ``environment'' to measure the system instead. This is most transparent in the context of quantum error correction represented in the Stinespring picture: As shown in \cite{SpeedLimit}, unless and until an error-correction operation is performed (using the outcomes of the measurements), the state of the measured degrees of freedom is always the maximally mixed state $\DensMat^{\,}_{\infty} \propto \ident$, which is simply a random classical bit (or ``dit'' for $\LocDim>2$), with all measurement outcomes equiprobable (for arbitrary measurements). Importantly, there is no physical significance to ``reading'' (or ``recording'') the measurement outcomes---they must be \emph{utilized}, or they have no effect on physically observable quantities \eqref{eq:n-point-expval}.

\subsection{Measuring ``charge'' operators has no effect}
\label{subsec:Observ Meas Sym}
Having ruled out detectable effects due to measurements in generic (i.e., structureless) hybrid circuits, we now investigate the fate of \eqref{eq:n-point-expval} in nonadaptive hybrid circuits with arbitrary conservation laws. We consider Abelian symmetries, and allow for symmetry-compatible constraints, but require that all configurations have associated dynamics. Importantly, we restrict to the measurement of ``charge'' operators---i.e., those that are diagonal in the charge (and/or computational) basis.

We note that chaotic time evolution enriched with one or more conservation laws admits late-time states $\DensMat^{\,}_{\rm av} (t)$ that are not trivially proportional to the identity. The reason is that information about conserved charges is protected against scrambling. In the presence of block structure, $\DensMat^{\,}_{\rm av} (t)$ generally realizes the maximally mixed state \emph{within} each symmetry (or charge) sector, or an ``absorbing'' (or ``steady'') state \cite{U1FRUC,ConstrainedRUC,ODea}. In general, such states are compatible with nontrivial phase structure, where experimentally observable quantities \eqref{eq:n-point-expval} may be nonzero, indicating order. \emph{A priori}, such circuits may be compatible with nontrivial orders due to measurements.

Without loss of generality, suppose that the projectors $\Proj{r}{(\alpha)}$ onto dynamical blocks \eqref{eq:GenProjGate} correspond to sums \eqref{eq:blockprojectorsum} over $\weight{}$-basis product states \eqref{eq:WeightOpDef}. For the hybrid protocol $\evo$ as a whole to respect the desired conservation laws, the circuit observables $\mobserv^{\,}_{t,\sigma,r}$ must act diagonally in the charge (and/or computational) basis, and thus commute with one another and with the projectors onto dynamical blocks \eqref{eq:GenProjGate}. Thus, the circuit observables $\set{\mobserv^{\,}_{t,\sigma,r}}$ must realize sums of $\weight{n}$ \eqref{eq:Normalized projectors}.

The blocks generically capture any Abelian symmetry, along with symmetry-compatible constraints \cite{ConstrainedRUC}. The $\weight{}$-basis circuit observables $\mobserv^{\,}_{t,\sigma,r}$ correspond to ``charge'' operators (e.g., as in \cite{UtkarshChargeSharp}). The transition matrices corresponding to symmetric time evolution and symmetry-compatible measurements (in the $\weight{}$ basis) are given in \eqref{eq:Tgate nice form Z} and \eqref{eq:Tmat meas elements nonadapt degen Z}, respectively. 

Because the observables $\mobserv^{\,}_{t,\sigma,r}$ commute with one another, any number of consecutive layers of measurements is equivalent to a single layer (up to possibly lifting degeneracies). Hence, it is sufficient to consider the product of the transition matrices corresponding to  a single round of measurements \eqref{eq:Tmat meas elements nonadapt degen Z} and one layer of time evolution  \eqref{eq:Tgate nice form Z}, in either order. Using \eqref{eq:Tgate result} and \eqref{eq:Tgate meas basis-indep form}, we have
\begin{align}
    \tgate^{\vpd}_{\rm eff} \, &= \, \tgate^{\rm (evo)}_{t,\lambda,r} \, \tgate^{\rm (meas)}_{t,\sigma,r'} \notag \\
    = \sum\limits_{\alpha} &\frac{1}{n^{\,}_{\alpha}} \sum\limits_{\vec{a},\vec{a}'\in \alpha} \sum\limits_{\mu=0}^{\Noutcome-1}  \sum\limits_{\vec{m},\vec{m}' \in \mu}  \opBKbkop{\nbasisop{\vec{a}\vec{a}}}{\nbasisop{\vec{a}'\vec{a}'}}{\nbasisop{\vec{m}\vec{m}'}}{\nbasisop{\vec{m}\vec{m}'}} \notag \\
    %
    = &\sum\limits_{\alpha} \frac{1}{n^{\,}_{\alpha}} \,\sum\limits_{\vec{a}\in \alpha} \opBKop{\nbasisop{\vec{a}\vec{a}}}{\nbasisop{\vec{a}'\vec{a}'}} 
    \, = \, \tgate^{\rm (evo)}_{t,\lambda,r} \, ,~
\end{align}
meaning that the measurement has no effect whatsoever. The reverse ordering of the measurement gate relative the time-evolution gate corresponds to the transpose of the above; since the transition-matrix gates are Hermitian, the result is the same in either order. 

The result above holds independent of the relative sizes of the clusters $r$ and $r'$, since the time-evolution layer $\lambda$ tiles all sites. Hence, every configuration $\vec{m},\vec{m}'$ in the measurement layer $\sigma$ is matched to the block-compatible configurations $\vec{a},\vec{a}'$ above, and thereby trivially absorbed into the time-evolution layer. Also note that the result above holds \emph{even} in the presence of trivial blocks $\beta$ \eqref{eq:KinProjGate}---i.e., for fine-tuned models in which constraint-violating configurations are noninteracting.

Additionally, any further layers of $\weight{}$-basis measurements do not change this result, as they are successively absorbed into---and trivialized by---the time-evolution layer. Thus, we conclude that projective measurements that commute with the generator(s) of any Abelian symmetry (along with the projectors encoding any constraints compatible with that symmetry) have no detectable effect---on average---compared to time evolution alone in the absence of outcome-dependent feedback. In Sec.~\ref{sec:adaptive}, we recover nontrivial results for adaptive circuits with charge measurements, and genuine phase transitions in constrained models without conservation laws.

\subsection{Other measurements ``undo'' structure}
\label{subsec:Observ Meas Other Pauli}
The scenarios considered thus far are the two most relevant to experiment \cite{og-MIPT,MIPT-exp,UtkarshChargeSharp}. We first ruled out the possibility that measurements \emph{of any type} have an observable effect compared to generic chaotic time evolution alone \cite{NahumOperator,RUCNCTibor}. We then ruled out the possibility of ``steering'' toward symmetry sectors in chaotic quantum systems with
Abelian conservation laws using measurements of charge operators (without outcome-dependent feedback). 

The results apply to all quantities of the form \eqref{eq:n-point-expval} in nonadaptive circuits upon averaging over measurements. As far as we are aware, there have been no attempts to investigate the measurement of ``charge-changing'' operators in chaotic quantum circuits with conserved quantities---most likely, this is because one na\"ively expects that measuring such operators simply spoils the symmetry of the time-evolution gates in the full hybrid protocol $\evo$, leading to a featureless state  $\DensMat^{\,}_{\rm av} (t) \propto \ident$ \cite{NahumOperator,RUCNCTibor}.

We now confirm that this is, indeed, the case for hybrid circuits acting on qubits ($\LocDim=2$), where the Weyl $\shift{}$ and $\weight{}$ operators reduce to the familiar Pauli matrices $\PX{}$ and $\PZ{}$, with $\PY{} = \ii \, \PX{} \PZ{}$. For $\LocDim>2$, the same arguments are expected to apply, as described in Sec.~\ref{sec:OneCopy} and conjectured in \cite{og-MIPT}. For concreteness, time evolution is generated by two-site, block-diagonal gates of the form \eqref{eq:GenProjGate}, with symmetry blocks generated by $\PZ{}$, and we consider projective measurements of $\PX{}$ and $\PZ{}$ operators.

\paragraph*{Qubit models.---} The unitary gates \eqref{eq:MeasUnitaryGate} corresponding to Pauli measurements are also Hermitian \cite{SpeedLimit}. Consider the single- and two-site physical density matrices
\begin{subequations}
\label{eq:Pauli Dens Mat Easy}
\begin{align}
    \DensMat^{\vpp}_{i} (t) \, &= \, \sum\limits_{\mu = 0}^3 \, C^{(t)}_{\mu} \, \Pauli{\mu}{i} \label{eq:Pauli Dens Mat Easy 1 site} \\
    \DensMat^{\vpp}_{i,j} (t) \, &= \, \sum\limits_{\mu,\nu = 0}^3 \, C^{(t)}_{\mu,\nu} \, \Pauli{\mu}{i}\Pauli{\nu}{j}~,~~\label{eq:Pauli Dens Mat Easy 2 site}
\end{align}
\end{subequations}
and now, measuring $\PX{i}$ and $\PX{i}\PX{j}$, respectively, gives
\begin{subequations}
\label{eq:Pauli Dens Mat Post X Meas Easy}
\begin{align}
    \DensMat^{\vpp}_{i} (t') \, &= \,  C^{(t)}_{0} \,\ident^{\vpp}_i +  C^{(t)}_{1} \, \PX{i}  \\
    \DensMat^{\vpp}_{i,j} (t') \, &= \,  C^{(t)}_{0,0} \, \ident^{\vpp}_{ij} +  C^{(t)}_{1,0}\,\PX{i}+  C^{(t)}_{0,1} \, \PX{j}  +  C^{(t)}_{1,1}\, \PX{i} \PX{j}~,~~
\end{align}
\end{subequations}
and taking $\PX{i,j} \to \PZ{i,j}$ in the expressions above gives the updates corresponding to measuring $\PZ{i}$ and $\PZ{i} \PZ{j}$. 

However, such $\PZ{}$ measurements can generically and trivially be absorbed into the $\PZ{}$-basis time-evolution gates, as established in Sec.~\ref{subsec:Observ Meas Sym}. As a result, any measurements of $\PZ{}$ have no effect on the density matrix, unless they appear between  $\PX{}$ measurements. However, we ignore this scenario, since such $\PZ{}$ measurements have no effect on the evolution of any observable or density matrix.

\paragraph*{Ising symmetry.---} Suppose that the two-site time-evolution gates \eqref{eq:GenProjGate} conserve the $\Ints^{\,}_2$ Ising parity $\mathcal{G}^{\,}_{i,j} = \PZ{i} \PZ{j}$. The unitary gate $\gate^{\,}_{i,j}$ contains two symmetry blocks, each with two states, corresponding to $\mathcal{G}^{\,}_{i,j} = \pm 1$, with associated projectors $\left( \ident \pm \PZ{i} \PZ{j} \right)/2$. The Haar-averaged update to \eqref{eq:Pauli Dens Mat Easy 2 site} is given by
\begin{equation}
    \label{eq:Pauli Dens Mat Ising Update}
    \overline{\DensMat^{\vpp}_{i,j} (t') } \, = \,\overline{C^{(t)}_{0,0}} \, \ident^{\,}_{i,j} + \overline{C^{(t)}_{3,3}} \, \PZ{i} \PZ{j} \, ,~~
\end{equation}
and taking \eqref{eq:Pauli Dens Mat Post X Meas Easy} with $\PX{}\to \PZ{}$ is consistent with the claim in Sec.~\ref{subsec:Observ Meas Sym}  that measuring any number of $\PZ{}$ operators before or after applying such a time-evolution gate has no effect compared to time evolution alone.

According to \eqref{eq:Pauli Dens Mat Post X Meas Easy}, measuring $\PX{i}$ or $\PX{j}$ \eqref{eq:Pauli Dens Mat Post X Meas Easy} before or after time evolving \eqref{eq:Pauli Dens Mat Ising Update} gives
\begin{equation}
    \label{eq:Pauli Dens Mat Ising and Meas Single X Update}
    \overline{\DensMat^{\vpp}_{i,j} (t'') } \, = \,\overline{C^{(t)}_{0,0}} \, \ident^{\,}_{i,j} \, , ~~
\end{equation}
which is equivalent to generic Haar-averaged time evolution \eqref{eq:Tevo Tmat no blocc} with neither conservation laws nor constraints. 

From \eqref{eq:Pauli Dens Mat Ising and Meas Single X Update}, we conclude that measuring $\PX{i} \PX{j}$ gives the same result; it is also straightforward to verify that measuring $\PX{i} \PX{j}$ before or after applying the gates $\gate^{\,}_{i,k}$ and $\gate^{\,}_{j,\ell}$ also leads to the maximally mixed state $\DensMat \propto \ident$ on all sites $i,j,k,\ell$. In other words, \emph{any} measurement involving charge-changing operators $\PX{j}$ applied before or after an Ising-symmetric gate $\gate^{\,}_{i,j}$ is equivalent to replacing $\gate^{\,}_{i,j}$ with a random $4 \times 4$ unitary. 

Essentially, $\PX{}$ measurements ``undo'' the Ising symmetry, replacing the nontrivial density matrix with the maximally mixed state $\DensMat \propto \ident$ on every measured site. Thus, although measurements cannot \emph{introduce} new universal behavior (compared to chaotic time evolution alone), they can \emph{negate} the universal properties of time evolution, instead realizing ``generic'' (i.e., featureless) evolution. This rules out the possibility of realizing nontrivial phases in Ising-symmetric systems using nonadaptive circuits with competing $\PX{}$- and $\PZ{}$-like measurements, and implies that the measurement-induced $\Ints^{\,}_2$ symmetry-breaking and symmetry-protected topological orders reported in \cite{hsieh} and \cite{barkeshli}, respectively, are  experimentally unobservable.

\paragraph*{Charge conservation.---} For a more nuanced example, suppose that the two-site time-evolution gates conserve the local $\U{1}$ charge $\mathcal{G}^{\,}_{i,j} = \PZ{i} +\PZ{j}$ (as depicted in Fig.~\ref{fig:U1gate}). The unitary gate $\gate^{\,}_{i,j}$ contains three symmetry blocks corresponding to $\mathcal{G}^{\,}_{i,j} = +1,-1,0$ (i.e., the configurations $\uparrow \uparrow$ / $\downarrow \downarrow$,  $\uparrow \downarrow$ \emph{and} $\downarrow \uparrow$), with associated projectors $\left( \ident^{\,}_i \pm \PZ{i} \right) \left( \ident^{\,}_{j} \pm \PZ{j} \right) /4$ and $\left( \ident^{\,}_{ij} - \PZ{i} \PZ{j} \right)/2$. The Haar-averaged update to \eqref{eq:Pauli Dens Mat Easy 2 site} is given by
\begin{align}
    \overline{\DensMat^{\vpp}_{i,j} (t') } \, &= \,\overline{C^{(t)}_{0,0}} \, \ident^{\,}_{i,j}  +\overline{C^{(t)}_{3,3}} \, \PZ{i} \PZ{j} \notag \\
    &~~+\frac{1}{2} \left( \overline{C^{(t)}_{0,3}}  + \overline{C^{(t)}_{3,0}}  \right) \left( \PZ{i} + \PZ{j} \right)\, ,~~ \label{eq:Pauli Dens Mat U1 Update}
\end{align}
where the first line is equivalent to the Ising update \eqref{eq:Pauli Dens Mat Ising Update}, and the new terms in the second line correspond to the two-state block $\uparrow \downarrow$, $\downarrow \uparrow$. Note that $\U{1}$-symmetric evolution mixes the operators $\PZ{i}$ and $\PZ{j}$. 

As in the Ising case, one can verify that measuring $\PZ{}$ operators immediately prior or subsequent to time-evolution layer $\lambda$ has no effect on average compared to time evolution alone. Additionally, measuring $\PZ{}$ in between $\PX{}$ measurements again has no effect. Hence,  we need only consider measuring the symmetry-incompatible operator $\PX{j}$, either before or after the $\U{1}$-symmetric time-evolution gate $\gate^{\,}_{i,j}$ that produces \eqref{eq:Pauli Dens Mat U1 Update}. 

Averaging over measurement outcomes and the Haar ensemble (in either order), the resulting density matrix is
\begin{align}
    \overline{ \DensMat^{\vpp}_{i,j} (t'') } \, &= \, \overline{C^{(t)}_{0,0}} \, \ident^{\,}_{i,j}  +\frac{1}{2} \left( \overline{C^{(t)}_{0,3}}  + \overline{C^{(t)}_{3,0}}  \right) \PZ{i} \,
    , ~~ \label{eq:Pauli Dens Mat U1 and Single X Meas Update}
\end{align}
meaning that measuring $\PX{j}$ annihilates $\PZ{j}$ terms in \eqref{eq:Pauli Dens Mat Easy 2 site}; however, the time-evolution gate $\gate^{\,}_{i,j}$ mixes $\PZ{j}$ with $\PZ{i}$ (with a factor of $1/2$), so that the $0,3$ coefficient survives.  If one instead measures the product operator $\PX{i} \PX{j}$, then only the trivial term remains ($\DensMat \propto \ident$), as in the $\Ints^{\,}_2$ Ising case \eqref{eq:Pauli Dens Mat Ising and Meas Single X Update}. The same result holds if one measures both $\PX{i}$ and $\PX{j}$ independently.

\paragraph*{Generalization.---} We now consider charge-changing measurements for systems with $\LocDim$ states per site and arbitrary Abelian symmetries. If constraints are present, we require that all blocks in \eqref{eq:GenProjGate} have corresponding dynamics (i.e., even constraint-violating configurations are assigned a $1 \times 1$ Haar-random unitary, which captures generic interactions). As always, we take the charge (and/or computational) basis to be $\weight{}$.

Note that an observable $\mobserv$ with \emph{any} $\shift{}$ content in its Weyl decomposition \eqref{eq:ManyBodyOpUnitaryBasis} constitutes a ``charge-changing'' operator. However, there is no generic expression for the spectral projectors \eqref{eq:ASpectralDecomp} of such an operator, given that we have fixed the $\weight{}$ basis. Given an operator
\begin{align}
    \mobserv^{\vpd}_{x} \, &= \, \sum\limits_{\vec{m},\vec{n}} \, a^{\vpp}_{\vec{m},\vec{n}} \, \ubasisop{\vec{m},\vec{n}} \, , ~~\label{eq:Gen Mobserv Weyl Decomp}
\end{align}
we can define a rotated operator
\begin{align}
    \widetilde{\mobserv}^{\vpd}_{x} \, &\equiv \, \mathcal{Z}^{\dagger}_x\mobserv^{\vpd}_{x} \,  \mathcal{Z}^{\vpd}_x \, = \, \sum\limits_{\vec{k}} \, \widetilde{a}^{\vpp}_{\vec{k}} \, \Shift{\vec{k}}{x} \, , ~~\label{eq:CC Op X only}
\end{align}
where $\mathcal{Z}$ comprises only $\weight{}$ operators, and the new coefficient can be written in the form
\begin{equation}
     \widetilde{a}^{\vpp}_{\vec{k}}  =  \sum\limits_{\vec{m},\vec{n}}  \, a^{\vpp}_{\vec{k},\vec{m}-\vec{n}}  \, \omega^{- \vec{m}\cdot \vec{k}} \, \tr{\mathcal{Z}^{\dagger}_{x}  \Weight{\vec{m}}{x} } \,  \tr{\Weight{-\vec{n}}{x}  \mathcal{Z}^{\vpd}_{x}} \, , 
\end{equation}
though we do not use this expression. Instead, we imagine measuring $\widetilde{\mobserv}$ \eqref{eq:CC Op X only} on cluster $x$, applying the channel $\mathcal{Z}^{\dagger}$ prior to measurement, and the channel $\mathcal{Z}$ after. Importantly, because $\mathcal{Z}$ only contains $\weight{}$ operators, the corresponding transition matrix preserves any $\shift{}$ content.

In the absence of $\beta$ blocks \eqref{eq:KinProjGate}, the unitary gate $\gate^{\,}_{t,\lambda,r}$ acting on cluster $r$ in layer $\lambda$ of time step $t$ corresponds to a transition-matrix gate with elements
\begin{align}
    \tgate^{(t,\lambda,r)}_{m,n;m',n'}  &=  \kron{m,0}\kron{m',0} \sum\limits_{\alpha}  \frac{1}{n^{\,}_{\alpha}}  \sum\limits_{a,a' \in \alpha}  \frac{\omega^{a' \cdot n' - a \cdot n}}{\LocDim^{\ell}} \, ,  \label{eq:evo Tgate alpha only}
\end{align}
where $\ell = \abs{r}$ and we have dropped the vector notation for presentation. Importantly, the gate  \eqref{eq:evo Tgate alpha only} annihilates any $\shift{}$ content, and the transition-matrix gates corresponding to both the $\mathcal{Z}$ and $\mathcal{Z}^{\dagger}$ channels does not change this. 

Suppose that only a single layer of charge-changing measurements occurs between any pair of time-evolution layers. In this case, the time-evolution transfer matrices ensure that there is no $\shift{}$ content in any operator (or density matrix) before or after the measurement channel. The transition-matrix gate corresponding to the measurement of $\widetilde{\mobserv}$ \eqref{eq:CC Op X only} has elements
\begin{align}
    \tgate^{(t,\sigma,r')}_{m,n;m',n'} &=  \kron{m,m'} \, \kron{n,0} \, \kron{n',0} \notag \\
    &+\kron{n,n'} \sum\limits_B  \sum\limits_{b \neq b' \in B} \frac{\omega^{b \cdot \left( m' - m \right)}}{\LocDim^{\abs{x}}}  \kron{n',b'-b} \, , \label{eq:cc meas Tgate}
\end{align}
and imposing the condition of no $\shift{}$ operators (due to the time-evolution and $\mathcal{Z}$ transition matrices), we set $m=m'=0$, and \eqref{eq:cc meas Tgate} becomes
\begin{align}
    \tgate^{(t,\sigma,r')}_{m,n;m',n'} \, &\approx \, \kron{m,m'} \, \kron{n,n'} \, \left( \kron{n,0} + \frac{\mathcal{I}^{\vpp}_{\widetilde{\mobserv}} (n)}{\LocDim^{\abs{x}}} \right) \, ,~~ \label{eq:cc meas Tgate eff} 
\end{align}
where $\mathcal{I}^{\vpp}_{\widetilde{\mobserv}} (n)$ is the integer
\begin{align}
    \mathcal{I}^{\vpp}_{\widetilde{\mobserv}} (n) \, &= \,\sum\limits_B  \sum\limits_{b \neq b' \in B} \,  \kron{n,b'-b} \, ,~~\label{eq:cc meas counter}
\end{align}
which counts the number of pairs of states $b \neq b'$ in each block $B$ satisfy $b'-b=n$, for a given $n=n'$. Hence, the effective transition-matrix gate for measurement of the charge-changing observable $\widetilde{\mobserv}$ \eqref{eq:CC Op X only}  is given by
\begin{align}
    \tgate^{\vpd}_{\widetilde{\mobserv},x} \, &= \, \opBKop{\ident}{\ident}^{\vpp}_x + \frac{1}{\LocDim^{\abs{x}}} \, \sum\limits_{\vec{n}\neq\vec{0}} \, \mathcal{I}^{\vpp}_{\widetilde{\mobserv}} (\vec{n})\, \opBKop{\weight{\vec{n}}}{\weight{\vec{n}}}^{\vpp}_{x} \, , ~~\label{eq:Charge-Changing Meas Eff Tgate}
\end{align}
which preserves any trivial operator (or density matrix) content $\propto \ident^{\,}_x$ and suppresses certain $\Weight{}{x}$ operators by a factor $\Order{1/\LocDim^{\abs{x}}}$, while annihilating others entirely. 

If there are multiple, overlapping measurements of charge-changing observables between two time-evolution layers, then we cannot enforce $m=m'=0$ on all intermediate measurements (on sites where overlaps occur). However, we note that the two terms in \eqref{eq:Charge-Changing Meas Eff Tgate} do not mix: The first term only allows the identity, and the second term annihilates the identity. All terms with $\shift{}$ content are annihilated by \emph{each} time-evolution transition-matrix layer; these operators may be repopulated by the measurement transition matrices, and must also contain nontrivial $\weight{}$ operator content. In the simple case where all measurements are nondegenerate, then only the identity survives, as the nontrivial term in \eqref{eq:Charge-Changing Meas Eff Tgate} vanishes. In the less simple case of degenerate measurements, but where $\mathcal{Z}=\ident$, the \emph{same} $\vec{n}$ must have nonzero $\mathcal{I}$ for \emph{all} measurements between any two time-evolution layers; intuitively, this is because the sequence of $\shift{}$-basis measurements ``lifts'' the degeneracies of each individual observable measured. Finally, if we allow for $\beta$ blocks, as in \eqref{eq:KinProjGate}, the time-evolution layers no longer annihilate all $\shift{}$ content, and the hybrid transition matrices are more complicated.

In the most general case, we find that the $\ident$ part of any operator (if present) survives the entire circuit unscathed. Importantly, this term is always present in any density matrix, with constant coefficient $\LocDim^{-\Nsite}$. The nontrivial terms (i.e., other than $\ident$) in any operator or density matrix are suppressed by a factor of $1/\LocDim$, every time a charge-changing measurement is made on a site where the operator or density matrix acts nontrivially. This suppression is present even in the most complicated cases above, and is captured by the qubit examples. Hence, we conclude that the trivial part $\ident$ of any operator $\observ$ or density matrix $\DensMat$ is unaffected by the hybrid evolution, while, in the presence of charge-changing measurements, all other terms are suppressed by a factor of roughly $\LocDim^{- \measrate \, t \, \Nsite}$, where $\measrate$ is the rate of charge-changing measurements.

In other words, we find that measuring observables that are not compatible with the block structure of the time-evolution protocol can only trivialize (i.e., ``undo'') the universal features of the time evolution itself. It is not possible for measurements to introduce new terms not permitted by time evolution without an adaptive protocol. 
The two limiting scenarios correspond to (\emph{i}) no modification to the physics corresponding to time evolution alone and (\emph{ii}) trivialization of the time-evolution physics to featureless chaotic evolution. At early times, $\DensMat$ may realize any state compatible with block-structured time evolution (\emph{i}), while at late times, $\DensMat$ reduces to the maximally mixed state, as would result from generic time evolution alone. As the measurement rate $\measrate$ is increased, one can tune from case (\emph{i}) to case (\emph{ii}) by suppressing all states other than the maximally mixed state with each additional measurement, without ever seeing a sharp transition. We also note that neither limiting case corresponds to ``new'' universal physics---measurements either preserve or trivialize the physics of time evolution. The  ``crossover'' between these two regimes is further smoothed by averaging over measurement locations, so that the weight of nonidentity terms falls off as $\LocDim^{-\measrate \, \Nsite \, t}$.

In other words, measurements can (\emph{i}) do nothing at all compared to time evolution; (\emph{ii})  trivialize the dynamical properties of time evolution; or (\emph{iii}) anything intermediate between these limits. However, there is no sharp ``transition'' as one tunes between these limits, and we conclude that measurements cannot lead to new universality classes nor transitions that can be witnessed using probes of the form \eqref{eq:n-point-expval} without feedback.

\section{Spectral form factors}
\label{sec:SFF}
We now study spectral properties of hybrid Floquet circuits with both projective measurements and chaotic unitary dynamics with and without block structure. Importantly, such a spectral analysis in the presence of measurements was not possible prior to our development of the unitary Stinespring formalism in Sec.~\ref{sec:Stinespring}, and particularly, the identification of that unitary with the time evolution of the system and measurement apparatus \cite{AaronDiegoFuture}. 

Historically, one of the most prominent examples of a sharp transition between area-law and volume-law scaling of entanglement entropy corresponds to the thermalization transition between ergodicity and many-body localization \cite{ETH1, ETH2, ETH3, mblarcmp, mblrmp}. However, such [de]localization transitions are generically manifest in experimentally tractable probes of the form \eqref{eq:n-point-expval}, and are always evident in \emph{spectral properties}, which generically distinguish chaos from its alternatives.

Here we consider the spectral form factor (SFF), an analytical diagnostic of spectral rigidity that quantifies the signature level repulsion  of chaotic quantum systems. We evaluate this quantity in the presence of measurements for the first time, identifying several possible extensions to hybrid dynamics. Of these, only a handful are physically sensible, and none show a transition as a function of measurement rate $\measrate$. As in the case of observables \eqref{eq:n-point-expval}, both in the absence of block structure and in the case of structure-preserving measurements (i.e., corresponding to ``charge'' operators), there is no dependence of the SFF on $\measrate$ whatsoever. In the case of structure-incompatible measurements (i.e.,  corresponding to ``charge-changing'' operators), the effects of block structure on the SFF are smoothly diluted as a function of $\measrate$ without a transition.

Even more striking is our finding that even variants of the SFF that is quadratic in the density matrix $\DensMat$ is equally blind to the effects of projective measurements. Such variants are based on temperature-dependent spectral form factors in thermal, Hamiltonian systems. Even with postselection (of the outcomes realized for the two copies of $\DensMat$), measurements have no meaningful effect on the spectral properties of the underlying chaotic dynamics.

Our analysis of the SFF provides compelling evidence that (\emph{i}) measurements do \emph{not} compete with chaotic dynamics as far as the spectrum of the underlying evolution operator is concerned, but merely destroy entanglement, and (\emph{ii}) nonlinearity  in the density matrix $\DensMat$---even when combined with postselection---is not sufficient to realize a transition as a function of the measurement rate $\measrate$.

\subsection{Chaos and spectral rigidity}
\label{subec:SFFintro}
Having established that dynamical signatures of quantum chaos, universality, and phase structure \eqref{eq:n-point-expval} are blind to the effect of measurements in Sec.~\ref{sec:OneCopy}, we now consider a separate signature of chaos: \emph{level repulsion}. Thermal correlations in chaotic systems cause the eigenvalues of the generator of dynamics (i.e., Hamiltonians and Floquet unitaries) to repel one another. In general, the eigenvalues of $\Floq$ \eqref{eq:HybridFloqDef} for a chaotic quantum system are expected to obey an RMT distribution \cite{BohigasChaos, RMT_SFF, YoshidaSFF, ShenkerRMT, CDLC1, U1FRUC, ConstrainedRUC}; this is also termed ``spectral rigidity'' \cite{CDLC1, U1FRUC}. 

A useful diagnostic of RMT spectral rigidity is the two-point spectral form factor (SFF) \cite{RMT_SFF, YoshidaSFF, ShenkerRMT, ProsenRMTChaosPRX, bertini2018exact, CDLC1, CDLC2, U1FRUC, SubirSFF, ConstrainedRUC, SamJohnFeynman}; 
for purely unitary Floquet dynamics the SFF is defined as
\begin{align}
    \label{eq:SFF def}
    K(t) \, &= \, \sum\limits_{m,n=1}^{\HilDim} \, e^{\ii \, t \, \left( \theta^{\,}_m - \theta^{\,}_n \right)} \, = \,  \abs{\tr{ \, \Floq^{\, t}\, }}^2 \, ,~~
\end{align}
where $\LocDim$ is the on-site Hilbert space dimension, $\Nsite$ is the number of sites, $\{ \theta^{\,}_m \}$ are the eigenphases of the Floquet unitary $\Floq$. The SFF \eqref{eq:SFF def} is  essentially the temporal Fourier transform of two-point correlations of the \emph{eigenvalue} density \cite{RMT_SFF}.  We then average \eqref{eq:SFF def} over an ensemble of statistically similar realizations of $\Floq$ (which we denote by an overline, $K^{\,}_{\rm av} (t) = \overline{K(t)}$).

Chaotic many-body quantum systems are insensitive to their initial state and microscopic details; hence, one expects that $K(t)$ \eqref{eq:SFF def} in such systems is well approximated by a \emph{random} evolution operator with the same symmetries. For fully generic models, we expect $\overline{K(t)} = K^{\,}_{\rm CUE} (t) = t$ (for $1 \leq t \leq \HilDim$) corresponding to the Circular Unitary Ensemble \cite{BnB,CDLC1,U1FRUC,ConstrainedRUC}. At $t=0$, $K(t)$ is trivially $\HilDim^2 = \LocDim^{2 \Nsite}$; for $t \geq \tau^{\,}_{\rm \small Heis} = \HilDim$, the Heisenberg time (equal to the inverse mean level spacing), $K(t) = \HilDim$. For $t \geq \tau^{\,}_{\rm \small Heis}$, the dominant contribution is the sum over all $m=n$ terms in \eqref{eq:SFF def}, and the off-diagonal contributions ($m \neq n$) are effectively random complex numbers that sum to zero when $t$ is larger than the mean inverse level spacing. 

This linear ramp for times $1 \leq t \leq \HilDim$ is a fingerprint of spectral rigidity, realized by generic Floquet circuits \cite{CDLC1}. Intuitively, in the linear ramp regime,  $K(t)$ \eqref{eq:SFF def} is dominated by paired ``Feynman histories'' \cite{SamJohnFeynman}, where at time $t$ there are $t$ possible pairings, so that $K(t) \sim t \left( 1 + \dots \right)$, where the $\dots$ terms are suppressed at long wavelengths and by ensemble averaging \cite{CDLC1,U1FRUC,ConstrainedRUC,SamJohnFeynman}.

In practice, typical systems do not show RMT behavior starting from $t=1$. To see this, we first note that noninteracting, integrable, and localized models do not show a linear ramp at all, because their eigenphases are \emph{uncorrelated} and thus do not repel. We find $K(t) = \HilDim$ for all times $t>0$ for such uncorrelated phases, as can be verified for certain noninteracting disordered models \cite{CDLC1,U1FRUC,ConstrainedRUC}. For a typical system, at early times one expects that interactions have not yet had a chance to produce long-wavelength thermal correlations, and the system is effectively noninteracting. Modeling the dynamics via single-site Floquet circuits predicts $K(t) \sim t^{\Nsite}$. However, as interactions entangle the system, these effective single-site blocks grow to size $\xi (t)$, with $\xi(0) = \Order{1}$. The SFF at time $t$ is then expected to be $K(t) = t^{\Nsite/\xi (t)}$. From this picture, we see that the onset of the linear ramp regime---associated with thermalization---occurs at the time $\tau^{\,}_{\rm \small th}$ (the ``Thouless time'' 
\cite{CDLC1, U1FRUC, ConstrainedRUC, ShenkerRMT}, named in analogy to the quantity describing disordered wires \cite{Thouless74, ThoulessThinWire}), defined by $\xi (\tau^{\,}_{\rm \small th}) = \Nsite$, so that $K(t) \sim t$ for $t > \tau^{\,}_{\rm \small th}$. Here, ``thermalization'' refers to the onset of RMT behavior and the loss of information about initial conditions, captured by a thermal density matrix and $K^{\,}_{\rm av} (t) = t$.

The Thouless time heralds the onset of thermalization: Maximally chaotic systems have $\tau^{\,}_{\rm \small th} =\Order{1}$, while nonthermal (e.g., localized) systems have $\tau^{\,}_{\rm \small th} > \HilDim$, so that there is no linear ramp regime. In generic systems, the scaling of the Thouless time $\tau^{\,}_{\rm \small th}  \sim \size^z$ is directly associated with linear-response correlators \cite{ConstrainedRUC}, where the exponent $z$ is the dynamical exponent. A $1\SpaceDim$ system with a $\U{1}$ conserved charge, e.g., has $z=2$ \cite{U1FRUC, ShenkerRMT, ConstrainedRUC}, from which we conclude that thermalization is delayed until information about the symmetry diffuses through the system, which requires $t \gtrsim \tau^{\,}_{\rm \small th} \propto \size^2$.

In fact, the Thouless time is directly related to linear-response correlation functions at infinite temperature \cite{YoshidaSFF, ShenkerRMT, ConstrainedRUC}. Using $\Floq^{\dagger}=\Floq^{-1}$, we rewrite \eqref{eq:SFF def} as 
\begin{align}
    K(t) 
    &= \, \sum\limits_{a,b=1}^{\HilDim} \, \matel{a}{\Floq^{-t}}{a} \hspace{-0.5mm} \matel{b}{\Floq^{\, t}}{b} \notag \\
    &= \, \sum\limits_{a,b=1}^{\HilDim} \,  \tr{ \, \BKop{b}{a} \, \Floq^{\, -t} \, \BKop{a}{b} \, \Floq^{\, t} } \, , ~~
    \label{eq:SFF def sum}
\end{align}
where the labels $a$ and $b$ correspond to any valid choice of orthonormal basis for $\Hilbert$, e.g. the Weyl $\weight{}$ basis \eqref{eq:WeightOpDef}. In the na\"ive operator basis of App.~\ref{app:NaiveBasis}, \eqref{eq:SFF def sum} becomes
\begin{align}
    K(t) \, &= \, \sum\limits_{a,b=1}^{\HilDim} \, \frac{1}{\HilDim} \, \tr{ \, \nbasisopdag{ab} \, \Floq^{\, -t} \, \nbasisop{ab} \, \Floq^{\, t} }\,  \notag \\
    &= \, \sum\limits_{a,b=1}^{\HilDim} \, \tr{ \, \nbasisopdag{ab} (0) \, \nbasisop{ab} (t) \, \DensMat^{\vpp}_{\infty} \, } \notag \\
    &= \,\sum\limits_{a,b=1}^{\HilDim}  \, \expval{ \, \nbasisopdag{ab} (0) \, \nbasisop{ab} (t) \, }^{\,}_{T=\infty}  \, , ~~\label{eq:SFF no meas corr}
\end{align}
in the Heisenberg picture of Floquet operator evolution, where in the last line, the correlation functions are evaluated in the infinite-temperature state $\DensMat^{\,}_{\infty} = \ident/\HilDim$. 

Importantly, \eqref{eq:SFF no meas corr} is an equal-weight sum over the temporal autocorrelation functions of all basis operators at infinite temperature---where \eqref{eq:SFF def} defines the SFF at infinite temperature (this is most natural given that Floquet evolution does not conserve energy and generically heats up to $T=\infty$, although it is possible to define the SFF at finite temperature as well, as in Sec.~\ref{subsec:SFF DM}).

\subsection{Extension to hybrid circuits}
\label{subsec:SFF hybrid}
The Stinespring formalism of Sec.~\ref{sec:Stinespring} provides for a straightforward extension of the SFF $K(t)$ \eqref{eq:SFF def} to hybrid quantum dynamics with both time evolution and projective measurements. For a spectrum to exist, the evolution must be periodic in time (i.e., Floquet), as outlined in Sec.~\ref{subsec:FloqCirc}. While the measurement protocol is periodic, the observed outcomes need not repeat in time. 

Na\"ively, one could extend the SFF to the \emph{isometric} measurements described in Sec.~\ref{subsec:SSIsometric}. Because the corresponding Floquet $\Floq$ is an isometry, it creates Stinespring kets $\ket{\bvec{m}}$ with each measurement (while $\Floq^{\dagger}$ creates Stinespring bras $\bra{\bvec{n}}$). The natural extension of \eqref{eq:SFF def sum} to isometric measurements is given by
\begin{align}
    K^{\vpd}_{\rm \small in} (t) &= \sum\limits_{\bvec{m}} \sum\limits_{a,b=1}^{\HilDim^{\,}_{\rm ph}}  \matel{a}{\Floq^{\, -t}}{a,\bvec{m}} \hspace{-0.5mm} \matel{b,\bvec{m}}{\Floq^{\, t}}{b} \,, ~\label{eq:SFF isomeas post}
\end{align}
where the label ``in'' reflects the fact that the two evolutions create Stinespring kets and bras pointed inward. Note that \eqref{eq:SFF isomeas post} is postselected in the sense that both evolutions realize the same outcome trajectory. 

However, upon Haar averaging (as described in Sec.~\ref{subsec:SFF Haar}), we find that $K^{\,}_{\rm \small in}(t) \ll t$. Yet, for maximally chaotic systems, one expects $K(t) \sim t$, while for less chaotic systems, one expects $K(t) > t$ (with $K(t)=\HilDim$ for maximally athermal systems). Thus, $K^{\,}_{\rm \small in}(t) \ll t$ instead suggests that the SFF is not well defined (i.e., because there is effectively no single operator that generates all time evolution upon absorbing measurements). If one includes the default Stinespring initial state $\bvec{0}$ by writing $\matel{a}{\cdots}{b} \to \matel{a,\bvec{0}}{\cdots }{b,\bvec{0}}$ in \eqref{eq:SFF isomeas post}, one can then absorb the outcome trajectory $\bvec{m}$  into the measurement \emph{unitaries} via 
\begin{align} 
\label{eq:SFF Project Meas Unitary}
    \measunitary^{\vpd}_{\sigma,r} \, , ~ \measunitary^{\dagger}_{\sigma,r} \,  \to \, \Proj{r}{(m^{\vpp}_{t,\sigma,r})}  \, , ~~
\end{align}
at which point the hybrid evolution is no longer unitary. More importantly, the evolution is no longer \emph{periodic}. In practice, the Haar averaging procedure only gives nonzero contributions for the small fraction of trajectories $\bvec{m}$ that are periodic, leading to $K^{\,}_{\rm \small in}(t) \ll t$ \eqref{eq:SFF isomeas post}, meaning that \eqref{eq:SFF isomeas post} is not a valid extension of the SFF to hybrid dynamics, because there is effectively no spectrum. 

If one reverses the order of $\Floq$ and $\Floq^{\dagger}$ in \eqref{eq:SFF isomeas post}, then the corresponding SFF is given by
\begin{align}
    K^{\vpd}_{\rm \small out} (t) &= \sum\limits_{\bvec{m},\bvec{n}} \sum\limits_{a,b=1}^{\HilDim^{\,}_{\rm ph}}  \matel{a,\bvec{m}}{\Floq^{\, t}}{a} \hspace{-0.5mm} \matel{b}{\Floq^{\, t}}{b,\bvec{n}} \,, ~\label{eq:SFF isomeas av}
\end{align}
where ``out'' reflects the fact that the evolutions spawn outward-facing Stinespring kets and bras. Note that including the default initial state and absorbing the trajectory into the evolution via \eqref{eq:SFF Project Meas Unitary} leads to independent sums over all measurement projectors, realizing the identity in place of each measurement, independent of the nature of the measurement. Hence, $K^{\,}_{\rm \small out}(t)$ \eqref{eq:SFF isomeas av} is trivially independent of measurements, and thus not a valid hybrid SFF. Finally, we note that if we do not sum \eqref{eq:SFF isomeas av} over trajectories, the resulting SFF is operator-valued (and acts on the Stinespring Hilbert space), and thus has no obvious interpretation as a spectral probe.

Essentially, there is no valid definition of the hybrid SFF \eqref{eq:SFF def} that results from considering \emph{isometric} measurements. Importantly, we note that these quantities all have analogues in which the measurement channel is unitary. Essentially, any SFF in which the outcome trajectory makes an explicit appearance is not a valid probe of spectral rigidity in the presence of measurements: These quantities either  (\emph{i}) fail to have a spectrum, (\emph{ii}) fail to be sensitive to measurements in any sense, or (\emph{iii}) fail to realize a scalar object. Thus, we conclude that the hybrid SFF must be defined in terms of unitary measurement channels, and that we must simply trace over all Stinespring registers. The resulting hybrid SFF is
\begin{align}
    \label{eq:Hybrid SFF def}
    K^{\vpd}_{\rm \small meas} (t) \, &= \, \trace\limits_{\rm dil} \left[ \, \Floq^{\, t} \, \right] \,  \trace\limits_{\rm dil} \left[ \, \Floq^{- t} \, \right] \, , ~~
\end{align}
which will prove to have all the required properties. The only adjustment required for $ K^{\vpd}_{\rm \small meas} (t) $ \eqref{eq:Hybrid SFF def} to be well behaved is that we must either (\emph{i}) enforce postselection on the two copies by including a ``swap'' operation between the two Stinespring Hilbert spaces or (\emph{ii}) explicitly include the channel that resets the Stinespring qubits to the default state $\ket{0}$. Though we primarily consider the former case (corresponding to postselection), the two possibilities give equivalent predictions in the cases of interest. Additionally, postselection is not a barrier since the SFF cannot be measured experimentally anyway.

\subsection{Ensemble averaging}
\label{subsec:SFF Haar}
The diagrammatic method for averaging over the unitary group is detailed in \cite{BnB}, and its application to the SFF is prescribed in \cite{CDLC1}. This procedure also generalizes to systems with block structure  \cite{U1FRUC, ConstrainedRUC}, which generally realize a $\size$-dependent Thouless time $\tau^{\,}_{\rm th} \sim \size^z \gg 1$ \cite{ConstrainedRUC}, where $\size$ is the linear size of the system ($\Nsite \sim \size^{\SpaceDim}$). We diagramatically average $K^{\vpd}_{\rm \small meas} (t)$ \eqref{eq:Hybrid SFF def} over the Haar ensemble prior to contending with the measurement gates.

The Haar-averaging procedure applies only to the unitaries $\Haar$ in the block-structured gates $\gate$ \eqref{eq:GenProjGate}. For block $\alpha$ with $n^{\,}_{\alpha}$ states,  $\Haar$ is drawn from the unitary group $\U{n^{\,}_{\alpha}}$ with uniform measure. Calculations for generic (i.e., ``featureless'') circuits realize from taking the limit of a \emph{single} block containing all $\LocDim^{\ell}$ states with  $\Proj{r}{(\alpha)} \to \ident^{\,}_r$. Regarding the various terms in \eqref{eq:HybridFloqDef}, we note that the shift operator $\Tshift$, unitary measurement gates $\measunitary$, and projectors onto blocks $\Proj{r}{(\alpha)}$ are not modified by the Haar-averaging procedure, and so we need only introduce placeholders to keep track of them while averaging the unitary gates these projectors sandwich \eqref{eq:GenProjGate} over the Haar ensemble.

\begin{figure}[t!]
    \centering
    \includegraphics[width=.85\columnwidth]{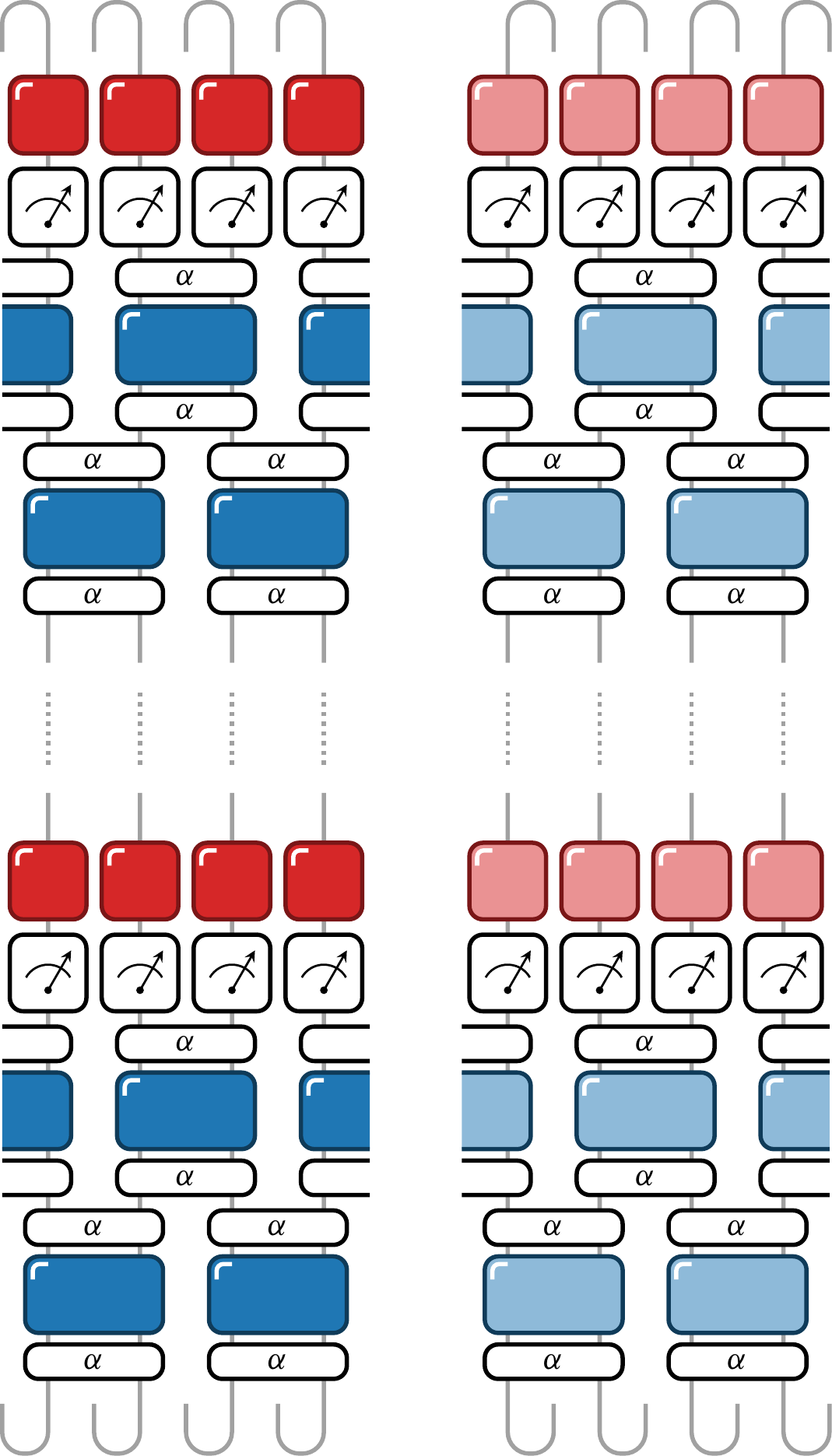}
    \caption{Diagrammatic depiction of the spectral form factor $K(t)$ \eqref{eq:SFF def} for a simple circuit corresponding to the Floquet operator depicted in Fig.~\ref{fig:FloqDiag}. The trace loop on the left involves the product of $t$ copies of $\Floq$, while the trace on the right is a product of $t$ copies of $\Floq^{\dagger}$. The loop on the right has been transposed (rotated by $180^{\circ}$) for convenience of pairing with the loop on the left.  
    }
    \label{fig:SFF start}
\end{figure}

In the absence of block structure, the Haar-averaging procedure for $K(t)$  involves summing over pairings of each of the $t$ Haar-random unitary gates $\Haar^{\,}_{s,\lambda,r}$ that appear in $\Floq^{\, t}$ with their counterparts $\Haar^{*}_{s,\lambda,r}$ in $(\Floq^{\, *})^t$, for $1 \leq s \leq t$. In practice, we replace $\matel{b,\bvec{n}}{\Floq^{\, - t}}{b,\bvec{n}}$ with $\matel{b,\bvec{n}}{(\Floq^{\, *} )^t}{b,\bvec{n}}^{T}$, as depicted in Fig.~\ref{fig:SFF start} for $\ell=2$ with a single round of measurements per time step. Note that the trace on the right---corresponding to $\Floq^*$---has been rotated by $180^{\circ}$ compared to the trace on the left. The unitaries $\gate^{\,}_{\lambda,r,\alpha}$ and their conjugates are distinguished by different shading, and the gates labeled $\alpha$ represent projectors onto dynamical blocks. Additionally, only the physical traces are depicted explicitly (via ``hooks''), and the Stinespring traces are hidden ``behind'' them.

The SFF is the sum of an exponential number of ``diagrams'' in which a given gate $\Haar$ in $\Floq^{\, t}$ is contracted with one of its partners $\Haar^{*}$ in $(\Floq^{\, *})^t$ \cite{BnB,CDLC1}. The diagrammatic method assigns a weight $V$ to each contraction, where the contractions and their weights are determined by the $t$-fold Haar channel \cite{YoshidaCBD}. A convenient approximation of $K(t)$ \eqref{eq:SFF def} corresponds to retaining only the leading-order ``Gaussian'' diagrams \cite{BnB,CDLC1}, which is equivalent to approximating the elements of each unitary gate as Gaussian-random complex numbers, leading to an equivalent of Wick's theorem, so that the corresponding diagrams only contain ``1-cycles'' \cite{BnB,CDLC1}. This approximation is equivalent to taking the ``large-$N$'' limit $\LocDim \to \infty$. 

Note that each non-Gaussian diagram is suppressed by $\LocDim^{-2 \, \ell}$ per ``violation'' of the Gaussian condition (at most $t$ per site). In general, there are far too many non-Gaussian diagrams to enumerate, let alone evaluate. It is also known that the $\Order{1/\LocDim^2}$ corrections  to the Gaussian approximation do not explain the SFF's plateau for $t \geq \HilDim$, nor do they predict a nontrivial Thouless time $t^{\,}_{\rm th} > 1$ in the absence of block structure \cite{CDLC1}. It also appears that the subleading corrections in $\LocDim^{-2}$ do not capture any universal features, as indicated by excellent agreement between the calculation of $K(t)$ with $\LocDim,\Nsite,\tfin \to \infty$ and numerical simulation for $\LocDim=2,\Nsite \sim 14, \tfin \lesssim 100$ in the presence of a $\U{1}$ conserved charge \cite{U1FRUC}.

Consider a calculation involving $N \times N$ Haar-random unitaries $\Haar^{\,}_{\alpha}$ and $\Haar^{\dagger}_{\alpha}$, each of which appears $n$ times (the Haar average is zero if there are different numbers of $\Haar^{\,}_{\alpha}$ versus $\Haar^{\dagger}_{\alpha}$). The weight associated with any Gaussian diagram for this calculation is $V^{\,}_n = N^{-n} + \Order{N^{-n-2}}$. 

In the context of the spectral form factor, in the absence of block structure, there are $t$ Gaussian diagrams (indexed $k \in [0,t-1]$) corresponding to the pairing of all Haar-random gates $\Haar^{\,}_{\lambda,r}$ in each time step $1 \leq s \leq t$ with their conjugates in time step $s+k$. The Gaussian condition (i.e., the limit $\LocDim \to \infty$) demands that \emph{all} gates share the same temporally shifted pairing (with shift $k$).

In the presence of block structure, the particular unitary $\Haar^{\,}_{\lambda,r,\alpha}$ realized by the gate $\gate^{\,}_{\lambda,r}$ \eqref{eq:GenProjGate} is determined by the state upon which $\gate^{\,}_{\lambda,r}$ acts. If the state belongs to the block $\alpha$, then $\gate^{\,}_{\lambda,r}$ acts as $\Haar^{\,}_{\lambda,r,\alpha}$.  The full expression for $K(t)$ is then a sum over all possible ``block trajectories'' of the Floquet circuits $\Floq$ and $\Floq^{\dagger}$, where a given Haar-random unitary appears at most $t$ times. 

However, the Haar average is only nonzero if there are equal numbers of a given Haar-random unitary $\Haar^{\,}_{\alpha}$ and its conjugate $\Haar^{\dagger}_{\alpha}$. More importantly, the Gaussian approximation requires that we pair $\Haar^{\,}_{\alpha}$ in time step $s$ with $\Haar^{\dagger}_{\alpha}$ in time step $s+k$ (for $k \in [0,t-1]$), for all time steps $s$ and all gates in each time step. Hence,  $\Floq^{\dagger}$ must realize the \emph{same} block trajectory $\vec{\alpha}$ as $\Floq$, up to a cyclic temporal shift (modulo $t$). As a result, each Gaussian diagram has an associated weight
\begin{align}
    V^{\,}_{\vec{\alpha}} \, &= \, \prod\limits_{s=1}^{t} \prod\limits_{\lambda=1}^{\ell} \prod\limits_{r \in \lambda} \, 
    \frac{1}{n^{\,}_{s,\lambda,r}} \, , ~~
    \label{eq:DiagramWeight} 
\end{align}
to leading order, where
\begin{equation}
    \vec{\alpha} \, = \, \left\{ \, \alpha^{\vpp}_{s,\lambda,r} \, \middle| \, 1 \leq s \leq t \, , \, 1 \leq \lambda \leq \ell \, , \, r \in \lambda \, \right\}  \label{eq:BlockTrajectory}
\end{equation}
labels all blocks in a given ``block trajectory,'' $n^{\,}_{s,r,\lambda}=\tr{\Proj{r}{(\alpha^{\,}_{s,r,\lambda})}}$ is  the number of states in the local block labeled $\alpha^{\,}_{s,r,\lambda}$, and we neglect subleading terms.

\begin{figure}[t]
    \centering
    \includegraphics[width=.85\columnwidth]{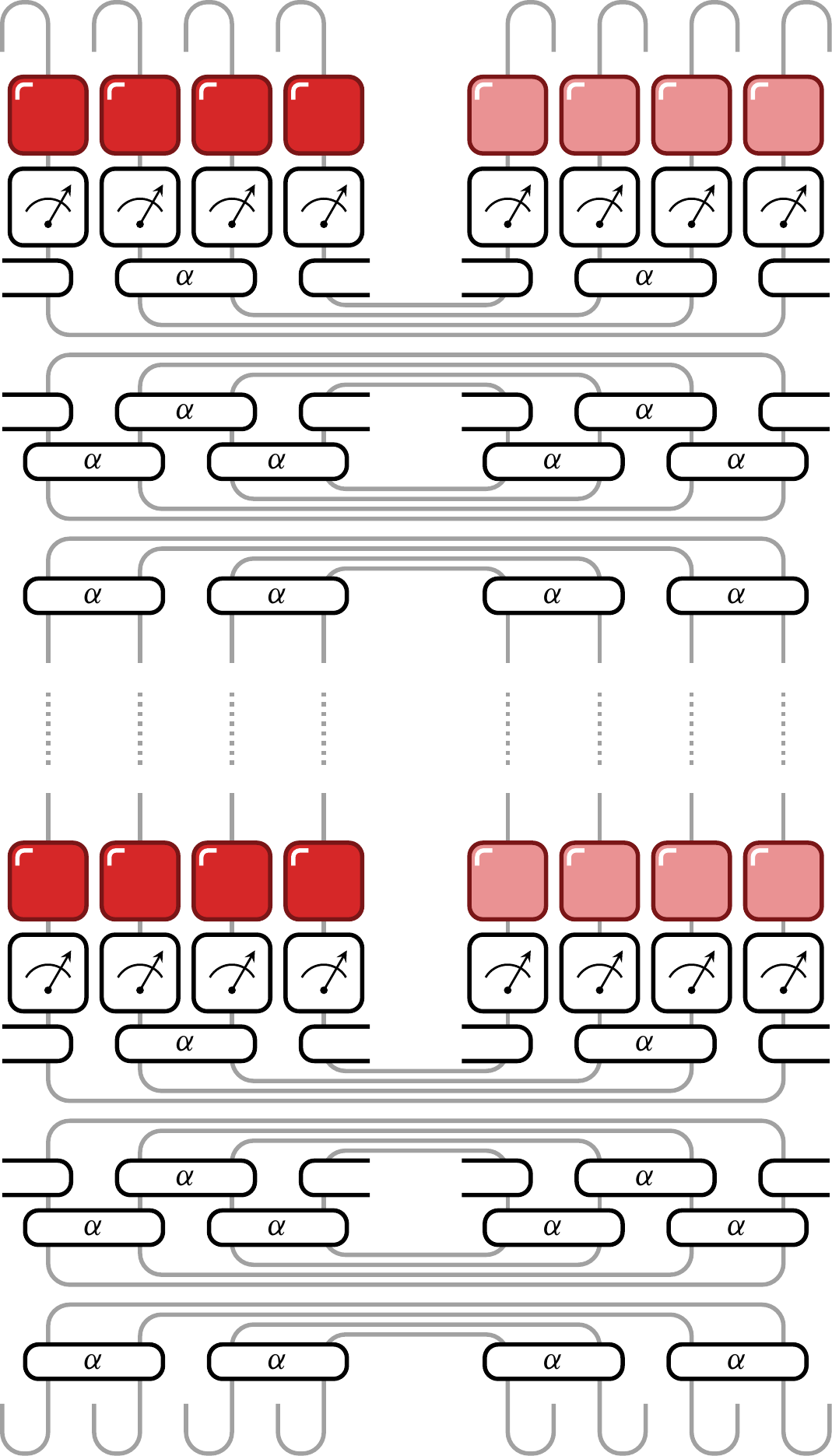}
    \caption{The SFF circuit diagram pictured in Fig.~\ref{fig:SFF start} after implementing the equal-time contractions. Closed loops correspond to traces, and we note that only the physical degrees of freedom are affected by the contractions---all Stinespring registers are affected only through the entangling action of the measurement unitaries $\measunitary^{\,}_{j}$  \eqref{eq:MeasFloquetGate}. 
    }
    \label{fig:SFF contracted}
\end{figure}

A total of $t$ Gaussian diagrams contribute to $K(t)$, all with equal weight \eqref{eq:DiagramWeight}, one of which is always the equal-time contraction depicted in Fig.~\ref{fig:SFF contracted}. The other $t-1$ diagrams can also be expressed as equal-time pairings, where $\Floq^{\dagger}$ has been shifted by $0<k<t$ time steps relative $\Floq$. Periodicity of $\Floq$ and cyclic invariance of the trace then ensure that all $t$ contracted diagrams (as depicted in Fig.~\ref{fig:SFF contracted}) are equivalent to one another.

Thus, to leading order, the SFF \eqref{eq:SFF def} corresponds to $t$ times the sum over block trajectories $\bvec{\alpha} = \{ \alpha^{\,}_{s,\lambda,r} \}$ \eqref{eq:BlockTrajectory} of the diagrammatic weight  $V^{\,}_{\vec{\alpha}}$ \eqref{eq:DiagramWeight} times  the equal-time Gaussian contraction of $\Floq$ with $\Floq^{\dagger}$. The corresponding algebraic expression is a sum of products of traces over the remaining objects in the circuit after Haar averaging (which includes the measurement unitaries, block projectors that enforce the ``block trajectory'',  temporal shift operators, and Stinespring reset channels). The different traces result from the Haar averaging, and join various block projectors and dilated operations into various ``trace loops'' that involve terms from both $\Floq$ and $\Floq^{\dagger}$.

Fig.~\ref{fig:SFF contracted} depicts the trace loops resulting from the equal-time contraction for a $1\SpaceDim$ hybrid circuit with $\ell=2$ layers of time evolution followed by a single layer of single-site measurements in each time step. The pointer dials in Fig.~\ref{fig:SFF contracted} reflect the measurement unitary $\measunitary$ \eqref{eq:MeasFloquetGate}, which acts on ``hidden'' Stinespring registers. The remaining gates (in red) reflect the cyclic temporal shift operator $\Tshift$ \eqref{eq:T shift}, which act only on the Stinespring qubits. These shift operations can be absorbed into the Stinespring part of the measurement operations (denoted by pointer dials), so that each measurement gate is coupled to a distinct Stinespring register (in both the $\Floq$ and $\Floq^{*}$ copies).

At this stage, we can eliminate the Stinespring part of the traces in \eqref{eq:Hybrid SFF def}, hidden in Figs.~\ref{fig:SFF start} and \ref{fig:SFF contracted}. First, consider the case in which we include the Stinespring reset channel prior to each measurement and absorb the temporal shifts $\Tshift$ into all measurement unitaries $\measunitary$. While the Haar averaging joins the two copies of the evolution into a single trace over \emph{physical} degrees of freedom, the Stinespring traces remain separate. For a given measurement gate $s,\sigma,r$, the two Stinespring traces correspond to
\begin{subequations}
    \label{eq:SFF reset SS trace loops}
    \begin{align}
        \trace\limits_{\tau,r} \left[ \, \Shift{m}{r,\tau} \, \sum\limits_{m'=0}^{\Noutcome^{\vpp}_{\tau,r}} \, \BKop{0}{m'}^{\vpp}_{r,\tau} \, \right] \, &= \, 1 \label{eq:SFF reset SS trace loop Floq} \\
        \trace\limits_{\tau,r} \left[ \,  \sum\limits_{n'=0}^{\Noutcome^{\vpp}_{\tau,r}} \, \BKop{n'}{0}^{\vpp}_{r,\tau} \, \Shift{-n}{r,\tau} \, \right] \, &= \, 1 \, , ~~\label{eq:SFF reset SS trace loop Floq Dag}
    \end{align}
\end{subequations}
where $m$ and $n$ denote the measurement projectors $\Proj{m,n}{r,0}$ (the physical part of the measurement unitary $\measunitary^{\,}_{\tau,r}$) in $\Floq$ and $\Floq^{*}$, respectively. In other words, the Stinespring traces always return one, leaving behind individual sums over projectors, which resolve the identity and vanish. 

Alternatively, consider the case in which we ``postselect'' by including a swap operation in the two traces, i.e.,
\begin{align}
    K^{\vpd}_{\rm \small post} \, &\equiv \, \left. \trace\limits_{\rm dil} \right._{\raisebox{1.5mm}{\text{\footnotesize{$(2)$}}}}  \left[ \, \hat{S}^{\vpp}_{\rm ss} \, \Floq^{\, t} \otimes \Floq^{-t} \, \right] \, , ~
    \label{eq:SFF umeas post}
\end{align}
where $\hat{S}$ is a swap operator. This operator satisfies
\begin{align}
\label{eq:Hilbert swap}
    \left. \trace\limits_{\rm ss} \right._{\raisebox{1.5mm}{\text{\footnotesize{$(2)$}}}} \left[ \hat{S} \, A^{\vpp}_{\rm ss} \otimes B^{\vpp}_{\rm ss} \, \right] \, = \, \trace\limits_{\rm ss} \left[ A^{\vpp}_{\rm ss} \,  B^{\vpp}_{\rm ss} \, \right] 
\end{align}
and, as a result, the two Stinespring traces for a given measurement unitary become a single trace given by
\begin{align}
    \label{eq:SFF post SS trace loop}
     \left. \trace\limits_{\rm ss} \right._{\raisebox{1.5mm}{\text{\footnotesize{$(2)$}}}} \left[ \, \Shift{-n}{r,\tau}  \, \Shift{m}{r,\tau} \, \right] \, = \, \kron{m,n} \, , ~~
\end{align}
where we omit the reset operations in this case. Hence, the swap operation $\hat{S}$ \eqref{eq:Hilbert swap} on the Stinespring Hilbert spaces enforces postselection in \eqref{eq:SFF umeas post}. 

We now consider this postselected variant of the hybrid SFF \eqref{eq:Hybrid SFF def}. It is straightforward to generalize the scenario depicted in Fig.~\ref{fig:SFF contracted} to arbitrary hybrid Floquet unitaries with any $\ell$, $\SpaceDim$, and number of measurement layers $\MeasRounds$ per time step. The projectors onto time-evolution blocks---denoted by rectangular boxes labeled $\alpha$---are constrained by the Haar-averaging procedure to realize the same ``block trajectory'' in both the $\Floq$ and $\Floq^{*}$ traces depicted in Fig.~\ref{fig:SFF start}. Hence, the block projectors at equal times in each copy correspond to the same block. Because they are idempotent \eqref{eq:BloccGateProjectorCompleteness}, the trace loops in Fig.~\ref{fig:SFF contracted} corresponding to consecutive layers of time-evolution gates $\lambda,\lambda+1$ (with no intervening measurements) can be simplified according to $\tr{\prod_{r,r'} P^2_r Q^2_{r'}}=\tr{\prod_{r,r'}  P^{\,}_r Q^{\,}_r}$, where $P^{\,}_r$ and $Q^{\,}_{r'}$ are placeholders for the projectors in the top and bottom layer of each such loop.

With $\ell$ time-evolution layers $\lambda$ per time step $s$ \eqref{eq:TevoStep}, we label the layer $s,\lambda$ as $\varsigma \, = \, \ell \, s + \lambda$. The SFF is then
\begin{align}
\label{eq:SFF m post Haar traces}
    \overline{K^{\vpp}_{\rm post} (t) } \, &= \, t\,  \sum\limits_{\vec{\alpha},\vec{m}} 
    \, V^{\vpp}_{\vec{\alpha}} \, \prod\limits_{\varsigma=1}^{\ell\, t} \,  \,  \tr{ \, \Proj{\varsigma, \vec{m}^{\,}_{\varsigma}}{\dagger} \, \Proj{\varsigma, \vec{m}^{\,}_{\varsigma}}{\vpd} \, } \, , 
\end{align}
where $V^{\vpp}_{\vec{\alpha}}$ is the diagrammatic weight \eqref{eq:DiagramWeight} for block trajectory $\vec{\alpha}$, $\vec{m}$ labels outcome trajectories, and the projectors above are defined by the time-ordered products
\begin{align}
\label{eq:SFF m post Haar loop projectors}
    \Proj{\varsigma; \vec{m}^{\,}_{\varsigma}}{\vpd} \, &= \, \Proj{}{(\alpha^{\vpp}_{\varsigma+1})} \, \Big[  \prod\limits_{\sigma=\varsigma}^{\varsigma+1} \, \Proj{}{(m^{\vpp}_{s,\sigma})}\, \Big] \, \Proj{}{(\alpha^{\vpp}_{\varsigma})} 
    \, , ~~
\end{align}
where the individual projectors above act on $\Hilbert^{\,}_{\rm ph}$ as either
\begin{subequations}
\label{eq:SFF projector layers def}
\begin{align}
    \Proj{}{(\alpha^{\vpp}_{\varsigma})} \, &= \, \bigotimes\limits_{r \in \lambda} \, \Proj{r}{(\alpha^{\vpp}_{s,\lambda,r})} \label{eq:SFF Tevo projector layer} \\
    \Proj{}{(m^{\vpp}_{s,\sigma})} \, &= \, \bigotimes\limits_{r \in \sigma} \, \Proj{r}{(m^{\vpp}_{s,\sigma,r})} \label{eq:SFF Tevo meas layer} \, , ~~
\end{align}
\end{subequations}
and the middle terms in \eqref{eq:SFF m post Haar loop projectors} result from the postselected Stinespring traces \eqref{eq:SFF post SS trace loop}, which ensure that the physical measurement projectors \eqref{eq:SFF Tevo meas layer} left over from the measurement unitaries in both copies of the evolution realize the same trajectory $\vec{m}$. The projector \eqref{eq:SFF m post Haar loop projectors} reflects both the block and outcome trajectories indicated in \eqref{eq:SFF m post Haar traces}. Note that \eqref{eq:SFF m post Haar loop projectors} allows for any number of measurement rounds  between time-evolution layers $\varsigma$ and $\varsigma+1$; and if $\varsigma = \ell t$ is the final time-evolution layer, then $\varsigma+1 = 1$ is the first time-evolution layer.


Returning to the contraction in Fig.~\ref{fig:SFF contracted}, there are two types of trace loops corresponding to $\varsigma = s,1$ and $\varsigma = s,2$. In the former case, the trace loop in \eqref{eq:SFF m post Haar traces} involves the time-evolution layers $s,1$ and $s,2$ in the same time step $s$, with no intervening measurements. In the latter case, the trace loop includes projectors corresponding to all  measurements that occur between time-evolution layers $s,2$ and $s+1,1$. In general, we allow for arbitrary numbers of measurement rounds between time-evolution layers $\lambda$.

\subsection{Generic circuits}
\label{subsec:SFF no sym}
The evaluation of \eqref{eq:SFF m post Haar traces} for hybrid circuits with generic time evolution (i.e., no block structure) is straightforward. The important simplifications to the equations in Sec.~\ref{subsec:SFF Haar} follow from replacing the block projectors \eqref{eq:GenProjGate} according to $\Proj{r}{(\alpha)} \to \ident^{\,}_r$, so that the single ``block'' contains all $\LocDim^{\ell}$ states. There is only one ``block trajectory'' in \eqref{eq:SFF m post Haar traces}, with weight $V \, = \, \LocDim^{-\ell \, \Nsite \, t} $ \eqref{eq:DiagramWeight}. We have
\begin{equation}
    \label{eq:SFF m post Haar traces no sym}
    K^{\,}_{\rm \small post} (t)\, = \, t \, \sum\limits_{\bvec{m}} \, \prod\limits_{\varsigma=1}^{\ell\, t} \, \frac{1}{\LocDim^{\Nsite}} \,  \tr{ \, \Proj{\varsigma, \vec{m}^{\,}_{\varsigma}}{\dagger} \, \Proj{\varsigma, \vec{m}^{\,}_{\varsigma}}{\vpd} \, } \, , ~~
\end{equation}
where the projectors in \eqref{eq:SFF m post Haar loop projectors} simplify to
\begin{align}
    \Proj{\varsigma, \vec{m}^{\,}_{\varsigma}}{\dagger} \, \Proj{\varsigma, \vec{m}^{\,}_{\varsigma}}{\vpd} \, &= \,  \prod\limits_{\sigma = \sigma^{\,}_{\MeasRounds}}^{\sigma^{\,}_1} \, \Proj{}{(m^{\vpp}_{s,\sigma})}   \prod\limits_{\sigma' = \sigma^{\,}_1}^{\sigma^{\,}_{\MeasRounds}} \, \Proj{}{(m^{\vpp}_{s,\sigma'})} \, , ~~
    \label{eq:SFF no sym proj factor}
\end{align}
and using idempotency  \eqref{eq:ProjectorHermitianIdempotent}, the middle two copies of $\Proj{}{(m^{\vpp}_{s,\sigma})}$ become a single $\Proj{}{(m^{\vpp}_{s,\sigma})}$. Summing  $\Proj{}{(m^{\vpp}_{s,\sigma})}$ over outcomes $m^{\vpp}_{s,\sigma^{\,}_1}$ resolves the identity by completeness \eqref{eq:ProjectorOrthoComplete}, removing this  measurement layer from \eqref{eq:SFF no sym proj factor}. We then repeat this procedure for the two copies of $\Proj{}{(m^{\vpp}_{s,\sigma+1})}$ in the middle of the product \eqref{eq:SFF no sym proj factor}, and so on. Summing \eqref{eq:SFF no sym proj factor} over all measurement trajectories resolves the identity, so that \eqref{eq:SFF m post Haar traces no sym} becomes
\begin{align}
    \overline{K^{\,}_{\rm \small post} (t)} \, &= \,  t \, \prod\limits_{\varsigma=1}^{\ell\, t} \, \frac{1}{\LocDim^{\Nsite}} \,  \sum\limits_{\vec{m}^{\,}_{\varsigma}} \tr{ \, \Proj{\varsigma, \vec{m}^{\,}_{\varsigma}}{\vpd} \, } \notag \\
    &= \, t \, \prod\limits_{\varsigma=1}^{\ell\, t} \, \frac{1}{\LocDim^{\Nsite}} \, \tr{ \, \ident \, } \notag \\
    &= \, K^{\vpp}_{\rm CUE} (t) \, = \, t \, , ~~\label{eq:SFF result no sym}
\end{align}
which is identical to the SFF for time evolution alone \cite{CDLC1}, which corresponds to the RMT prediction for unitary gates of any size drawn from the CUE \cite{BnB,CDLC1}. Importantly, the fact that the measurements disappear on average is not trivial, as is the case for alternative definitions of the SFF other than \eqref{eq:SFF umeas post}. Thus, on average, measurements have no effect on spectral properties in generic chaotic models with neither symmetries nor constraints.

\subsection{Enriched circuits}
\label{subsec:SFF sym commute}
We now evaluate \eqref{eq:SFF m post Haar traces} for hybrid circuits in which the projectors onto dynamical blocks \eqref{eq:GenProjGate} and 
measurement outcomes \eqref{eq:ASpectralDecomp} commute. This corresponds to measuring operators in the charge (or computational) basis, which we take to be the Weyl $\weight{}$ basis, without loss of generality. We again consider the projectors in \eqref{eq:SFF m post Haar traces}, which correspond to closed loops in the diagram in Fig.~\ref{fig:SFF contracted}. 

Importantly, because the projectors in \eqref{eq:SFF m post Haar loop projectors} mutually commute, we can write them in any order. Using idempotency of the projectors, we then write
\begin{align}
\label{eq:SFF commuting sym proj factor}
    \Proj{\varsigma, \vec{m}^{\,}_{\varsigma}}{\dagger} \, \Proj{\varsigma, \vec{m}^{\,}_{\varsigma}}{\vpd} \, &= \, \Proj{}{(\alpha^{\vpp}_{\varsigma+1})} \, \Big[  \prod\limits_{\sigma = \sigma^{\,}_{1}}^{\sigma^{\,}_{\MeasRounds}} \, \Proj{}{(m^{\vpp}_{s,\sigma})} \, \Big] \, \Proj{}{(\alpha^{\vpp}_{\varsigma})}  \, , ~~
\end{align}
and the trace of this quantity can be written as
\begin{align}
\label{eq:SFF charge meas pre outcome avg}
    \inprod{a^{\prime}_{\varsigma+1}}{m^{\,}_{s,\sigma^{\,}_{\MeasRounds}}} \hspace{0.5mm} \matel{m^{\,}_{s,\sigma^{\,}_{\MeasRounds}}}{\, \cdots \,}{{m^{\,}_{s,\sigma^{\,}_{1}}}} \hspace{-0.5mm} \inprod{{m^{\,}_{s,\sigma^{\,}_{1}}}}{a^{\,}_{\varsigma}} \, ,~~
\end{align}
so that each $\BKop{m}{m}$ term appears exactly once in the SFF \eqref{eq:SFF m post Haar traces}. Summing over measurement outcomes gives $\sum_m \BKop{m}{m} = \ident$  by completeness of the measurement projectors \eqref{eq:ProjectorOrthoComplete}. Averaged over outcomes, \eqref{eq:SFF charge meas pre outcome avg} reduces to $\inprod{a^{\prime}_{\varsigma+1}}{a^{\vpp}_{\varsigma}}$, which is exactly the same result that obtains in the absence of measurements \cite{U1FRUC,ConstrainedRUC}.

For concreteness, \eqref{eq:SFF commuting sym proj factor} allows the SFF \eqref{eq:SFF umeas post} to be written in the standard form \cite{U1FRUC, ConstrainedRUC}
\begin{align}
\label{eq:SFF result evo version}
    \overline{K (t) } \, &= \, t \, \tr{ \, \Tmat{}{\, t} \,} \, , ~~
\end{align}
i.e., $t$ times the [physical] trace of the $t$th power of the time-evolution \emph{transition matrix} for a single period,
\begin{align}
    \tmat \, &\equiv \, \prod\limits_{\lambda=1}^{\ell} \, \tmat^{\vpp}_{\lambda} \, = \, \prod\limits_{\lambda=1}^{\ell} \, \bigotimes\limits_{r \in \lambda} \, \tgate^{\vpd}_{\lambda,r} \,, ~~
    \label{eq:SFF Transition Matrix Layer}
\end{align}
where the individual time-evolution gates are given by
\begin{align}
    \tgate^{\vpd}_{\lambda,r} \, &\equiv \, \sum\limits_{\alpha} \,\frac{1}{n^{\vpp}_{\alpha}} \, \sum\limits_{a,b \in \alpha} \, \BKop{a}{b}^{\vpp}_r \, ,~~
    \label{eq:SFF Transition Matrix Gate}
\end{align}
which is identical to the superoperator gate $ {\tgate}$ responsible for the time evolution of operators \eqref{eq:Tgate result}; the only difference is that the $\LocDim^{\ell}$ diagonal operators $\nProj{}{(a,b)}$ have been replaced by the $\LocDim^{\ell}$ \emph{states} $\ket{a,b}$ in \eqref{eq:SFF Transition Matrix Gate} \cite{U1FRUC,ConstrainedRUC}. 

At the same time, the sum over measurement outcomes---enforced by the postselection condition \eqref{eq:SFF post SS trace loop}---ensures that the measurements drop out of the transition matrix $\tmat$ \eqref{eq:SFF Transition Matrix Layer}, as in the evaluation of $n$-point functions \eqref{eq:n-point-expval} in Sec.~\ref{sec:OneCopy}. The fact that the measurement projectors commute with the projectors onto dynamical blocks (and each other) ensures that every pair of projectors onto the outcomes of a given measurement can be placed sequentially, and combined into a single projectors. Summing that projector over all outcomes resolves the identity. As a result, the transition matrix $\tmat$ in \eqref{eq:SFF result evo version} is identical to the measurement-free transition matrix. Thus, symmetry-compatible (and/or constraint-compatible) measurements have no discernible effect on the spectral rigidity of chaotic, enriched quantum dynamics, even in the case of the postselected spectral form factor \eqref{eq:SFF umeas post}.

\subsection{Charge-changing measurements}
\label{subsec:SFF charge changing}
We now comment on the measurement of charge-changing operators, where the projectors onto measurement outcomes $\vec{m}$ do not commute with the projectors onto symmetry and dynamical sectors $\vec{\alpha}$ (and may not commute with one another). As a result, the pairs of projectors \eqref{eq:SFF m post Haar loop projectors} onto a given measurement outcome cannot trivially be brought together in the trace loops in \eqref{eq:SFF m post Haar traces}, merged into a single projector, and summed over outcomes to resolve the identity. In general, attempting to do so result in (\emph{i}) the generation of a superposition of terms in the product operator \eqref{eq:SFF commuting sym proj factor}, likely with an $\Order{\LocDim^{-\measrate \, \Nsite}}$ suppression and (\emph{ii}) the association of an $\vec{m}$- and $\vec{\alpha}$-dependent phase factor $\omega^{f(\alpha,m)}$. Each of these complicates the sum over outcomes and interpretation of the resulting trace term(s) in \eqref{eq:SFF m post Haar traces}. 

It does not appear possible to recover generic results for the postselected SFF \eqref{eq:SFF umeas post} in the case of charge-changing measurements. Meanwhile, the SFF with the reset operation included---and independently summed over outcome trajectories of the two copies of the evolution operators---is trivially insensitive to the effects of measurement. This also provides evidence that the postselected SFF \eqref{eq:SFF umeas post} is not trivial \emph{by construction}, but that the symmetry-compatible (and/or constraint-compatible) measurements genuinely have no effect, on average, on the spectral properties of enriched chaotic dynamics (nor do measurements have any effect on the spectral properties of featureless chaotic dynamics). What is clear is that charge-changing measurements destroy the conservation laws of the underlying time evolution (as seen in Sec.~\ref{subsec:Observ Meas Other Pauli}). However, it is not clear whether the resulting SFF is corresponds to generic chaotic time evolution \eqref{eq:SFF result no sym}, and it does not appear possible to extract a Thouless time.

\subsection{Including the density matrix}
\label{subsec:SFF DM}
While we have thus far considered Floquet circuits with discrete time-translation symmetry, it is perhaps more common to define the SFF for Hamiltonian dynamics \cite{YoshidaSFF, SubirSFF} with \emph{continuous} time-translation invariance,
\begin{equation}
    \label{eq:Ham SFF}
    K(t) \, = \, \sum\limits_{m,n=1}^{\HilDim} \, e^{\ii \, t \, \left( E^{\,}_m - E^{\,}_n \right)} \, = \,\abs{\tr{ \, e^{-\ii \, t\, \Ham} \,}}^2 \, , ~~
\end{equation}
where $\{ E^{\,}_m\}$ are the eigenstates of the Hamiltonian $\Ham$. 

While Floquet systems do not conserve energy, and generically heat up to infinite temperature $\beta \to 0$, in the case of Hamiltonian dynamics \eqref{eq:Ham SFF}, one can also define a temperature-dependent spectral form factor \cite{SubirSFF}. Using the inverse temperature $\beta = T^{-1}$, one first defines the finite-temperature partition function,
\begin{equation}
\label{eq:Partition Func}
    Z (\beta ) \, \equiv \, \tr{ \, e^{- \beta \, \Ham} \,}  ~,~~\text{where}~~\DensMat^{\vpp}_{\beta} = \frac{e^{-\beta \, \Ham}}{Z (\beta)} \, ,~~
\end{equation}
so that the corresponding finite-temperature SFF is
\begin{align}
    \label{eq:SFF with beta 1}
    K^{\vpp}_{\rm th} ( t, \beta ) = \HilDim^2 \,  \overline{  Z (\beta + \ii \, t ) \, Z (\beta - \ii \, t)  \, / \, Z (\beta)^2  }  \, ,~
\end{align}
where the overline denotes ensemble averaging; in some conventions \cite{SubirSFF}, the denominator and numerator in \eqref{eq:SFF with beta 1}  are separately averaged\footnote{The ``finite-temperature'' extension of the operator inner product is $\opinprod{A}{B}^{\,}_{\beta} = \tr{A^{\dagger} \sqrt{\DensMat^{\,}_{\beta}} B \sqrt{\DensMat^{\,}_{\beta}}}$ \cite{XiaoAndyDensity, CLYbounds}. Then $K^{\,}_{\rm th} (t,\beta/2) = \sum_{a,b=1}^{\HilDim} \, ( \,\nbasisop{a,b}(0)\, |\, \nbasisop{a,b}(t) \, )^{\,}_{\beta}$ is a sum over (valid but nonstandard) correlation functions of basis operators at inverse temperature $\beta$.} Also note that \eqref{eq:SFF with beta 1} correctly reproduces \eqref{eq:Ham SFF} in the limit $\beta \to 0$ ($T \to \infty$), where $ Z (0) = \HilDim$ and $Z (\beta \pm \ii \, t ) \to  e^{\mp \ii \, t \, \Ham}$.  

However, we are primarily concerned with the SFF in the context of Floquet dynamics, where there is no (extensive) conserved energy. Accordingly, temperature is ill defined, and effectively infinite. However, we can instead view $\DensMat^{\vpp}_{\beta}$ in \eqref{eq:Partition Func} as the late-time density matrix $\DensMat^{\vpp}_{\beta} \to \DensMat (\tau)$, which realizes $\DensMat^{\vpp}_{\beta}$ for Hamiltonian dynamics at sufficiently late times $\tau$ (more chaotic systems requires shorter times $\tau$ to equilibrate). 

Making this replacement in \eqref{eq:SFF with beta 1} gives
\begin{align}
    K^{\vpp}_{\rm \small st}  (t,\tau) 
    &\to \HilDim^2 \, \overline{ \tr{ \, \DensMat (\tau) \, \evo (t) \,} \, \tr{ \, \DensMat (\tau) \, \evo^\dagger (t) \,} } \notag \\
    &=  \HilDim^2 \,  \overline{ \abs{ \tr{ \DensMat (0) \, \evo (t) \,} }^2} \, ,~~ \label{eq:SFF dens mat 0}
\end{align}
which is independent of $\tau$ for \emph{either} Hamiltonian ($\evo (t) = e^{-\ii \, t \, \Ham}$) or Floquet ($\evo (t) = \Floq^t$) evolution, since $\DensMat(\tau) = \evo(t) \DensMat(0) \evo^\dagger(t)$. We recover the standard form \eqref{eq:SFF def} upon taking $\DensMat(0) \to \DensMat^{\,}_{\infty} = \ident / \HilDim$, and more generally, a state-dependent SFF given by
\begin{align}
    K^{\vpp}_{\rm \small st}  (t, \DensMat^{\vpp}_0 ) \, &\equiv \, \HilDim^2 \,  \overline{ \abs{ \tr{ \DensMat^{\vpp}_0 \, \evo (t) \,} }^2} \, ,~~ \label{eq:SFF dens mat 1}
\end{align}
which we now evaluate in the presence of measurements.

Upon including the initial  density matrix via \eqref{eq:SFF dens mat 1}, the $t$ different Gaussian diagrams that contribute to \eqref{eq:SFF m post Haar traces} correspond to the $t$ different relative positionings of the two copies of $\DensMat^{\,}_0$ in $\Floq$ versus $\Floq^{*}$. The $t-1$ terms in which the two copies of $\DensMat^{\,}_0$ appear at different times always contain two trace loops with one copy of $\DensMat^{\,}_0$ each. Importantly, the two trace loops with $\DensMat^{\,}_0$ can be treated as before: Both in the absence of block structure and in the presence of structure-compatible measurements, cyclic invariance of each trace loop allows for the projectors corresponding to a given measurement to be brought together and summed to the identity, independent of the presence $\DensMat^{\,}_0$. The remaining term (with two copies of $\DensMat^{\,}_0$ in the same trace loop) cannot be treated in this way. 

However, this term is equivalent to its measurement-free analogue in the absence of block structure, or if the measurements and initial density matrix $\DensMat^{\,}_0$ commute with the block structure. 
In fact, even for unphysical choices of $\DensMat^{\,}_0$ that are not compatible with the block structure, we still find that $K^{\,}_{\measrate}(t) = K^{\,}_0 (t) + \Order{1}$ is asymptotically independent of the measurement rate $\measrate$.

Thus, we find that $K^{\,}_{\measrate} (t,\DensMat^{\,}_0) = K^{\,}_{0} (t,\DensMat^{\,}_0)$ (the measurement-free SFF) in featureless chaotic dynamics and in the case where the measurements and $\DensMat^{\,}_0$ are compatible with any symmetries and/or constraints of the time evolution. More generally, one should consider \eqref{eq:SFF dens mat 0} with $\DensMat (\tau)$ replaced by some density matrix of interest, i.e.,
\begin{align}
    K^{\vpp}_{\rm st}  (t,\DensMat) \,  &\equiv \, \HilDim^2 \,  \overline{ \abs{ \tr{ \DensMat \, \evo (t) \,} }^2} \, ,~~ \label{eq:SFF dens mat}
\end{align}
where, e.g., $\DensMat \propto \exp( - \mu \sum_j \charge^{\,}_j )$ corresponds to some conserved quantity. In this case, the analysis of \eqref{eq:SFF dens mat 1} holds, but now, $\DensMat$ is guaranteed to commute with any symmetries of the underlying time evolution. As a result, the hybrid, state-dependent SFF \eqref{eq:SFF dens mat} for any measurement rate $\measrate$ is exactly equal to its measurement-free ($\measrate=0$) counterpart. Hence, we conclude that even the postselected, quadratic-in-$\DensMat$ hybrid spectral form factor has only asymptotically vanishing and nonuniversal dependence on the measurements; in the cases of greatest interest, there is no effect whatsoever due to measurements.

\section{Measurement-induced entanglement transitions do not separate distinct phases of matter}
\label{sec:can't work}
In Secs.~\ref{sec:OneCopy} and \ref{sec:SFF}, we found that measurements generically have no experimentally observable effect on the universal properties of the underlying chaotic time evolution, except for possibly destroying symmetries of the underlying dynamics. In other words, measurements either have no effect whatsoever, or trivialize universal features present without measurements. In the former case, no quantity depends on the measurement rate $\measrate$; in the latter case, the dynamics may exhibit a crossover as a function of $\measrate$ from symmetric to featureless. 

The generic indifference of correlation functions, linear response functions, and spectral statistics to the inclusion of projective measurements distinguishes the measurement-induced entanglement transition (MIET) \cite{og-MIPT, FisherMIPT1, chan, FisherMIPT2, ChoiQiAltman, RomainHolographic2, RUCreview, ZabaloMIPT, GullansHuseOP, GullansHusePRX, MIPT-ATA, UtkarshChargeSharp, hsieh, barkeshli, MIPT-exp} from all known conventional and topological phase transitions. Perhaps surprisingly, the results of Secs.~\ref{sec:OneCopy} and \ref{sec:SFF} further distinguish the MIET from \emph{thermalization} transitions, which are also associated with a transition from area- to volume-law scaling of entanglement entropy. 

These results suggest that (\emph{i}) projective measurements do \emph{not} compete with chaotic time evolution \emph{except} in the context of trajectory-resolved measures of entanglement, (\emph{ii}) relatedly, standard diagnostics of phase structure do not undergo transitions as a function of $\measrate$, (\emph{iii}) measurements can only \emph{remove} universal properties of the underlying chaotic time evolution as diagnosed by such probe, and (\emph{iv}) functional dependence of a quantity on the density matrix $\DensMat$ is not the crucial factor in whether a transition as a function of $\measrate$ is possible.

Together, these findings raise the question of whether the MIET and similar transitions \cite{og-MIPT, FisherMIPT1, chan, FisherMIPT2, ChoiQiAltman, RomainHolographic2, RUCreview, ZabaloMIPT, GullansHuseOP, GullansHusePRX, MIPT-ATA, UtkarshChargeSharp, hsieh, barkeshli, MIPT-exp} are  transitions between distinct phases of matter in any physically meaningful---and historically consistent---sense of the term. We now argue that they are not: We explicitly enumerate widely recognized criteria for what constitutes a phase of matter, establish that no postselected probe can fulfill these criteria, and  further comment on various claims and proposed workarounds from the literature on MIETs, arguing that none fulfill the criteria for a phase of matter. We also distinguish certain ``classifier'' transitions from physical phase transitions, before discussing the requirements for a quantity to be sensitive to the measurement rate $\measrate$, finding that the measurement outcomes must be \emph{utilized}. Finally, we identify outcome-dependent feedback (adaptive dynamics) as the \emph{only} route to realizing genuine measurement-induced \emph{phase} transitions.

\subsection{What is a phase of matter?}
\label{subsec:what is a phase}
Before discussing the measurement-induced entanglement transition itself, we first outline the physical conditions that phases of matter must fulfill. We identify a pair of physical criteria on the types of probes that can identify phases of matter, which apply to \emph{all} known  phases of matter (conventional or topological, both in and out of equilibrium). At the same time, relaxing either of these criteria removes all physical meaning associated with the concept of a phase of matter. 

The criteria follow from simple requirements that have always been associated with the term ``phase of matter,'' along with the fact that phases of matter are \emph{only} well defined in the thermodynamic limit. The first requirement is that phases of matter be \emph{physical}---meaning that their physical existence and experimental detection is possible in principle (i.e., cannot be ruled out). The second requirement is that knowledge of the phase  realized by a particular sample or material must furnish predictions about the ``universal'' and physically detectable properties of that sample (i.e., dynamical and/or thermal properties that are not sensitive to microscopic details).

We now formally enumerate the two physical criteria that enforce the foregoing requirements constraining genuine phases of matter. Namely, a phase of matter must be detectable (\emph{i}) using experimentally measurable quantities whose detection requires resources (e.g., run time) that scale at most polynomially in the number of degrees of freedom $\Nsite$ and (\emph{ii}) the existence and detection of this property must be robust to microscopic detail.

The first criterion (\emph{i}) simply requires that the detection of any genuine phase of matter be possible \emph{in principle} even in the thermodynamic limit. Note that one can ascribe universal properties not yet realized in any experiment to a legitimate phase of matter; however, the same cannot be said for properties that provably cannot be observed in the thermodynamic limit. The latter condition is captured by the requirement that the resources required to detect the phase  scale at most polynomially in system size $\Nsite$ (e.g., the total experimental run time must satisfy $t \leq \text{poly}(\Nsite)$), where the polynomial has bounded degree that does not scale with $T$, $\Nsite$, or the linear dimension $\size$ (this is consistent with historical examples).

The second criterion (\emph{ii}) requires that phases of matter \emph{not be fine tuned}. Otherwise, any instance of any sample would qualify as a phase, stripping the term of meaning. At the same time, knowledge of the phase of the sample would not afford predictions about universal properties particular to that sample. Genuinely universal features would be shared by other ``phases'' (suggesting that other, microscopically distinct samples lie in the same phase), while nonuniversal features may not manifest in every experimental shot due to fluctuations and microscopic variations (suggesting that the phase itself lacks the required predictive property). Hence, this criterion requires that a notion of universality exist, which must be robust to \emph{some} class of perturbations and microscopic variations, and must specifically include sample-to-sample and shot-to-shot variations in experiments. 

Importantly, any proposed phase of matter whose detection does not satisfy the foregoing criteria lacks at least one of the required features of a phase of matter. If the signatures of a proposed phase of matter cannot be observed at all, then there is no physical meaning to those signatures, and they cannot be associated with a phase of matter. On the other hand, suppose that the signatures of a proposed phase require a time $t \sim \exp(\Nsite)$ to observe in experiment. Then those signatures can only be observed far away from the thermodynamic limit, where phases of matter are not well defined\footnote{We stress that experiments performed far from the thermodynamic limit have no bearing on whether a phase of matter exists.}; in the thermodynamic limit, observation of a single sample in the phase would be impossible, so the phase does not exist in any physically meaningful sense. Additionally, if the signatures of a proposed phase of matter are not robust to microscopic detail, then every instance of every object could be considered a phase of matter. More importantly, however, without robustness, observation of the phase may destroy the phase itself, so that knowledge of a sample's ``phase'' no longer affords universal predictions.

A key distinction between genuine phases of matter (as constrained above) and properties of particular quantum states follows from the notion of dynamical stability. While numerous phases of matter can be distinguished using a single experimental probe (e.g., a ferromagnet is distinguished by measuring the magnetization, a spin liquid can be diagnosed using transport, scattering experiments~\cite{KnolleFieldGuide}, or, more recently, nonlinear spectroscopy~\cite{nck}, and numerous electronic phases are characterized via transport), other conventional phases may require multiple measurements (and statistics) to distinguish. 

A useful example is the \emph{spin glass} phase \cite{SpinGlassReview,VincentSG}, whose temporally long-ranged order is diagnosed, e.g., via autocorrelation functions. Importantly, this long-range order is a property of the entire spin glass \emph{phase}, and is not sensitive to the exact microscopic state of the system (i.e., generic states within the phase give equivalent predictions). Thus, it is possible to perform the multiple measurements required to diagnose spin-glass order on a single sample using independent shots, without worrying whether disturbances to the state (necessary to determine the phase of the sample) spoil the associated order. 

In contrast, properties of particular states of the system are highly sensitive to such disturbances, and do not admit robust diagnostics as described above. For example, in diagnosing the entanglement of a quantum state---even using ancilla probes---after making the first probe measurement, one must then prepare an \emph{identical} many-body quantum state before performing the next measurement. While each ``shot'' of the experiment may require a time $t \sim \text{poly}(\Nsite)$, the total number of shots---and hence, the total run time---scales as $\exp(\Nsite)$, which violates criterion (\emph{i}). Moreover, the requirement that each sample is prepared and evolved identically violates the robustness criterion (\emph{ii}). We also note that, while the effects of measurements are manifest (in some sense) along particular trajectories $\bvec{n}$, there are no meaningful predictions to be made (as discussed in  Sec.~\ref{subsec:NoTrajectories}); thus, an exponentially large ensemble of shots is required. Thus, we conclude that the need for multiple measurements on a \emph{single} sample is acceptable in a diagnostic of phase structure; the need for a large number of macroscopic samples in identical quantum many-body states is not.

\subsection{Postselection is unphysical}
\label{subsec:no postselect}
As discussed in Sec.~\ref{sec:OneCopy}, the ``standard'' probes of phase structure compatible with the two criteria of Sec.~\ref{subsec:what is a phase} are completely blind to the measurement-induced entanglement transition. This is also true of the experimentally undetectable spectral form factor discussed in Sec.~\ref{sec:SFF}, which is related to correlations functions. In fact, far from showing any sharp feature (as is associated with a transition between distinct phases), in the hybrid protocols relevant to the literature, these quantities  are independent of the measurement rate $\measrate$, on average. 

However, two classes of quantities have been reported in the literature to show a sharp transition at some critical $\measrate^{\,}_c$ \cite{RUCreview}: measures of entanglement \cite{og-MIPT, chan, FisherMIPT1, FisherMIPT2, RomainHolographic2, ChoiQiAltman, UtkarshChargeSharp, hsieh, barkeshli, MIPT-exp, MPAF_cross, GullansHuseOP} and variances of two-point functions \cite{UtkarshChargeSharp,hsieh,barkeshli}. Importantly, these two types of probes do not generally diagnose the same transition---i.e., the charge-sharpening transition is not the MIET \cite{UtkarshChargeSharp}. Note that we do not consider either type of quantity herein, and instead refer the reader to the literature for technical details related to these quantities.  

Rather, we investigate whether the MIET constitutes a transition between two genuine phases of matter, as codified in Sec.~\ref{subsec:what is a phase}. We find that there is no physical sense in which it does so; more generally, we find that \emph{no postselected probe can differentiate distinct phases of matter} in any conventional sense of the term. We now discuss two separate issues---one practical and one conceptual---that render any postselected quantity incompatible with the identification of phase structure.

The \emph{practical} issue with postselection is generally acknowledged in the MIET literature \cite{RUCreview}. Postselected quantities require $\exp(\Nsite)$ resources to measure in an actual experiment. Specifically, postselected quantities are impossible to measure using independent experimental ``shots'' performed on an individual sample; instead, one must coherently interfere the outcomes of multiple independent experiments, or alternatively, prepare multiple identical copies of the sample, subject them to identical hybrid-circuit protocols, and ensure that they realize identical outcomes for \emph{all} measurements in the protocol. 

This requires (\emph{a}) an unknown level of precision in the application of time-evolution gates and (\emph{b}) postselection of measurement outcomes. Even ignoring the former requirement, the latter entails an exponential number of experimental shots, so that the run time is exponential in the spacetime volume $t^{\,}_{\rm run} \sim \exp ( \Nsite \, T)$).

This is often termed the ``postselection problem'' 
\cite{RUCreview}. Consider a $\SpaceDim$-dimensional system with $\Nsite \sim \size^{\SpaceDim}$ $\LocDim$-state qudits, measurement rate $\measrate$, and maximum circuit depth $\tfin$ per shot. In the thermodynamic limit---the only regime in which phases of matter are well defined---both $\Nsite$ and  $\tfin$ diverge. Evaluation of a postselected quantity requires $ \LocDim^{\measrate \,\Nsite \, \tfin}$ shots on average, which is experimentally impossible in the thermodynamic limit (e.g., 4000 shots for $\LocDim=2$, $\Nsite=8$ \cite{MIPT-exp}). We also rule out quantum state tomography as a viable practical probe for the same reason \cite{RUCreview}. 

While \cite{og-MIPT} refers to this postselection problem as a ``severe statistical challenge,'' in reality, it is a genuine \emph{impossibility}: The probability of measuring \emph{any} postselected quantity in \emph{any} finite time is identically zero in the thermodynamic limit. Thus, postselected quantities necessarily violate criterion (\emph{i}) for probes of phase structure, which requires that the phase to which a sample belongs can be determined experimentally in polynomial time (in $\size$), even as the sample size approaches the thermodynamic limit. Additionally, postselected probes almost certainly violate  criterion (\emph{ii}), as they are not robust to shot-to-shot variations (e.g., in the application of time-evolution gates) and are fundamentally fine tuned. While the \emph{existence} of a postselection problem is acknowledged in the MIET literature, the \emph{extent} of that problem has neither been acknowledged nor successfully remedied in the MIET literature, as far as we are aware.

Additionally, we now point out a \emph{conceptual} problem, which does not appear to be recognized in the literature, and which persists \emph{even if the practical postselection issue is remedied}, as far as we are aware. One of the requirements of a phase of matter discussed in Sec.~\ref{subsec:what is a phase} is that knowledge of the phase to which a sample belongs must furnish predictions about that sample. 

Now, suppose that we knew the ``entanglement phase'' of a particular sample (i.e., a particular hybrid circuit protocol and initial state, along with a late-time many-body state of the system whose entanglement entropy has known area- or volume-law scaling). This would only allow for the prediction of postselected quantities like entanglement measures and variances of correlation functions, neither of which can be experimentally observed---much less validated---in any finite time for thermodynamically large systems. Meanwhile, the quantities \eqref{eq:n-point-expval} that \emph{can} be measured in finite time are indifferent to the ``entanglement phase.'' Thus, knowing that a sample realizes some ``phase'' that is only accessible to postselected quantities affords no experimentally testable predictions about the behavior of that sample in the thermodynamic limit, meaning that any such ``phase'' is incompatible with the defining property of a phase of matter. 

We now briefly remark on several proposals for mitigating the postselection problem. For example, the ``scalable'' ancilla probe of \cite{GullansHuseOP} parametrically reduces the number of measurements that require postselection; however, the measurement outcomes within the light cone emanating from the ancilla qubit must be postselected, meaning that the number of shots remains exponentially divergent in the thermodynamic limit. This manifestly violates criterion (\emph{i}) for probes of phase structure. Likewise, the ancilla probe of \cite{UtkarshChargeSharp} only reduces the measurement overhead \emph{per shot}; however, this is not the limiting issue with postselection, and the total cost remains exponential in the full spacetime volume of the hybrid circuit. In other words, none of the ``scalable'' probes of the MIET  reported in the literature scale to the thermodynamic limit, which is the only meaningful notion of scalability. Additionally, the dual-unitary protocol of \cite{DU-MIPT} is fine tuned, violating criterion (\emph{ii}), and requires postselection of measurements on the spacetime boundary, which is unlikely to scale and does not afford useful predictions. It is also unclear whether experimental realizations exist \cite{RUCreview}. Finally, proposals that require concurrent classical simulations with the \emph{same} measurement outcomes also fail to remedy the practical and conceptual issues (e.g., the recent experiment \cite{MIPT-exp} using the ancilla probe of \cite{GullansHuseOP}). 

Hence, we conclude that (\emph{i}) the postselection problem is not merely a statistical challenge, but impossible in the only limit in which phases of matter can be defined; (\emph{ii}) in addition to being experimentally impossible to evaluate in thermodynamically large systems, postselection requires an unknown degree of fine tuning; (\emph{iii}) in addition to the foregoing practical issues, even knowing which postselected probe a particular sample realizes affords no practically testable predictions about that sample, which is a defining property of a phase of matter; and (\emph{iv}) there is no means by which to circumvent these practical and conceptual issues: Any quantities that involve postselection cannot diagnose phase structure.

\subsection{Distinction from ``classifier'' transitions}
\label{subsec:classifiers}
We also distinguish between phases of matter and, e.g., computational complexity classes \cite{Vazirani}. It is commonplace in the literature on, e.g., computer science to refer to sharp transitions between regimes with different notions of complexity and/or simulability as ``phase transitions.'' However, there is no sense in which these transitions amount to transitions between distinct phases of matter.

In particular, knowledge of a system's phase of matter allows for quantitative predictions about observable outcomes of experiments performed on that system. By contrast, knowledge of a problem's computational complexity class allows for predictions about whether a given instantiation of the problem can be solved in a certain amount of time. There are several works in the MIET literature that draw parallels between computational complexity (or simulability) classes and the MIET itself. However, such transitions do not provide a way to detect the MIET in practice, and in no way imply that the MIET realizes a genuine transition between distinct phases of matter. 

For example, the ``learnability transition'' of \cite{learnability} is a simulability transition. Knowledge of the learnability class of the problem predicts \emph{whether} a classifier external to the system can determine the system's total charge after some number of measurements. Not only that, but the learnability transition (for the classifier) does not coincide with the MIET---even if there were a proven and direct relation between the two transitions, the learnability transition still would not ``detect'' the MIET or classify the phase of the system in any meaningful sense. In other words, ``learnability'' is a property of the classifier that reflects  computational complexity; \emph{a priori}, it does not afford any predictions about the observable properties of the system accessible using a single sample. Although the learnability probe does not require postselection---and thereby avoids the issues detailed in Sec.~\ref{subsec:no postselect}---the learnability transition \cite{learnability} does not imply a transition between distinct phases of matter. Analogous considerations apply to the neural-network decoders of \cite{HosseinNN}.

\subsection{Necessity of feedback}
\label{subsec:Need Feedback}
We now seek to identify valid probes of measurement-induced phase structure. It is instructive to consider why postselected quantities like entanglement measures see a measurement-induced transition while standard diagnostics of the form \eqref{eq:n-point-expval} do not. The results of Sec.~\ref{subsec:SFF DM} reveal that nonlinear dependence of a quantity on the density matrix $\DensMat$ is not sufficient to guarantee that that quantity sees a measurement-induced transition. Rather, the reason postselected quantities and certain decoders see a transition is that the measurement outcomes are \emph{utilized}. However, both postselected quantities and decoders utilize the outcomes in a manner that is incompatible with genuine phase structure.

However, a recent cross-entropy diagnostic \cite{MPAF_cross} using concurrent classical simulations successfully resolves the practical issues---and partially remedies the conceptual issues---that afflict other measures of entanglement entropy. Importantly, the cross-entropy probe \cite{MPAF_cross} does \emph{not} require postselection: The classical Clifford simulation uses a different initial state and allows for arbitrary measurement outcomes. In this sense, it also enjoys greater robustness to shot-to-shot variations. More importantly, the computational cost of the classical ``decoding'' procedure scales polynomially---rather than exponentially---in $\Nsite$, meaning that this classically assisted cross-entropy probe can, in principle, be scaled to the thermodynamic limit. However, the cross-entropy protocol is only viable when the time-evolution gates are restricted to Clifford operations, and to the best of our knowledge does not survive generic perturbations, so fails criterion (\emph{ii}) for probes of genuine phases of matter.

We note that, in general, such classical post processing (or ``decoding'') \emph{may be} compatible with nontrivial and scalable probes of phase structure, as noted in \cite{Sam-meas,JongYeonDecode}. In principle, it is possible to identify probes of the MIET that utilize the outcomes for classical decoding, thereby avoiding the trivial results observed in Sec.~\ref{sec:OneCopy}. However, to our knowledge, no such classical prescription has managed to fulfil all of our criteria for a genuine phase of matter, especially as the transition is approached. In particular, it seems likely that the classical decoding procedure of \cite{Sam-meas} become exponentially costly (in system size) as the the measurement rate $\measrate$ approaches the critical value $\measrate^{\,}_c$.

More generally, our analysis in Sec.~\ref{sec:OneCopy} using the Stinespring representation of measurements demonstrates (by contradiction) that the outcomes of measurements must be utilized prior to extracting expectation values. If they are not utilized, hybrid dynamics with any measurement rate $\measrate>0$ are equivalent to chaotic time evolution alone with the same combined symmetries. Hence, it would appear that classical post processing is only viable when applied to the full density matrix for a given outcome trajectory, as with the cross-entropy probe of \cite{MPAF_cross}. However, this is impossible in the thermodynamic limit without fine tuning (e.g., to Clifford circuits). 

Thus, we expect that genuine phases of matter require \emph{quantum} decoding of measurement outcomes. This involves quantum feedback in the form of adaptive gates, which are conditioned on prior measurement outcomes, thereby avoiding the trivial results of Sec.~\ref{sec:OneCopy}. However, because the quantum state is modified by the quantum decoding, any genuine transition realized using such quantum feedback is unlikely to coincide with the MIET. Nonetheless, we now investigate whether measurement transitions of \emph{any} type can realize in adaptive protocols.

\section{Adaptive protocols}
\label{sec:adaptive}
Having ruled out genuine measurement-induced phases of matter in large swaths of systems, we now consider the remaining class of candidates. Essentially, the $n$-point functions \eqref{eq:n-point-expval} considered in Sec.~\ref{sec:OneCopy} and the SFF \eqref{eq:SFF def} considered in Sec.~\ref{sec:SFF} fail to show nontrivial effects due to measurements not because these quantities are linear in the density matrix, but because the measurement outcomes are not subsequently \emph{utilized}. As discussed, such protocols are tantamount to allowing the ``environment'' to measure the system, which has the same effect as ensemble-averaged chaotic time evolution alone. Including symmetries and/or constraints does not alter this conclusion, but merely introduces the possibility of using measurements to undo the block structure itself. Postselected quantities such as entanglement entropies use the outcomes of measurements in an unphyiscal manner; as discussed in Sec.~\ref{sec:can't work}, postselected quantities cannot distinguish phases of matter, by definition of the latter. 

The leading---and ostensibly \emph{only}---remaining possibility is the use of \emph{adaptive} hybrid protocols (see, e.g., \cite{SthitadhiSteer, GornyiMeas, TomMIPT, preselect, MIPT_wormhole, JongYeonDecode, ODea}), which utilize the outcomes of circuit measurements in the \emph{same} experimental shot via active feedback, avoiding the issues associated with postselection while also potentially realizing nontrivial effects due to measurements. The defining feature of adaptive circuits is that certain gates (or possibly measurements) are conditioned on the outcomes of prior measurements. It is known from \cite{SpeedLimit} that such adaptive gates are crucial to quantum error correction: In the context of quantum teleportation, e.g., the state is only transferred once the error-correction gate is applied; prior to that, the target qubit's state is simply the maximally mixed state~\cite{SpeedLimit}. 

We first define adaptive protocols and the constituent gates in Sec.~\ref{subsec:AdaptiveGates}. In Sec.~\ref{subsec:Generic Adaptive} we consider featureless circuits, finding that no observables are robust to any time evolution, precluding order. In Sec.~\ref{subsec:Discrete Adaptive} we consider chaotic evolution enriched with discrete symmetries, finding that no observables are robust to a full time step of chaotic evolution. In Sec.~\ref{subsec:Continuous Adaptive} we consider chaotic evolution enriched with continuous symmetries, finding that local order is possible and robust to time evolution. We showcase this via numerical simulation of a 1D chain of qubits involving $\U{1}$-symmetric two-qubit gates that conserves the magnetization $\sum_j \PZ{j}$ and measurements of $\PZ{j}$ followed by the operation $\PX{j}$ if the outcome is 1. This model steers toward the absorbing state \cite{Absorbing, GarrahanAbsorb, HayeAbsorbing} $\ket{\bvec{0}}$ for \emph{any} nonzero measurement rate $\measrate$ starting from the maximally mixed state. Hence, there is no  transition: While adaptive measurements replace $\PZ{j} \to \ident$ in the Heisenberg picture, the unitary evolution conserves the number of $\PZ{}$ operators, meaning there is no competition.

Finally, in Sec.~\ref{subsec:Constrained Adaptive} we consider chaotic circuits enriched with kinetic constraints only, which dynamically privilege operators in the constraint ($\weight{}$) basis but do not conserve their number, allowing for competition between adaptive measurements and chaotic evolution. We note that the same features are present in deterministic evolution without continuous symmetries, and possibly in Hamiltonian evolution with no other symmetries. We confirm that this competition leads to a genuine, measurement-induced absorbing-state transition by considering a 1D quantum East model \cite{QuantumEast2020}, in which a unitary is applied to site $j$ only if the East neighbor $j+1$ is in the state $\ket{1}$, and with the same adaptive protocol as in the $\U{1}$ example. We find a sharp transition in the directed-percolation universality class \cite{HayeAbsorbing} with critical measurement rate $\measrate^{\,}_{\rm c} \approx 0.038$. We expect this finding is generic to constrained Haar-random or deterministic chaotic time evolution with an appropriate choice of adaptive measurement protocol. Such models capture all known examples of genuine measurement-induced phase transitions \cite{TomMIPT, MIPT_wormhole, preselect, ODea}.

\subsection{Adaptive gates}
\label{subsec:AdaptiveGates}
We now define the crucial component of adaptive hybrid circuits: the feedback gate. Note that we do not consider outcome-dependent measurements, which are generally more complicated and no better suited to realizing large classes of states\footnote{Such protocols may be better suited to the preparation of ``magic'' states \cite{BravyiMagic}, but are not necessary to realize phase transitions.} \cite{SpeedLimit, GornyiMeas, SthitadhiSteer}. Consider a gate $\adapt$ conditioned on the outcome of measuring the observable $\mobserv^{\,}_{\tau,r}$ \eqref{eq:ASpectralDecomp} with $\Noutcome$ unique eigenvalues on cluster $r$ of measurement round $\tau=(t,\sigma)$ \eqref{eq:Combined SS label},
\begin{align}
\label{eq:Adaptive Gate}
    \Adapt{r',\tau';r,\tau} \, &\equiv \, \sum\limits_{m=0}^{\Noutcome} \, \RotGate{m;r',0} \otimes \SSProj{r,\tau}{(m)} \, ,~~
\end{align}
which acts on the physical cluster $r'$ in circuit layer $\tau'=t',\sigma'$ as the physical unitary $\rotgate^{\,}_{m;r',0}$ if the measurement of $\mobserv^{\,}_{\tau,r}$ (in some previous layer $\tau=t,\sigma$) resulted in the $m$th eigenvalue $\eig{m}$. The expression \eqref{eq:Adaptive Gate} can be extended to be conditioned on the product of multiple outcomes stored in a set of Stinespring registers, but is otherwise generic (note that the Stinespring registers are ``read only''). 

The example protocols we consider all act on systems of qubits ($\LocDim=2$), and we restrict to the measurement of Pauli-string observables for convenience. In this case, every measured observable $\mobserv$ has two unique eigenvalues $\pm1$, and the measurement unitary is always of the form
\begin{align}
    \label{eq:Pauli meas unitary}
    \Umeas{\mobserv} \, &= \, \frac{1}{2} \left( \ident + \mobserv \right) \otimes \SSid{\rm ss} + \frac{1}{2} \left( \ident - \mobserv \right) \otimes 
    \Shift{}{\rm ss} \, ,~~
\end{align}
and the outcome-dependent gates \eqref{eq:Adaptive Gate} act as
\begin{align}
    \label{eq:Qubit Adaptive Gate}
    \Adapt{r',\tau';r,\tau} \, &\equiv \, \RotGate{0;r',0} \otimes \SSProj{r,\tau}{(0)} + \RotGate{1;r',0}\otimes \SSProj{r,\tau}{(1)} \, ,~~
\end{align}
andd importantly, if both $\rotgate^{\,}_{0,1}$ commute with some observable $\observ$ of interest (at the Heisenberg time immediately prior to the feedback channel $\adapt$), the combination of measurement and feedback acts trivially on $\observ$. 

We now consider quantities of the form \eqref{eq:n-point-expval} evaluated in adaptive hybrid circuits involving  featuring maximally chaotic time evolution, measurements, and outcome-dependent unitary operations \eqref{eq:Adaptive Gate}. In particular, we consider order parameters of the form
\begin{align}
    \expval{\observ(t)} \, &\equiv \, \Emean{ \, \overline{ \matel{\psi (t) }{\observ}{\psi (t)}} \, } \notag \\
    &= \, \Emean{ \, \matel{\psi^{\,}_0}{\, \overline{\evo^{\dagger} (t,0) \, \observ \, \evo (t,0) }\, }{\psi^{\,}_0} \,} \, , ~~
    \label{eq:Adaptive OP}
\end{align}
that vanish under time evolution alone (and thus, in any \emph{non}adaptive hybrid circuit, as established in Sec.~\ref{sec:OneCopy}), ensuring that the corresponding order $\expval{\observ (t)} \neq 0$ is unique to adaptive hybrid protocols. We restrict to Hermitian observables $\observ$ with $\tr{\observ}=0$, and require that the criteria of Sec.~\ref{subsec:what is a phase} defining a phase of matter all be satisfied.

\subsection{Generic adaptive protocols}
\label{subsec:Generic Adaptive}
We first consider the fate of order parameters \eqref{eq:Adaptive OP} in adaptive hybrid protocols where the underlying time evolution is ``generic'' (i.e., the unitary gates do not have block structure). We consider the hybrid evolution of the order parameter $\observ$ \eqref{eq:Adaptive OP} in the Heisenberg-Stinespring picture. If the first channel encountered in the Heisenberg picture corresponds to featureless, Haar-random time evolution, then the update to $\observ$ is
\begin{align}
    \observ \, &\to \, \overline{ \gate^{\dagger}_{t,\lambda,r} \, \observ \, \gate^{\vpd}_{t,\lambda,r}} \, = \, \frac{1}{\LocDim^{\Nsite}} \, \ident \, \tr{\observ} \, = \, 0 \, , ~~\label{eq:Adapt Gen Tevo Update 1}
\end{align}
since $\observ$ has no identity component. Any term in $\observ$ proportional to $\ident$ evolves trivially both in time and under measurements, and has expectation value one. Since any observable can be written in the form $\observ = \tr{\observ} \, \ident + \dots$ (where the terms $\dots$ are all traceless), we simply ignore the part of any $\observ$ proportional to $\ident$. However, \eqref{eq:Adapt Gen Tevo Update 1} shows that, in the Schr\"odinger picture, the nontrivial part of \emph{any} of observable $\observ$ \eqref{eq:Adaptive OP} evaluated immediately following generic time evolution, on average, has vanishing expectation value (i.e., $\expval{\observ(t)}= \tr{\observ}$ is trivial).

If, instead, the first channel encountered in the Heisenberg picture corresponds to a projective measurement \emph{without} feedback, we know from Sec.~\ref{sec:OneCopy} that this measurement has no average effect on $\observ$. The only remaining option is that the first gate encountered corresponds to a feedback gate $\adapt$ \eqref{eq:Adaptive Gate}, in which case
\begin{align}
    \observ \, &\to \, \AdaptDag{r,t} \, \observ \, \Adapt{r,t} \notag \\
    &= \, \sum\limits_{m,n=0}^{\Noutcome-1} \, \RotGateDag{m;r,0} \, \observ\,  \RotGate{n;r,0} \otimes \SSProj{\MSites}{(m)} \, \SSProj{\MSites}{(n)} \notag \\
    &= \sum\limits_{m=0}^{\Noutcome-1} \, \RotGateDag{m;r,0} \, \observ \, \RotGate{m;r,0} \otimes \SSProj{\MSites}{(m)} \, , ~~\label{eq:Adapt Gen Adapt Update 1}
\end{align}
for a single adaptive gate acting on the physical cluster $r$ and conditioned on the product of outcomes in the Stinespring set $\MSites$\footnote{If the measurements all have $\Noutcome$ eigenvalues, e.g., then $m$ in \eqref{eq:Adapt Gen Adapt Update 1} reflects $\sum_{j \in \MSites} m^{\,}_j$ modulo $\Noutcome$. In other scenarios, more care may be required,  
but $\SSProj{\MSites}{(m)}$ can always be defined.}. If the next operation (after conjugation by $\adapt$) corresponds to generic time evolution, then we find
\begin{align}
    \observ \, &\to \, \tr{ \AdaptDag{r,t} \, \observ \, \Adapt{r,t} } \frac{1}{\LocDim^{\Nsite}} \, \ident \, = \, 0 \, , ~~
\end{align}
and, as before, a measurement for which the outcome is not utilized has no effect. The two remaining options for what comes next are (\emph{i}) another adaptive gate $\adapt'$ or (\emph{ii}) the measurement channels upon which $\adapt$ was conditioned. We note that (\emph{i}) can trivially be absorbed into $\adapt$, leaving (\emph{ii}). Conjugating \eqref{eq:Adapt Gen Adapt Update 1} by the appropriate measurement channels and evaluating in the default initial state of the Stinespring registers, we find
\begin{align}
     \observ \, &\to \, \Umeasdag{\mobserv} \, \AdaptDag{t,r} \, \observ \, \Adapt{t,r} \, \Umeas{\mobserv} \notag \\
     &= \, \sum\limits_{a,b,m=0}^{\Noutcome-1} \, \Proj{r',0}{(a)} \, \RotGateDag{m;r,0} \, \observ \, \RotGate{m;r,0}  \Proj{r',0}{(b)} \notag \\
     &~~~~~\times \, \matel{0}{\SShift{-a}{r',\tau} \, \SSProj{r',\tau}{(m)} \, \SShift{b}{r',\tau}}{0} \notag \\
     &= \, \sum\limits_{a,b,m=0}^{\Noutcome-1} \, \Proj{r',0}{(a)} \, \RotGateDag{m;r,0} \, \observ \, \RotGate{m;r,0}  \Proj{r',0}{(b)}\,  \matel{a}{\SSProj{r',\tau}{(m)}}{b} \notag \\
     &= \, \sum\limits_{m=0}^{\Noutcome-1} \, \Proj{r',0}{(m)} \, \RotGateDag{m;r,0} \, \observ \, \RotGate{m;r,0}  \Proj{r',0}{(m)}  \, , ~~\label{eq:Adapt Gen Full Update 1}
\end{align}
in the case where $\adapt$ is based on the outcome of measuring a single observable $\mobserv$ on cluster $r'$ in layer $\tau=t',\sigma'$. Extending the above to more general cases is straightforward \cite{SpeedLimit, AaronYifanFuture}, and the result generically takes the form \eqref{eq:Adapt Gen Full Update 1}. 

Additional sequences of measurement channels and adaptive gates conditioned on the recorded outcomes only lead to additional updates of the form \eqref{eq:Adapt Gen Full Update 1}. At some point, however, a layer of generic Haar-random gates tiling all sites will be encountered. Suppose this happens immediately after the measurement of $\mobserv$ (in the Heisenberg picture. In this case, \eqref{eq:Adapt Gen Full Update 1} becomes
\begin{align}
    \observ \, &\to \, \frac{1}{\LocDim^{\Nsite}} \, \ident \, \sum\limits_{m=0}^{\Noutcome-1} \, \tr{\RotGateDag{m;r,0} \, \observ \, \RotGate{m;r,0}  \Proj{r',0}{(m)}} \, , ~\label{eq:Adapt Gen Full Update Tevo 1}
\end{align}
which can indeed capture nontrivial order.

To see this, consider a system of qubits ($\LocDim=2$). Suppose $\observ = \mobserv = \PZ{j}$ are both the Pauli $\PZ{}$ operator, and define the outcome-dependent gate \eqref{eq:Qubit Adaptive Gate} to be
\begin{equation}
    \adapt \, = \, \ident \otimes \SSProj{j,\tau}{(0)} + \PX{j} \otimes \SSProj{j,\tau}{(1)} \,, ~~\label{eq:Adapt Gen Qubit Adaptive Gate}
\end{equation}
so that \eqref{eq:Adapt Gen Full Update Tevo 1} becomes
\begin{align}
    \PZ{j} (t') \, &= \, \frac{1}{2} \, \ident \, \sum\limits_{m=0,1} \, \trace_j \, \left[ \, \Shift{m}{j} \, \Weight{}{j} \,\Shift{m}{j} \, \frac{1}{2} \left( \ident + (-1)^m \, \Weight{}{j} \right) \, \right] \notag \\
    &= \, \frac{1}{4} \, \ident \, \sum\limits_{m=0,1} \, \trace_j \, \left[ \, (-1)^m \, \Weight{}{j} \, \left( \ident + (-1)^m \, \Weight{}{j} \right) \, \right] \notag \\
    &= \, \frac{1}{4} \, \ident \, \sum\limits_{\pm} \, \tr{ \ident^{\vpp}_j \pm \PZ{j} } \, = \, \frac{1}{2} \, \ident \,, ~~\label{eq:Adapt Gen Full Update Qubit Examp}
\end{align}
which means that $\expval{\PZ{j}(t)} = 1$, since no subsequent gates in the Heisenberg picture modify $\ident$ \eqref{eq:Adapt Gen Full Update Qubit Examp}. 

However, several remarks are in order. We first note that $\expval{\PZ{j}(t)} = 0$ in the absence of \emph{either} measurements or the outcome-dependent operation $\adapt$ \eqref{eq:Adapt Gen Qubit Adaptive Gate}, which means that such ``order'' is only possible in hybrid circuits with outcome-dependent gates. We also note that any observable $\observ$ (with $\tr{\observ}=0$) can be engineered to have a nonzero expectation value in some adaptive protocol (and zero expectation value in nonadaptive circuits). Importantly, this order is extremely fine tuned: The only scenario in which $\expval{\observ(t)}$ \eqref{eq:Adapt Gen Full Update Tevo 1} is nontrivial (e.g., $\expval{\PZ{j}(t)} \neq 0$)  is if no unitary time evolution occurs between the measurement of $\mobserv$, the application of the gate $\adapt$ conditioned on the outcome of measuring $\mobserv$, and the measurement of $\observ$. Additionally, no observables other than $\observ$ will realize nontrivial order (i.e., if we instead measure $\PZ{i}$ on some site $i \neq j$, we find $\expval{\PZ{i}(t)}=0$). By contrast, a \emph{local} order parameter $\observ^{\,}_j$ should satisfy $\expval{\observ^{\,}_j (t)} \neq 0$ for \emph{all} sites $j$.

The only way to ensure that such a local order parameter $\expval{\PZ{j}} \neq 0~\forall \, j$ is nontrivial is to measure every single site in the system, correctly apply outcome-dependent gates to all sites, and immediately evaluate $\observ$.  In other words, any order in generic, adaptive hybrid protocols is highly fine tuned, and incompatible with the definition of a phase of matter given in Sec.~\ref{subsec:what is a phase}. Essentially, in the presence of generic Haar-random evolution, the only ``privileged'' observable $\observ$ is the trivial one $\observ=\ident$; the nonzero expectation value of $\ident$ does not imply order. 

Hence, we conclude that observing measurement-induced order corresponding to a genuine phase of matter requires either (\emph{i}) introducing block structure or (\emph{ii}) avoiding maximally chaotic ensembles (or using deterministic time evolution). The latter is beyond the scope of this work; in the remainder, we investigate the possibility of a measurement-induced phase transition in adaptive protocols with various types of block structure.

\subsection{Discrete symmetries}
\label{subsec:Discrete Adaptive}
We now consider measurement-induced order in hybrid adaptive circuits in which the underlying time evolution has some structure, corresponding to a global discrete symmetry. Familiar examples from statistical mechanics include Ising, clock, and Potts models. While featureless Haar-random dynamics only privilege the trivial operator $\ident$, other operators are preserved in the presence of a discrete Abelian symmetry. The question is whether this leads to a robust order parameter $\expval{\observ(t)}\neq 0$ \eqref{eq:Adaptive OP}.

Suppose that the global discrete symmetry corresponds to the finite group $\mathcal{G}$ with elements $\{g^{\,}_1 , g^{\,}_2 , \dots ,  \}$, where the number of elements $\abs{\mathcal{G}}$ is finite. The symmetry group $\mathcal{G}$ is \emph{generated} by the conserved charges $\mathfrak{g}$ according to
\begin{equation}
    \label{eq:finite group generators}
    \mathcal{G} \, = \, \mathrm{span} \left\{ \mathfrak{g}^{\,}_1, \dots, \mathfrak{g}^{\,}_n \right\} \, , ~~
\end{equation}
where $n < \abs{\mathcal{G}}$ is the number of unique generators, and the  elements $g^{\,}_k$ of $\mathcal{G}$ (including $\ident$) correspond to all unique products of the various ``generators'' $\mathfrak{g}^{\,}_k$. Note that we also restrict to Abelian symmetries, where the symmetry generators (and elements) commute---i.e., $\com{g^{\,}_k}{g^{\,}_{k'}}=0~\forall \, k,k'$. 

As discussed in Secs.~\ref{sec:hybridcirc} and \ref{sec:OneCopy}, each unitary gate $\gate^{\,}_r$ in the circuit must commute with all elements of $\mathcal{G}$. Generally speaking, each $\ell$-site gate $\gate^{\,}_r$ (with $\ell=\abs{r}$) must commute with every generator $\mathfrak{g}^{\,}_k$ for any choice of cluster $r$; this is accomplished by identifying the unique ``local'' expressions of each generator $\mathfrak{g}^{\,}_k$, which may require a particular gate size. Each unitary gate $\gate^{\,}_r$ realizes a sum over symmetry blocks labelled $\alpha$ \eqref{eq:GenProjGate}, where each block $\alpha$ corresponds to a particular value of the \emph{local} charge $\mathfrak{g}^{\,}_{k,r}$ on cluster $r$, for each of the $n$ independent charges in \eqref{eq:finite group generators}.

Depending on the gate size $\ell = \abs{r}$ and the order $\abs{\mathcal{G}}$ of the symmetry group $\mathcal{G}$, any number of operators $\observ$ may survive a \emph{single} layer $\lambda$ of Haar-averaged, symmetric time evolution. These operators $\observ$ act on each cluster $r \in \lambda$ as an element of $g \in \mathcal{G}$, restricted to the cluster $r$ (i.e., $g^{\,}_r$). If the operator $\observ$ is exactly an element of $\mathcal{G}$, then it also survives all subsequent layers (note that the gates in each layer are staggered to tile all sites). However, most of the operators $\observ$ that survive a single layer of time evolution realize different elements $g$ on each cluster $r$.

For example, in a system of $\Nsite$ qubits ($\LocDim=2$) with a $\Ints^{\,}_2$ ``Ising'' symmetry generated by the $\PZ{}$-parity operator
\begin{equation}
    \mathfrak{g} \, = \, \prod\limits_{j=1}^{\Nsite} \, \PZ{j} \, , ~~\label{eq:Qubit Ising Generator}
\end{equation}
with $\mathfrak{g}^2=\ident$, the unitary gate $\gate^{\,}_{j,j+1}$ must commute with $\PZ{j}\PZ{j+1}$. In this way, each local Ising gate preserves the \emph{local} $\PZ{}$ parity $\PZ{j}\PZ{j+1}$ (i.e., in the Schr\"odinger picture, $\gate^{\,}_{j,j+1}$ only mixes the states $\ket{00}$ with $\ket{11}$ \emph{or} $\ket{01}$ with $\ket{10}$). However, the \emph{full} unitary circuit only preserves the \emph{global} $\PZ{}$ parity $\mathfrak{g}$ \eqref{eq:Qubit Ising Generator}. For example, suppose that the operator $\observ = \PZ{j}\PZ{j+1}$ survives a given layer of evolution. The next layer applies Ising-symmetric gates to the clusters $(j-1,j)$ and $(j+1,j+1)$; those gates respectively update the operators the operators $\ident \PZ{j}$ and $\PZ{j+1} \ident$, neither of which is preserved by the $\Ints^{\,}_2$ symmetry, so $\observ$ is annihilated by the second layer and $\expval{\observ (t)} = 0$. 

Hence, the only observables $\observ$ that have can realize nontrivial expectation values $\expval{\observ(t)} \neq 0$ under Ising-symmetric dynamics are (\emph{i}) $\observ = \prod_{j=1}^{\Nsite} \, \PZ{j}$, (\emph{ii}) observables that can be connected to $\ident$ using measurements and feedback, and (\emph{iii}) those that can be connected to $\mathfrak{g}= \prod_{j=1}^{\Nsite} \, \PZ{j}$ using measurement and feedback. The first case (\emph{i}) is guaranteed to have nonzero expectation value in \emph{any} $\PZ{}$-basis state, independent of whether measurements are made or feedback applied; hence $\observ = \mathfrak{g}$ does not diagnose measurement-induced properties. The second case (\emph{ii}) is identical to the case considered in Sec.~\ref{subsec:Generic Adaptive} without symmetries, and hence does not take advantage of the discrete symmetry. The final case (\emph{iii}) does require adaptive dynamics and utilizes the discrete symmetry; however, this case requires as much fine-tuning as in the generic case (\emph{ii}), and one must correctly measure and apply outcome-dependent rotations and to all sites of the system without intervening time evolution.

To see that this is not robust, imagine measuring $\PZ{j}$ on all sites $j$ with some amount of noise, so that  the operator actually measured is $(1-\epsilon) \PZ{j} + \epsilon \PX{j}$. Taking the product over all sites and expanding the measured operator onto Pauli strings, we find that the overlap with the intended operator $\mathfrak{g}= \prod_{j=1}^{\Nsite} \, \PZ{j}$ is $(1-\epsilon)^{\Nsite}$. For the measured operator to have $\Order{1}$ overlap with the intended operator requires $\epsilon \, \Nsite \ll 1$, which means that $\epsilon \to 0$ in the thermodynamic limit. Moreover, such a protocol merely uses measurements to replace the time-evolved state of the system with the desired state $\ket{\bvec{0}}$ or $\ket{\bvec{1}}$; it is not possible to steer gradually toward the desired state, and \emph{any} intervening time evolution results in failure.

Returning to generic systems with $\LocDim$ arbitrary, we now argue that, in the presence of a generic discrete symmetry group $\mathcal{G}$, no operators other than the symmetry elements $g \in \mathcal{G}$ survive \emph{all} time-evolution layers in a given time step. By definition, the gates must preserve all operators $g \in \mathcal{G}$l however, if operators $\observ \notin \mathcal{G}$ could survive all layers, this would imply that either (\emph{i}) there exists another symmetry that ought to belong to $\mathcal{G}$, (\emph{ii}) that the symmetry is actually continuous, or (\emph{iii}) that the dynamics trivially preserve a large number of operators. Each of these violates the assumption of a global discrete symmetry with nontrivial time evolution; hence only the elements of $\mathcal{G}$ survive all layers of time evolution.

In Sec.~\ref{subsec:Generic Adaptive}---in the context of generic Haar-random circuits---we found that a nontrivial order parameter $\expval{\observ(t)} \neq 0$ could only be realized if the Heisenberg evolution of $\observ$ converted $\observ$ to $\ident$ prior to \emph{any} time evolution. In the Schr\"odinger picture, the equivalent requirement is that the appropriate measurements and outcome-conditioned rotations be applied immediately prior to evaluation of $\expval{\observ}$ at time $t$, without intervening time evolution. In the case of discrete symmetries, the same arguments hold; however, one can also instead convert $\observ$ to $g \in \mathcal{G}$ in the Heisenberg picture using measurements and feedback. In the presence of the discrete symmetry group $\mathcal{G}$, all such operators $g$ are invariant under Haar evolution; hence, these operators may have nontrivial expectation values independent of measurements and feedback. However, those expectation values are determined by the initial state (and only $\ident$ is nontrivial in \emph{every} initial state). 

In other words, discrete Abelian symmetries are not sufficient to realize \emph{robust} order $\expval{\observ(t)} \neq 0$ unique to adaptive dynamics, let alone a phase transition. While it is possible to realize $\expval{\observ(t)} \neq 0$ in chaotic dynamics with discrete symmetries, this either (\emph{i}) is possible in the absence of measurements and/or outcome-dependent gates or (\emph{ii}) requires the same amount of fine tuning as in generic circuits. The only distinction compared to the generic case of Sec.~\ref{subsec:Generic Adaptive} is the existence of finitely many other ``paths'' to nontrivial orders (corresponding to the nontrivial elements $g \neq \ident \in \mathcal{G}$), provided that the initial state belongs to a particular sector of the discrete charge.

\subsection{Continuous symmetries}
\label{subsec:Continuous Adaptive}
We next consider the possibility of order in hybrid adaptive circuits enriched by a \emph{continuous} symmetry. Familiar examples include models with $\U{1}$ conserved charge (e.g., total $\PZ{}$ spin), models that conserve higher moments of charge (e.g., dipole), and those with continuous rotational invariance. We also allow for the possibility that the symmetry group $\mathcal{G}$ also contains discrete generators; the only requirement for description in terms of block-diagonal gates \eqref{eq:GenProjGate} is that $\mathcal{G}$ be Abelian.

The generators of the continuous part of the symmetry group $\mathcal{G}$ correspond to operators of the form
\begin{equation}
    \label{eq:continuous sym generators}
    \mathfrak{g} \, = \, \sum\limits_{j=1}^{\Nsite} \, \mathfrak{g}^{\vpp}_j \, , ~~
\end{equation}
where $\mathfrak{g}^{\,}_j$ is \emph{Hermitian}, so that the corresponding elements of the group $\mathcal{G}$ can be written as, e.g.,
\begin{equation}
\label{eq:continuous sym element}
    g \left( \theta \right) \, = \, \exp \left( - \ii \, \theta \, \mathfrak{g} \right) \in \mathcal{G} \, ,~~
\end{equation}
which is unitary, and depends on the continuous parameter $\theta$. We allow for multiple such generators $\mathfrak{g}$. In the case where \eqref{eq:continuous sym element} spans $\mathcal{G}$, the symmetry group  $\mathcal{G}$ is the Lie group $\U{1}$, parameterized by 
$\theta \in [0,2\pi)$. 

Each time-evolution gate $\gate^{\,}_{r}$ preserves the local charge
\begin{equation}
\label{eq:continuous local charge}
    \mathfrak{g}^{\vpp}_r \, = \, \sum\limits_{j \in r} \, \mathfrak{g}^{\vpp}_j \, , ~~
\end{equation}
for each distinct continuous generator $\mathfrak{g}$ \eqref{eq:continuous sym generators}, and for any discrete generators (as outlined in Sec.~\ref{subsec:Discrete Adaptive}). This implies that $\mathfrak{g}^{\,}_j$ is a privileged \emph{local} operator, in contrast to discrete symmetries, in which only the global generators are invariant under Haar-random evolution, on average.

As a result, there are $\Order{\Nsite}$ local operators that survive any number of layers of symmetric time evolution. In general, evolution under $\gate^{\,}_r$ must preserve the number of $\mathfrak{g}^{\,}_j$ operators in cluster $r$, but it may move them around. In the maximally chaotic case, a given number of symmetry operators $\mathfrak{g}^{\,}_j$ is uniformly distributed among the $\ell$ sites in the cluster $r$. This means that any observable $\observ$ proportional to some number of charge operators is preserved under dynamics, and a finite amount of that operator remains in place. As a result, we expect that models with continuous symmetries are robust to intervening time evolution and are further compatible with \emph{local} order parameters $\expval{\observ^{\,}_j (t)} \neq 0~\forall \, j$.

For example, in a system of $\Nsite$ qubits ($\LocDim=2$) with a $\U{1}$ symmetry corresponding to conserved total magnetization $\mathfrak{g} = \sum_{j=1}^{\Nsite} \, \PZ{j}$, the unitary gate $\gate^{\,}_{j,j+1}$ must commute with $\PZ{j}+\PZ{j+1}$. In other words,  $\gate^{\,}_{j,j+1}$ only mixes the states $\ket{01}$ and $\ket{10}$ (while leaving $\ket{00}$ and $\ket{11}$ invariant). In terms of operators, the gate $\gate^{\,}_r$ preserves the number of Pauli $\PZ{j}$ operators on cluster $r$. Since the Haar-random dynamics privilege $\PZ{j}$ operators (the local symmetry charge), any observable $\observ$ \eqref{eq:Adaptive OP} that acts nontrivially only as $\PZ{}$ is potentially robust to chaotic time evolution, while all other observables fail to realize robust order parameters for the reasons outlined in Sec.~\ref{subsec:Generic Adaptive}. The most natural choice of \emph{local} order parameter is
\begin{align}
    \observ^{\,}_j \, &= \, \PZ{j} \, ,~~\label{eq:adapt U1 OP Z}
\end{align}
though, in general, products of $\PZ{}$ operators are also robust. Under chaotic time evolution with $\U{1}$ conservation of \eqref{eq:continuous sym generators}, every gate $\gate^{\,}_{j,j+1}$ leaves $\PZ{j}$ in place with probability $1/2$, and hops $\PZ{j}$ to $\PZ{j+1}$ with probability $1/2$,
\begin{align}
    \gate^{\dagger}_{i,j} \, \PZ{j} \, \gate^{\vpd}_{i,j} \, &= \, \frac{1}{2} \left( \PZ{i} + \PZ{j} \right) \, , ~~\label{eq:U1 Z SSEP}
\end{align}
while $\ident$ and $\PZ{i}\PZ{j}$ are unaltered by the gate $\gate^{\,}_{i,j}$.

In other words, under $\U{1}$-symmetric time evolution acts on $\PZ{}$ operators as a Markovian symmetric simple exclusion process (SSEP) \cite{Schutz_SSEP, U1FRUC}, which for a single $\PZ{j}$ operator results in a random walk. Thus, the weight of the observable $\observ (0) = \PZ{j}$ \eqref{eq:adapt U1 OP Z} spreads diffusively in time $t$ from the site $j$ under time evolution; for times $t \ll \Order{\Nsite^2}$, the time-evolved operator $\observ (t)$ \eqref{eq:adapt U1 OP Z} still has significant overlap with the original operator  $\PZ{j}$.

We now consider the measurement and outcome-dependent channels. For simplicity, consider the standard measurement-based protocol in which one measures $\PZ{j}$ with some rate $\measrate$, and subsequently applies the operator $\PX{j}$ if the outcome is $-1$ (and does nothing otherwise). The corresponding dilated channels are given by
\begin{subequations}
    \label{eq:adaptive U1 dilated channels}
    \begin{align}
        \measunitary^{\vpd}_{j,\tau} \, &= \, \frac{1}{2} \left( \ident + \PZ{j,0} \right) \, \SSid{j,\tau} +\frac{1}{2} \left( \ident - \PZ{j,0} \right) \, 
        \Shift{}{j,\tau} \label{eq:adaptive U1 Z meas} \\
        &\Adapt{j'\tau';j,\tau} \, = \, \ident^{\vpd}_{j,0} \, \SSProj{j,\tau}{(0)} + \PX{j,0} \, \SSProj{j,\tau}{(1)} \label{eq:adaptive U1 X rot} \, , ~
    \end{align}
\end{subequations}
and we define the infinite-temperature diffusive ``kernel''
\begin{align}
    f \left( r, t \right) \, &\equiv \, \frac{1}{\HilDim} \, \trace_{\rm ph} \, \left[ \, \PZ{j+r} \, \evo^{\dagger} (t) \,\PZ{j} \evo (t) \,  \right] \label{eq:U1 diff ker def} \\
    &= \, 4^{-t} \, \begin{pmatrix} 2\, t - 1 \\ \floor{t -\abs{r}/2} \end{pmatrix} \label{eq:U1 diff ker form} \, ,~~
\end{align}
where $\evo (t)$ captures $\U{1}$-symmetric evolution by $t$ time steps, and $\abs{r} \to \abs{r}+1$ above if $r \leq 0$. 

For simplicity, suppose that the adaptive update \eqref{eq:adaptive U1 X rot} and the measurement of $\PZ{j}$ \eqref{eq:adaptive U1 Z meas} occur without intervening time evolution. If these two gates are separated by $r'$ sites and $t'$ time steps, their combined effect is suppressed by the finite factor $f(r',t')$ \eqref{eq:U1 diff ker def}; thus, applying the operations sequentially leads to a stronger signature \eqref{eq:Adaptive OP}. In the Heisenberg evolution of $\PZ{j}$ \eqref{eq:adapt U1 OP Z}, suppose that the first adaptive sequence is encountered after $t$ time steps and on site $j+r$. At this point, we have
\begin{align}
    \expval{\PZ{j}(t)}^{\vpp}_{\infty} \, 
    &\geq \, f(r,t) \, , ~~\label{eq:U1 OP lower bound}
\end{align}
for any site $j$; subsequent adaptive gates only increase this lower bound by providing additional routes to $\expval{\PZ{j}(t)} \neq 0 $ evaluated in the infinite-temperature state.

This suggests that the quantity $\expval{\PZ{x}(t)} = \Order{1}$ for all sites $x$ and times $t$, and may therefore be a valid local order parameter. However, we note that such ``order'' $\expval{\PZ{x}(t)} = \Order{1}$ can also realize in measurement-free dynamics provided that the initial state has definite $\U{1}$ charge. Thus, \eqref{eq:U1 OP lower bound} only meaningfully captures the effects of measurements and feedback when evaluated in the infinite-temperature state $\DensMat^{\,}_{\infty}$. However, this confirms that it is possible to realize nontrivial expectation values (or correlation functions) in the presence of measurements, even in the maximally mixed state. In fact, not only is $\expval{\PZ{x}(t)}^{\,}_{\infty} = \Order{1}$  possible in the presence of measurements----it is \emph{only} possible in the presence of measurements (combined with outcome-dependent feedback). 

Although the ``order'' \eqref{eq:U1 OP lower bound} is local and robust to the  $\U{1}$-symmetric time evolution, it does not distinguish two phases of matter. Essentially, $\expval{\PZ{x}(t)}^{\,}_{\infty} = \Order{1}$ \eqref{eq:U1 OP lower bound} implies that $\PZ{x}(t) \to \ident$,  meaning $\Proj{x}{(0)}(t) \to \ident$ under the adaptive $\U{1}$-symmetric circuit (for all sites $x$ and times $t$). Intuitively, the combination of measurement and feedback projects the state $\ket{\psi}$ of the qubit $y$ onto a $\PZ{y}$-basis state, and then ensures that state is $\ket{0}$. Thus, at sufficiently late times, \emph{any} initial state (including the maximally mixed state $\DensMat^{\,}_{\infty}$) is ``steered'' toward the state $\ket{\bvec{0}}$ by the hybrid circuit, up to a subextensive number of defects (so that the density of qubits in the state $1$ vanishes in the thermodynamic limit).

Note that $\ket{\bvec{0}}$ is an \emph{absorbing} state \cite{Absorbing, GarrahanAbsorb, ODea} of the $\U{1}$-symmetric time evolution. Since $\ket{\bvec{0}}$ belongs to its own charge sector, it is stationary under time evolution---once reached, the system remains in the state $\ket{\bvec{0}}$. However, a transition is not possible because, for \emph{any} measurement rate $\measrate>0$, the $\U{1}$-symmetric hybrid protocol described above reaches the absorbing state $\ket{\bvec{0}}$ (up to a vanishing density of defects) in finite time $t$.

\begin{figure}[t]
    \centering
    \includegraphics[width=\linewidth]{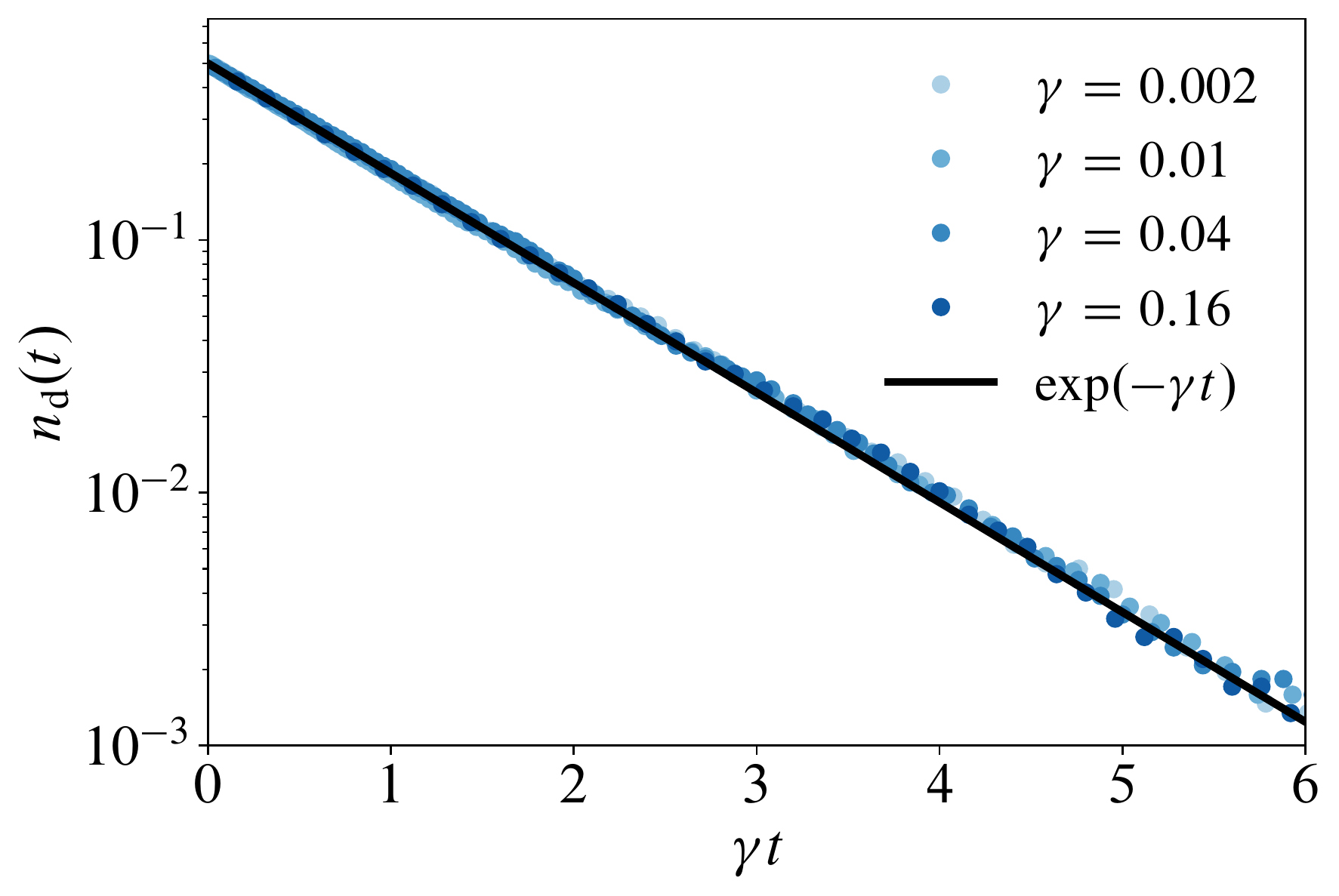}
    \caption{The density $n^{\,}_\text{d}(t,\size)$ of down spins  \eqref{eq:U1 defect density}  versus time (rescaled by the measurement rate $\measrate$) for a $1\SpaceDim$ circuit comprising $\mathsf{U}(1)$-symmetric, two-site Haar-random gates interspersed with $Z$-basis measurements, followed by $X$ if the outcome is down \eqref{eq:adaptive U1 dilated channels}. The time $t$ required to remove all but a vanishing density of defects $\ket{1}$ from the chain is independent of system size, and is a function of the measurement rate per site $\measrate$ alone, as shown by the collapse of the various curves. While the above data correspond to system of size $\size=256$, the time scale $t$ to reach $n^{\,}_\text{d}(t,\size) =0$ is independent of $\size$.}
    \label{fig:U1-defect-decay}
\end{figure}

To see that there is no sharp transition as a function of measurement rate $\measrate > 0$, we numerically simulate the $\U{1}$-symmetric adaptive circuit described above on a $1\SpaceDim$ chains of $\Nsite=\size$ qubits. 
In particular, we numerically evaluate the time- and $\size$-dependent ``defect density''
\begin{equation}
    \label{eq:U1 defect density}
    n^{\vpp}_{\rm d} (t,\size) \, = \, \frac{1}{\size} \sum\limits_{j=1}^{\size} \, \frac{1}{2} \, \left(\, 1  - \expval{\,  \PZ{j} (t) \, }^{\vpp}_{\infty} \,  \right)  \,, ~~
\end{equation}
in the infinite-temperature initial state $\DensMat^{\,}_{\infty}$. In Fig.~\ref{fig:U1-defect-decay}, we plot $n^{\,}_{\rm d}$ \eqref{eq:U1 defect density} as a function of rescaled time $\measrate \, t$ for various $\measrate$ . Though the data in Fig.~\ref{fig:U1-defect-decay} correspond to $\size=256$, other system sizes realize the same scaling form. 

We simulate the evolution of local observables $\observ$ in the Heisenberg picture by sampling the ensemble- and outcome-averaged dynamics of the adaptive $\U{1}$ evolution. This is achieved by mapping the operator evolution to a classical stochastic process, which we then sample. Under a  two-site $\U{1}$-symmetric gate acting on sites $j$ and $j+1$, the operators $\ident$ and $\PZ{j} \PZ{j+1}$ are stationary,$\PZ{j}$ and $\PZ{j+1}$ are updated according to \eqref{eq:U1 Z SSEP}, and all operators with $\PX{}$ or $\PY{}$ content are annihilated. Additionally, we use the fact that $\PZ{j} \to \ident$ under the combination of the measurement and feedback gates \eqref{eq:adaptive U1 dilated channels}.

As a result, we need only consider operators $\observ$ of the form $\prod_{i}Z_i^{b_i}$, which we represent as classical bit strings $\bvec{b}=b_1 b_2 \cdots b_{\size}$. The bit string tracks the number and locations of all $\PZ{}$ operators in $\observ$, which are updated classically according to the rules enumerated above. In simulating \eqref{eq:U1 defect density}, the initial operator $\PZ{j}(0)$ corresponds to an initial bit string with a single nonzero entry $b_j = 1$, with all other $b_i = 0$ for $i \neq j$. The $\U{1}$-symmetric time evolution simply hops this nonzero bit between neighboring sites via \eqref{eq:U1 Z SSEP} or leave it in place, realizing a SSEP \cite{Schutz_SSEP, U1FRUC}. The combination of measurements and feedback leave any bit $b_i=0$ unchanged, but change $b_j=1$ to $b_j=0$, realizing the bit string $\bvec{b}=\bvec{0}$ (i.e., the identity $\ident$).

Note that only the identity operator survives the infinite temperature trace over initial states. Hence, $\expval{\PZ{j}(t)}$ is governed by the survival probability of the nonzero bit, which should decay as $\exp(-\measrate t)$. This expectation is confirmed in Fig.~\ref{fig:U1-defect-decay}, which shows that the density of defective spins \eqref{eq:U1 defect density}  decays exponentially with a rate $\gamma^{-1}$. Thus, the system \emph{always} reaches a state with a vanishing density of defects $n^{\,}_{\rm d} (t^{\,}_*,\size) \to 0 $ over a time scale $t^{\,}_* \sim \measrate^{-1}$, since the number of defective spins can only decrease under dynamics. Hence, a sharp transition is not possible, since for any system size $\size$ and any $\measrate>0$, there is only one regime: For some fixed time scale $t \gtrsim \measrate^{-1}$, \emph{every} initial state realizes the absorbing state $\ket{\bvec{0}}$ up to a vanishing density of defects $n^{\,}_{\rm d} (t,\size) = \order{1}$ \eqref{eq:U1 defect density}. 

Importantly, these conclusions generalize to arbitrary continuous symmetries with commuting local generators \eqref{eq:continuous sym generators}, and to arbitrary lattices in any spatial dimension $\SpaceDim \geq 1$ with any local Hilbert space dimension $\LocDim$. In any such model, there is an extensive number of local operators that are privileged under symmetric, chaotic time evolution, while all other observables decay rapidly. As a result, only the local symmetry operators $\mathfrak{g}^{\,}_j$  are robust to time evolution, and thereby candidates for a local order parameter \eqref{eq:Adaptive OP}. Moreover, it is \emph{always} possible to engineer adaptive hybrid dynamics \eqref{eq:adaptive U1 dilated channels} that ``steer'' any initial state into a particular state, provided that the target state (\emph{i}) stationary under the time-evolution part of the hybrid circuit and (\emph{ii}) has uniform charge on all sites (other states require nonlocal circuits to target). 

However, such a protocol always reaches the target absorbing state for any nonzero rate $\measrate>0$ of measuring symmetry operators (and applying outcome-dependent gates thereafter). Essentially, the guaranteed success of such protocols---and corresponding absence of a phase transition---stem from the fact that there is no competition between the measurements (and adaptive gates) and the symmetric time evolution. While the combination of measurements and feedback generically replace, e.g., $\weight{}$-type operators with $\exp(2 \pi \ii k / \LocDim) \, \ident$, the chaotic dynamics preserve the number of $\weight{}$ operators (owing to the symmetry). As a result, there is no competition with the effect of measurement. Thus, in adaptive dynamics with Haar-random time evolution that preserves a continuous symmetry, (\emph{i}) it is possible to realize nontrivial expectation values of local symmetry operators from \emph{any} initial state, (\emph{ii}) such ``order'' corresponds to the ability to steer toward uniform absorbing states in finite time for \emph{any} nonzero rate $\measrate > 0$ of measuring local symmetry operators, but (\emph{iii}) it is \emph{not possible} to realize a sharp transition as a function of $\measrate$.

\subsection{Constrained models}
\label{subsec:Constrained Adaptive}
Thus far, we have established that (\emph{i}) the absence of block structure is incompatible with robust order $\expval{\observ (t)} \neq 0$ and (\emph{ii}) maximally chaotic models with continuous symmetries, while compatible with robust order, do not admit a sharp transition as a function of the measurement rate $\measrate>0$. In the former case, the absence of dynamically privileged operators means that there is no choice of order parameter $\observ$ \eqref{eq:Adaptive OP} robust to \emph{any} amount of chaotic time evolution, on average. In the latter case, the fact that all of the dynamically privileged operators are conserved in number means that there is no competition to the combination of measurements and adaptive gates \eqref{eq:Adaptive Gate}, precluding a transition as a function of  $\measrate$.

Fortunately, kinetically constrained models \cite{RitortKCM, garrahan2010kinetically, KCMRydberg, Valado_2016, GarrahanLectures, QuantumEast2020, ConstrainedRUC} without conservation laws generically admit dynamically privileged operators, whose number is \emph{not} conserved. Hence, nonconserving constrained models are, in principle, compatible with genuine phase transitions as a function of the measurement rate $\measrate>0$. We establish that such transitions are possible in the context of a quantum East model \cite{QuantumEast2020, ConstrainedRUC} combined with measurements and feedback.

The quantum East model is defined on a $1\SpaceDim$ chain of qubits $\LocDim=2$; time evolution is generated by two staggered layers of two site gates that act as
\begin{align}
    \label{eq:East gate}
    \gate^{\vpd}_{j,j+1} \, &= \, \ident^{\vpd}_j \otimes \Proj{j+1}{(0)} + \Haar^{\vpd}_j \otimes \Proj{j+1}{(1)} \, , ~~
\end{align}
which applies the $2 \times 2$ Haar-random gate $\Haar$ to qubit $j$ if its ``East'' neighbor $j+1$ is in the state $\ket{1}$, and acts trivially otherwise. Note that the state $\ket{\bvec{0}}$ is an absorbing state of the dynamics generated by \eqref{eq:East gate}. 

We intersperse the chaotic dynamics generated by $\gate^{\,}_{j,j+1}$ \eqref{eq:East gate} with $\PZ{}$-basis projective measurements, followed by outcome-dependent spin flips, 
\begin{subequations}
    \label{eq:adaptive East dilated channels}
    \begin{align}
        \measunitary^{\vpd}_{j,\tau} \, &= \, \frac{1}{2} \left( \ident + \PZ{j,0} \right) \, \SSid{j,\tau} +\frac{1}{2} \left( \ident - \PZ{j,0} \right) \, 
        \Shift{}{j,\tau} \label{eq:adaptive East Z meas} \\
        &\Adapt{j'\tau';j,\tau} \, = \, \ident^{\vpd}_{j,0} \, \SSProj{j,\tau}{(0)} + \PX{j,0} \, \SSProj{j,\tau}{(1)} \label{eq:adaptive East X rot} \, , ~
    \end{align}
\end{subequations}
which is identical to the measurement protocol used in the $\U{1}$-symmetric example \eqref{eq:adaptive East dilated channels} in Sec.~\ref{subsec:Continuous Adaptive}. As in the $\U{1}$ case, we apply the adaptive gate $\adapt$ immediately following measurement for maximum effect.

\begin{figure*}[t]
    \centering
    \subfigure[\label{subfig:L-collapse}]{\includegraphics[height=5.5cm]{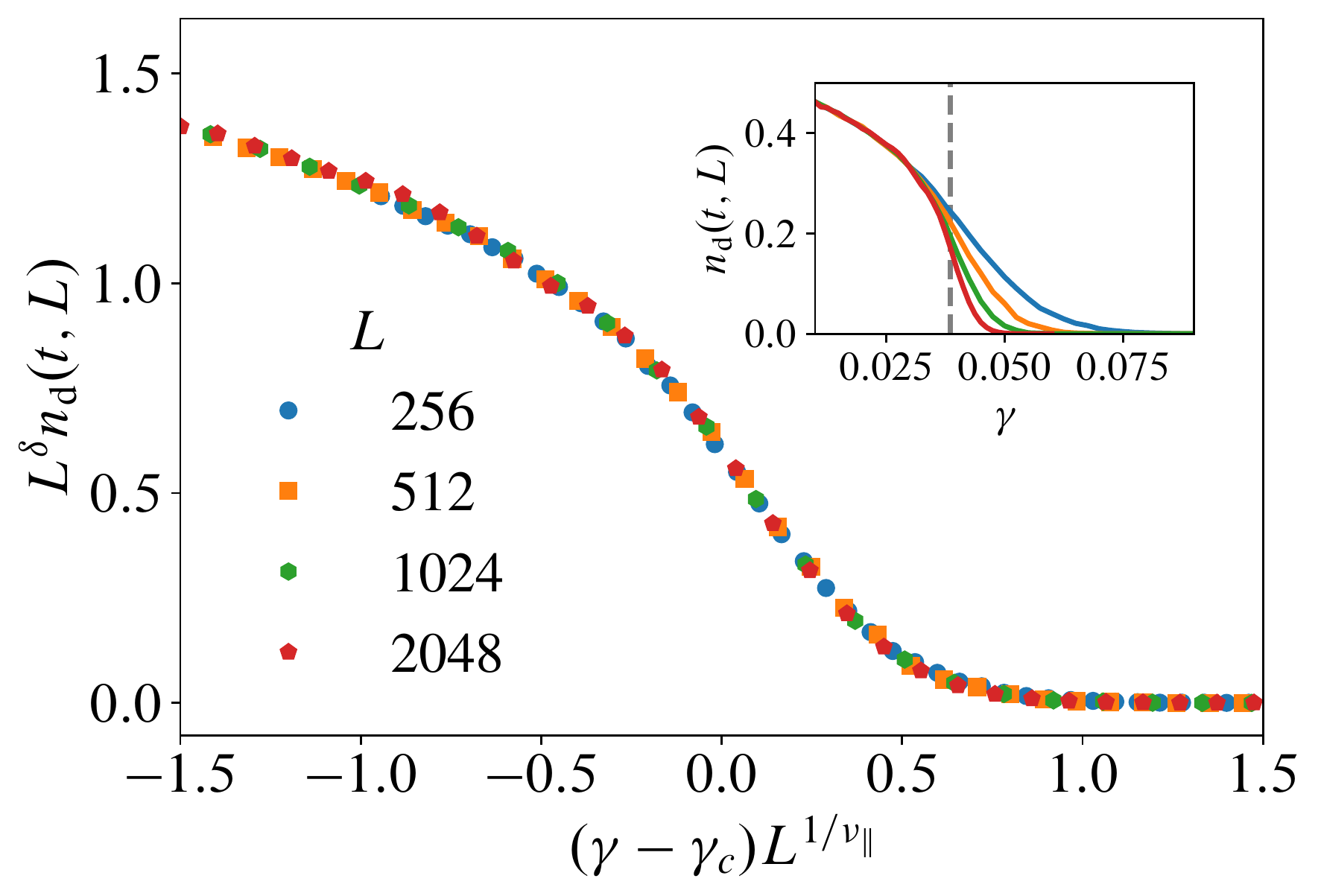}}
    \hspace{1cm}
    \subfigure[\label{subfig:t-collapse}]{\includegraphics[height=5.5cm]{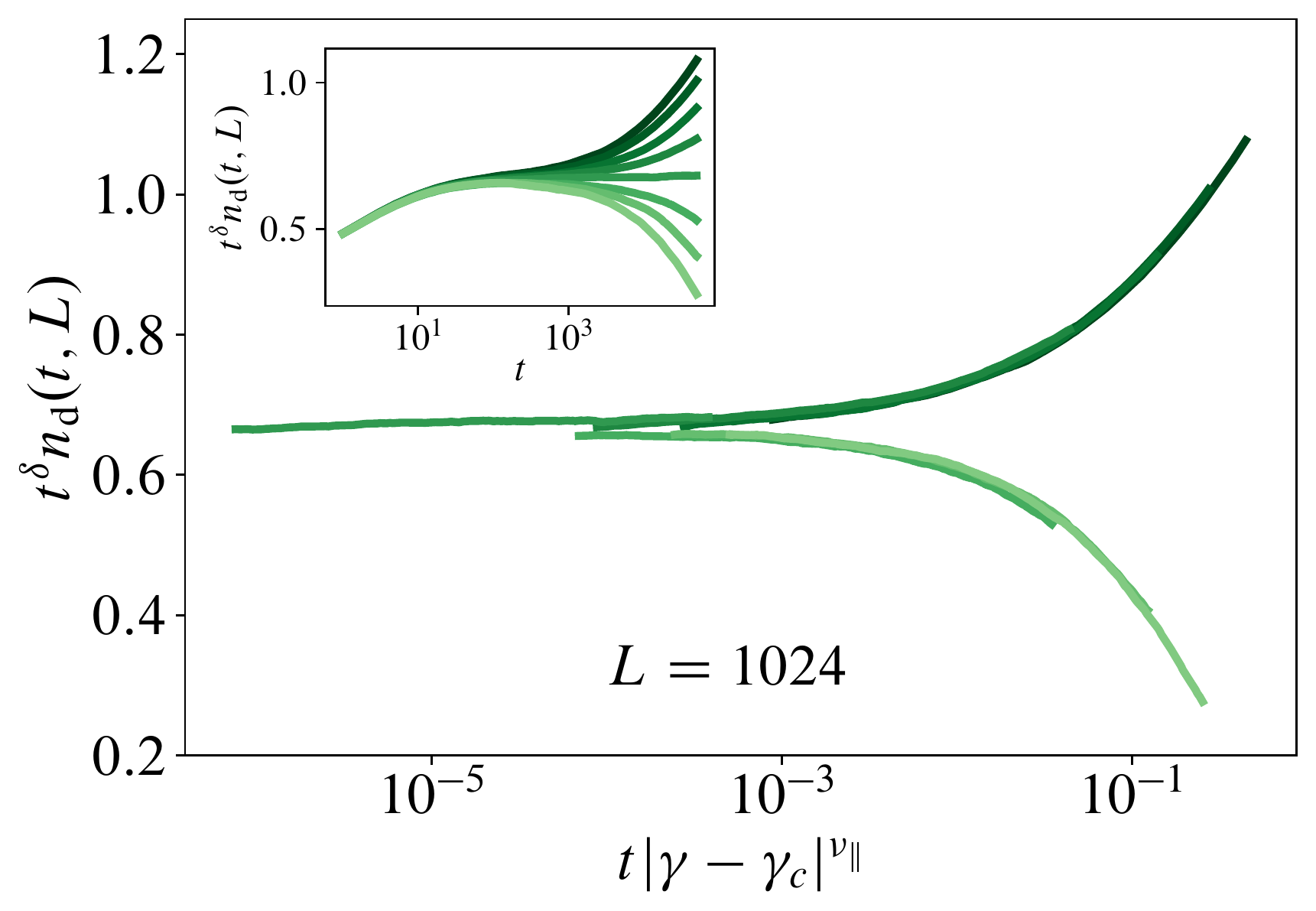}}
    \caption{{\bf Left}: Scaling collapse of the the density $n^{\,}_\text{d}(t,\size)$ of down spins \eqref{eq:East defect density} evaluated at a time $t=2\size$. The data exhibit a high-quality collapse for a critical measurement rate $\measrate^{\,}_{\rm c} \approx 0.038$ and using the exponents $\delta=0.159$ and $\nu^{\,}_{\|}=1.73$ consistent with directed percolation. The inset depicts $n^{\,}_\text{d}(t,\size)$ \eqref{eq:East defect density} without rescaling. {\bf Right}: scaling collapse of $n^{\,}_\text{d}(t,\size)$ for fixed system size $\size=1024$ as a function of time for various measurement rates $\measrate \in (0.03708, 0.03913)$.
    \label{fig:east}}
\end{figure*}

Evolving operators in the Heisenberg picture, the East gate $\gate^{\,}_{j,j+1}$ \eqref{eq:East gate} leads to the Haar-averaged updates 
\begin{subequations}
\label{eq:East Pauli basis updates}     
\begin{align}
    \ident^{\vpd}_j \otimes \Weight{n}{j+1} \, &\to \, \ident^{\vpd}_j \otimes \Weight{n}{j+1} \label{eq:East Pauli basis trivial update} \\
    \PZ{j} \otimes \Weight{n}{j+1} \, &\to \, \PZ{j} \otimes \frac{1}{2} \left( \ident + \PZ{j+1} \right) 
    \label{eq:East Pauli basis nontrivial update} \, , ~~
\end{align}
\end{subequations}
in the Pauli operator basis, and
\begin{subequations}
\label{eq:East Proj basis updates}
\begin{align}
    \Proj{j}{(n)} \otimes \Proj{j+1}{(0)} \, &\to \, \Proj{j}{(n)} \otimes \Proj{j+1}{(0)} \label{eq:East Proj basis trivial update} \\
    \Proj{j}{(n)} \otimes \Proj{j+1}{(1)} \, &\to \, \frac{1}{2} \ident^{\vpd}_j \otimes \Proj{j+1}{(1)} \label{eq:East Proj basis nontrivial update}  \, ,~~
\end{align}
\end{subequations}
in the (na\"ive) projector basis. As in previous examples, the combination of measuring $\PZ{j}$ and applying the outcome-dependent gate $\adapt$ \eqref{eq:adaptive East dilated channels} acts as 
\begin{equation}
    \label{eq:East adapt meas update}
    \ident^{\vpd}_j \, , ~\PZ{j} \, \to \, \ident^{\vpd}_j \, , ~~
\end{equation}
in the Heisenberg-Stinespring picture (averaged over measurement outcomes and projected onto the default state of the apparatus) while annihilating $\PX{}$ and $\PY{}$.

To see that the adaptive hybrid East model detailed above admits a transition as a function of $\measrate$, we again turn to numerical simulation. As with the $\U{1}$ example, we use the update rules \eqref{eq:East Pauli basis updates} and \eqref{eq:East adapt meas update} to sample the evolution of local operators $\PZ{j}(t)$ in the Heisenberg  picture using a stochastic classical Markov evolution of the bit string with $b_j=1$ with all other $b_i=0$. Sites with $b_i = 1$ will be referred to as either defective or ``active.''

As in the previous example, we compute the quantity 
\begin{equation}
    \label{eq:East defect density}
    n^{\vpp}_{\rm d} (t,\size) \, = \, \frac{1}{\size} \sum\limits_{j=1}^{\size} \, \frac{1}{2} \, \left(\, 1  - \expval{\,  \PZ{j} (t) \, }^{\vpp}_{\infty} \,  \right)  \,, ~~
\end{equation}
which vanishes if the adaptive dynamics (locally) realize the target absorbing state $\ket{\bvec{0}}$ up to a vanishing fraction of defects (i.e., a subextensive number of spins down in the thermodynamic limit $\size \to \infty$).
Note that $n^{\,}_{\rm d}(t,\size) \to 0$ corresponds to $\expval{\,  \PZ{j} (t) \, }^{\vpp}_{\infty}  = 1-o(1)$, signalling order. 

The numerical results for $n^{\,}_{\rm d}(t,\size)$ \eqref{eq:East defect density} under adaptive East dynamics are depicted in Fig.~\ref{fig:east}. In contrast to the $\U{1}$ example of Sec.~\ref{subsec:Continuous Adaptive}, the East updates \eqref{eq:East Pauli basis updates} may spawn additional nonzero bits (i.e., $\PZ{}$ operators) to the right of existing nonzero bits. Hence, although the combination of measurements and feedback removes defects, the dynamics generated by time evolution can add them back. Thus, the density of defective spins $n^{\,}_{\rm d}(t,\size)$ \eqref{eq:East defect density} with respect to the state $\ket{\bvec{0}}$ is no longer a nonincreasing function of time (though $\ket{\bvec{0}}$ remains an absorbing state since $Z$s cannot spawn spontaneously).  Since the system exhibits competition between ``survival'' and ``extinction'' (of $Z$'s), with no additional symmetries, it is expected to belong to the directed percolation (DP) universality class~\cite{HayeAbsorbing}. Since we probe the probability that a cluster grown from a single seed remains active, we expect $n^{\,}_{\rm d}(t, L)$ to obey the scaling form~\cite{HayeAbsorbing}
\begin{equation}
    n^{\vpp}_{\rm d}(t, \size) \sim t^{-\delta} f\left((\measrate-\measrate^{\,}_{\rm c}) \, t^{1/\nu_{\|}}, t^{1/z}/\size \right)
    \, , 
    \label{eqn:scaling}
\end{equation}
which we plot in Fig.~\ref{subfig:L-collapse} as a function of $\measrate-\measrate^{\,}_{\rm c}$ at fixed time $t = 2 \size$, for $\size = 2^n$ with $n=8,9,10,11$. We observe a high-quality scaling collapse consistent with the Ansatz \eqref{eqn:scaling}, using exponents $\nu^{\,}_{\|} = 1.73$ and $\delta = 0.159$ corresponding to the DP universality class (the DP exponent $\nu^{\,}_{\perp}$ describes spatial correlations, which are not relevant to this analysis) \cite{HayeAbsorbing}. Meanwhile,  Fig.~\ref{subfig:t-collapse} shows a plot of $n^{\,}_{\rm d}(t,\size)$ as a function of time $t$ for various $\measrate$ close to $\measrate^{\,}_{\rm c}$ at fixed system size $L=1024$. 

The two panels of Fig.~\ref{fig:east} show that, for $\measrate > \measrate^{\,}_{\rm c}$, the absorbing state $\ket{\bvec{0}}$ is reached rapidly in time, while for $\measrate < \measrate^{\,}_{\rm c}$ (and for sufficiently large system sizes $\size$), there remains a nonzero density $n^{\,}_{\rm d}$ of defective (or ``active'') sites. In the vicinity of the critical point, the density decays as $n^{\,}_{\rm d}(t, \size) \sim t^{-\delta}$ until a time $t \sim |\measrate-\measrate^{\,}_{\rm c}|^{-\nu_{\|}}$ corresponding to the the correlation length, which is depicted prior to rescaling in the inset of Fig.~\ref{subfig:t-collapse}. 
The collapses establish a critical measurement rate of $\measrate^{\,}_c \approx 0.038$.

The numerical simulation of the adaptive quantum East model establishes that measurement-induced transitions are possible in maximally chaotic models with kinetic constraints but without continuous symmetries. Note that a similarly small critical measurement rate $\measrate^{\,}_c$ recovers for a distinct  adaptive constrained model \cite{ODea}. 

Importantly, the adaptive absorbing-state MIPT of  \cite{ODea} is distinct from the MIET \cite{og-MIPT, FisherMIPT1, chan, FisherMIPT2, ChoiQiAltman, RomainHolographic2, RUCreview, ZabaloMIPT, GullansHuseOP, GullansHusePRX, MIPT-ATA, UtkarshChargeSharp, hsieh, barkeshli, MIPT-exp}. Additionally, while \cite{ODea} allows for independent rates of projective measurements and feedback, we note that this is not physically meaningful for physically observable quantities of the form \eqref{eq:n-point-expval}. As we have shown, any measurement without outcome-dependent feedback has no effect on dynamics, on average; at the same time, feedback without a corresponding measurement is just a unitary gate. This is consistent with the numerical data reported in \cite{ODea}.

We expect that such absorbing-state MIPTs can be realized in generic adaptive hybrid dynamics without continuous symmetries, involving either deterministic time evolution or Haar-random evolution subject to constraints. In either case, the order parameter \eqref{eq:Adaptive OP} must not correspond to a local generator of a continuous symmetry, or competition with the measurements (generally of the same type of operator as the order parameter) is impossible. The existence of nonconserved operators that are not dynamically trivial is guaranteed in the case of deterministic dynamics (without continuous symmetries). In the Haar-random case, the order parameter must be diagonal in the computational basis (in which the constraint is expressed) to ensure robustness to time evolution. These findings extend to models on arbitrary  $\LocDim$-state qudits on any graph in any number of spatial dimensions $\SpaceDim$. 

We expect that such transitions represent the \emph{only} genuine measurement-induced phase transitions given maximally chaotic dynamics. However, we expect that the constrained dynamics considered herein generally capture the universal features of deterministic (and/or submaximal) chaotic time evolution. Any such MIPT requires the utilization of measurement outcomes, with outcome-dependent unitary operations designed to ``steer'' toward a particular many-body state being the most promising, scalable, and generally applicable strategy. Finally, our identification of an MIPT captured by an  order parameter of the form $\expval{ \PZ{j}(t) }$ establishes that nonlinearity of a quantity in the density matrix $\DensMat(t)$ is not \emph{necessary} to observe measurement-induced phase structure.

\section{Conclusion} 
\label{sec:conclusion}
We have explored the generic effects of projective measurements on phase structure and dynamical universality in chaotic quantum dynamics. Our exhaustive study fully characterizes the landscape of physically observable hybrid quantum dynamics, and sharply constrains the existence of genuine (i.e., physical) measurement-induced phases of matter. Our results apply to generic chaotic quantum models acting on $\LocDim$-state qudits on any graph, in any spatial dimension, and with arbitrary (Abelian) combinations of symmetries 
and/or kinetic constraints. 

Crucially, we find that (\emph{i}) nonlinearity in the density matrix $\DensMat(t)$ is neither necessary nor sufficient for a quantity to detect nontrivial effects due to measurements; (\emph{ii}) the utilization of outcomes via adaptive feedback, postselection, or classical decoding is the crucial ingredient to realize a measurement-induced transition (observable or otherwise); (\emph{iii}) any transition that only manifests in postselected quantities is \emph{not} a transition between distinct phases of matter in \emph{any} physically meaningful or historically consistent sense; (\emph{iv})  projective measurements have no effect on the underlying chaotic spectrum, distinguishing the measurement-induced entanglement transition from thermalization (localization) transitions, which also manifest in entanglement entropy; (\emph{v}) genuine measurement-induced phases of matter can only be realized in adaptive hybrid protocols---in which gates are conditioned on prior measurement outcomes---in which the underlying time evolution lacks continuous symmetries and is either deterministic or constrained.

Our findings are made possible---or otherwise facilitated---by the representation of measurements we develop in Sec.~\ref{sec:Stinespring}. Following the Stinespring Dilation Theorem \cite{Stinespring}, measurement channels act unitarily on a dilated Hilbert space that includes the measurement apparatus. Our crucial insight is the identification of this unitary with the time evolution of the combination of the system and apparatus \cite{AaronDiegoFuture}, allowing for the evolution of operators in the Heisenberg picture and diagnosis of spectral properties in the presence of measurements.

In Sec.~\ref{sec:OneCopy}, we established that standard diagnostics of phase structure (e.g., expectation values of observables, $n$-point correlation functions, and both linear and nonlinear response) are blind to the average effects of measurements without feedback. We prove this result for the first time, on all time scales and in the presence of arbitrary symmetries and/or constraints. At most, measurements can continuously spoil the structure of the underlying time evolution without realizing a sharp transition. Essentially, on average and at late times, both projective measurements and chaotic time evolution drive a system toward a thermal mixed state (possibly within some symmetry and/or Krylov sector), so that only the structure preserved by their combination is important. At short times, the combination of chaotic evolution and measurements without feedback is equivalent to the former alone, indicating that there is no competition between measurements and chaotic evolution in the absence of feedback.

In Sec.~\ref{sec:SFF} we considered spectral properties in the presence of measurements for the first time. Our analysis of the spectral form factor (SFF) is only possible using the Stinespring formalism. Prior to the MIET, the most prominent example of a transition from area- to volume-law scaling of entanglement entropy corresponded to thermalization transitions \cite{mblrmp, mblarcmp, ETH1, ETH2, ETH3}; that transition also manifests in spectral statistics and the SFF. However, we find that measurements have no effect on the spectrum of the underlying time evolution. This holds even when the SFF is defined to be quadratic in $\DensMat(t)$ and even if the outcomes for the two copies of  $\DensMat(t)$ are postselected. This further establishes that nonlinearity in  $\DensMat(t)$ is not sufficient to detect a measurement-induced transition.

In Sec.~\ref{sec:can't work} we apply a physically reasonable and historically consistent definition of a ``phase of matter'' to measurement-induced transitions reported in the literature \cite{og-MIPT, FisherMIPT1, chan, FisherMIPT2, ChoiQiAltman, RomainHolographic2, RUCreview, ZabaloMIPT, GullansHuseOP, GullansHusePRX, MIPT-ATA, UtkarshChargeSharp, hsieh, barkeshli, MIPT-exp}. We find that postselected quantities are plainly unphysical and cannot define (or distinguish) genuine phases of matter. Thus, the measurement-induced entanglement transition, charge-sharpening transition, and other analogous measurement-induced transitions reported in the literature are \emph{not} transitions between distinct phases of matter in any physically meaningful or historically consistent sense. We also find that \emph{all} attempts to circumvent the postselection problem fail to meet the requirements for a phase of matter. We further distinguish classifier transitions \cite{learnability, HosseinNN} from transitions between phases of matter. Noting that nonlinearity in $\DensMat$ is unrelated to realizing MIPTs, we identify utilization of outcomes as the crucial ingredient. However, we note that classical utilization of the measurement outcomes via classical postprocessing does not appear to be sufficiently scalable to diagnose genuine phase structure.

In Sec.~\ref{sec:adaptive}, we consider the remaining option, corresponding to adaptive hybrid circuits, which use the outcomes of measurements to determine future gates. We find that adaptive protocols cannot realize MIPTs in maximally chaotic models without structure or with only discrete symmetries, as no observables are robust to time evolution. We then find that continuous symmetries are compatible with robust order and successful ``steering'' of quantum states, but not a sharp phase transition. To see this, we consider a $\U{1}$-symmetric hybrid model in $1\SpaceDim$, which successfully realizes the absorbing state $\ket{\bvec{0}}$ from \emph{any} initial state for any nonzero measurement rate $\measrate>0$. 

Finally, in Sec.~\ref{subsec:Constrained Adaptive}, we identify the only class of models that appears compatible with genuine measurement-induced transitions between distinct phases of matter. This corresponds either to deterministic time evolution or Haar-random evolution subject to kinetic constraints. In the latter case, the order parameter and projective measurements must act diagonally in the basis in which the constraint is defined. It is also essential that the evolution does not have any continuous symmetries, or that the order parameter (and/or measured observables) not be conserved themselves (i.e., we expect that continuous time translation symmetry is not an obstacle to realizing an MIPT). We provide evidence of a genuine MIPT in the context of an adaptive quantum East model \cite{QuantumEast2020}, in which measurements steer toward the absorbing state $\ket{\bvec{0}}$ starting from arbitrary initial states. We find a critical measurement rate of $\measrate^{\,}_{\rm c} \approx 0.038$, and a high-quality scaling collapse consistent with a directed percolation transition \cite{HayeAbsorbing}. These findings are consistent with concurrent results \cite{TomMIPT, preselect, ODea, JongYeonDecode, MIPT_wormhole} on adaptive dynamics, and our analysis straightforwardly extends to arbitrary models on generic systems. We expect that absorbing-state transitions in adaptive hybrid protocols (which realize ``steering'') are the only physically meaningful example of a measurement-induced phase transition.

\section*{Acknowledgements}
We thank Andy Lucas for useful discussions, and thank M.P.A. Fisher, Jong Yeon Lee, Andy Lucas, and Andrew Potter for feedback on this manuscript. AJF is supported in part by a Simons Investigator Award via Leo Radzihovsky. OH and RN are supported by
the Air Force Office of Scientific Research under award number FA9550-20-1-0222. RN acknowledges the support of the Simons Foundation through a Simons Fellowship in Theoretical Physics.  OH and RN acknowledge the hospitality of Stanford University, and RN further acknowledges the hospitality of the KITP 
during the completion of this work. The KITP is supported in part by the National Science Foundation under Grant No. NSF PHY-1748958.

\appendix
\renewcommand{\thesubsection}{\thesection.\arabic{subsection}}
\renewcommand{\thesubsubsection}{\thesubsection.\arabic{subsubsection}}
\renewcommand{\theequation}{\thesection.\arabic{equation}}

\section{Operator gymnastics}
\label{app:OperatorGym}
Here we provide some supporting details regarding operator spaces and bases relevant to the main text.

\subsection{Operator space}
\label{app:OpSpace}
A quantum ``state'' is realized by an appropriately normalized element of the physical Hilbert space $\Hilbert$ with dimension $\HilDim$---i.e., a vector space defined over the complex field $\Comps$ with $\HilDim$ linearly independent basis vectors $\left\{ \ket{n} \right\}$, imbued with an inner product satisfying $\inprod{a}{b}=\inprod{b}{a}^{\ast}$ and $\inprod{m}{n} = \kron{m,n}$ for the basis vectors. 

The operators $\opket{\observ}$ that act on the state space $\Hilbert$ are elements of the \emph{operator space} $\operatorname{End} \left( \Hilbert \right)$. The set $\operatorname{End} \left( \Hilbert \right)$ of linear maps on $\Hilbert$ is itself a  Hilbert space with dimension $\HilDim^2$ and an inner product satisfying $\opinprod{A}{B}=\opinprod{B}{A}^{\ast}$. 

By convention, the operator inner product is given by the standard (Frobenius) norm
\begin{equation}
    \label{eq:OpInner}
    \opinprod{A}{B} \, = \, \frac{1}{\HilDim} \tr{ A^{\dagger} B } \, , ~~
\end{equation}
which manifestly obeys $\opinprod{A}{B}=\opinprod{B}{A}^{\ast}$, where $\HilDim = \tr{\ident}$ is the dimension of the underlying 
state space. Given this norm, the $\HilDim^2$  basis operators satisfy
\begin{equation}
    \label{eq:BasisOpInner}
    \opinprod{\Pauli{\mu}{\,}}{\Pauli{\nu}{\,}} \, = \, \kron{\mu,\nu} \, , ~~
\end{equation}
where $\mu,\nu \in \left[ 0, \HilDim^2 -1 \right]$ label the $\HilDim^2$ many-body basis operators, where $\mu=0$ corresponds to the identity. 

This basis is both orthonormal and complete, i.e.,
\begin{equation}
    \label{eq:OpCompleteness}
    \superident \, = \, \sum\limits_{\mu=0}^{\HilDim^2-1} \opBKop{\Pauli{\mu}{\,}}{\Pauli{\mu}{\,}} \, ,~~
\end{equation}
realizes the superidentity $\superident \opket{A} = \opket{A}$, which maps any operator onto itself. Using these relations, any operator can be written in the basis-dependent form
\begin{equation}
    \label{eq:GenOpDecomp}
    A \, = \, \sum\limits_{\mu=1}^{\HilDim^2} \, A^{\,}_{\mu} \, \Pauli{\mu}{\,} \, \longleftrightarrow \, \opket{A} \, = \, \sum\limits_{\mu=1}^{\HilDim^2} \, A^{\,}_{\mu} \opket{\Pauli{\mu}{\,}} \,, ~~
\end{equation}
where the coefficients are given  by
\begin{equation}
    \label{eq:OpDecompCoef} 
    A^{\,}_{\mu} \, = \, \opinprod{\Pauli{\mu}{\,}}{A} \, = \, \frac{1}{\HilDim} \tr{ \left( \Pauli{\mu}{\,} \right)^{\dagger} A } \, ,~~
\end{equation}
much as a vector can be decomposed in a given basis according to $\ket{v} = \sum_n v^{\,}_n \ket{n}$ with $v^{\,}_n = \inprod{n}{v}$. In the case of time-dependent operators, all time dependence is carried by the coefficients \eqref{eq:OpDecompCoef}, as the basis 
is static.

In general, our interest lies in local lattice models, in which the physical Hilbert space factorizes over a collection of sites as $\Hilbert = \Hilbert_1^{\otimes \Nsite}$, where $\Hilbert^{\vps}_1 = \Comps^{\LocDim}_{\,}$ is the on-site Hilbert space. We then form a \emph{many-body} operator basis as the set of Kronecker products over sites of the single-site basis operators. In other words, \eqref{eq:GenOpDecomp} becomes
\begin{equation}
    \label{eq:LocOpDecomp}
    A  \, = \, \sum\limits_{\bvec{\mu}} \, a^{\,}_{\bvec{\mu}} \, \Pauli{\bvec{\mu}}{}\, \equiv \, \sum\limits_{\bvec{\mu}} \, A^{\,}_{\bvec{\mu}} \, \bigotimes\limits_{j=1}^{\Nsite} \Pauli{\mu^{\,}_j}{j} , ~~
\end{equation}
where the length-$\Nsite$ vector $\bvec{\mu} = \left( \mu_1 , \dots , \mu_{\Nsite} \right)$ stores which basis operator acts on each site, with $\mu^{\,}_j \in \left[0,\LocDim^2-1 \right]$. For a system of qubits ($\LocDim=2$), the on-site operator space is spanned by the Pauli matrices $\ident$, $X$, $Y$, and $Z$; the many-body operator basis is then the set of all possible Pauli string operators $\Pauli{\bvec{\mu}}{}$.

\subsection{The na\"ive basis}
\label{app:NaiveBasis}
A simple choice of operator basis---which we term the ``na\"ive basis''---is specified straightforwardly by the matrix elements of operators as realized by a particular basis for the underlying Hilbert space $\Hilbert$ for \emph{states}. 

We define the on-site orthonormal basis operators
\begin{equation}
    \label{eq:NaiveBasisOp}
     \nbasisop{mn} \, = \, \LocDim^{1/2} \, \BKop{m}{n} \, ,~~
\end{equation}
where $0 \leq m,n \leq \LocDim -1 $ (in some orthonormal basis for the single-site Hilbert space $\Hilbert^{\,}_j$) label the $\LocDim^2$ unique basis operators for a given site. The basis operators \eqref{eq:NaiveBasisOp} are orthonormal under the operator inner product \eqref{eq:OpInner}, as can be verified by direct inspection:
\begin{align}
    \opinprod{\nbasisop{k\ell}}{\nbasisop{mn}} \, &= \, \frac{1}{\LocDim} \, \tr{ \,  \LocDim^{1/2} \, \BKop{\ell}{k} \,  \LocDim^{1/2} \, \BKop{m}{n} \,} \notag \\
    &= \, \kron{k,m} \, \kron{\ell,n} \, , ~~\label{eq:NaiveBasisOrtho}
\end{align}
and completeness of the basis follows from linear independence: i.e., the only solution to
\begin{equation}
    \sum\limits_{m,n=0}^{\LocDim-1} \, a^{\vpp}_{m,n} \, \nbasisop{mn} \, = 0 \, ,~~
\end{equation}
is the trivial solution $a^{\,}_{m,n}=0$ for all $m,n$.

Following \eqref{eq:GenOpDecomp}, generic single-site operators can be expressed in this basis as
\begin{align}
    A \, &= \, \sum\limits_{m,n=0}^{\LocDim-1} \, A^{\vpp}_{m,n} \, \nbasisop{mn} \, , ~~
    \label{eq:op in naive basis}
\end{align}
where the coefficient
\begin{equation}
    A^{\vpp}_{m,n} \, = \, \matel{m}{A}{n} \, ,~~
    \label{eq:op coeff in naive basis}
\end{equation}
is simply the matrix element of the operator $A$ 
in the chosen on-site state-space basis. The many-body operator basis is simply the Kronecker product over sites of the single-site operator basis  \eqref{eq:LocOpDecomp}.

\subsection{The Weyl basis}
\label{app:UnitaryBasis}
The Weyl basis is useful, unitary extension of the Pauli-matrix basis for $\LocDim = 2$ to generic on-site dimension $\LocDim>2$,
\begin{equation}
    \label{eq:UnitaryBasisOp}
    \ubasisop{mn} \, = \shift{m} \weight{n} \, ,~~
\end{equation}
where the Weyl operators $\shift{\,}$ and $\weight{\,}$ are defined by
\begin{subequations}
\label{eq:ClockOpDefs}
\begin{align}
    \shift{\,} \, &\equiv \sum\limits_{k=0}^{\LocDim-1} \, \BKop{ k+1}{k} \label{eq:ShiftOpDef} \\
    \weight{\,} \, &\equiv \sum\limits_{k=0}^{\LocDim-1} \, \omega^k \, \BKop{k}{k } \label{eq:WeightOpDef} \, , ~~
\end{align}
\end{subequations}
which generalize the Pauli $\PX{}$ and $\PZ{}$. Above, note that the ket labels are defined modulo $\LocDim$ (i.e., $\ket{k + \LocDim} \cong \ket{k}$) and $\omega$ is the $\LocDim$th root of unity, 
\begin{equation}
    \label{eq:omega}
    \omega \, \equiv \, e^{2 \pi \ii / \LocDim} \, ,~~
\end{equation}
and we also note the general relations
\begin{subequations}
\begin{align}
    \ident \, &= \, \shift{\LocDim} \, = \, \weight{\LocDim}   \label{eq:ClockQ} \\
    0 \, &= \, \tr{\weight{\,}} \, = \, \tr{\shift{\,}}  \label{eq:ClockTrace} \\
    \shift{\dagger} \, &= \, \shift{-1} \label{eq:ShiftDag} \\
    \weight{\dagger} \, &= \, \weight{-1} \label{eq:WeightDag} \, ,~~
\end{align}
\end{subequations}
so that $\weight{\,}$ and $\shift{\,}$ are both traceless and \emph{unitary}, and reduce to the $\PX{}$ and $\PZ{}$ Pauli matrices for $\LocDim=2$. These two operators obey the \emph{multiplication rule}
\begin{equation}
    \label{eq:ClockMultRule}
    \weight{\,} \shift{\,} \, = \, \omega \, \shift{\,} \weight{\,} \, , ~~
\end{equation}
and corollary relations 
follow from 
$\weight{m} \shift{n} \, = \, \omega^{mn} \, \shift{n} \weight{m}$ (note that $\omega \to -1$ for $\LocDim = 2$). We verify that $\ubasisop{mn}$ \eqref{eq:UnitaryBasisOp} forms an orthonormal basis by checking \eqref{eq:BasisOpInner}: 
\begin{align}
    \opinprod{\ubasisop{jk}}{\ubasisop{mn}} \, &= \, \frac{1}{\LocDim} \, \tr{ \weight{-k} \shift{-j} \shift{m} \weight{n} } \notag \\
    &= \, \frac{1}{\LocDim} \, \tr{ \shift{m-j} \weight{n-k} } \notag \\
    &= \, \kron{j,m} \kron{k,n} \, ,~~\label{eq:UnitaryBasisInner}
\end{align}
and completeness follows from linear independence.

Single-site operators are decomposed in this basis via
\begin{equation}
    \label{eq:SingleSiteOpUnitaryBasis}
    A \, = \, \frac{1}{\LocDim} \, \sum\limits_{m,n=0}^{\LocDim-1} \, \tr{ \weight{-n} \shift{-m} A } \, \shift{m} \weight{n} \, , ~~
\end{equation}
or, in the operator ket notation,
\begin{align}
    \label{eq:SingleSiteOpUnitaryKetForm}
    \opket{A} \, &= \, \sum\limits_{m,n=0}^{\LocDim-1} \, \opket{\ubasisop{mn}} \opinprod{\ubasisop{mn}}{A} \, = \, \sum\limits_{m,n=0}^{\LocDim-1} \, A^{\vpd}_{m,n} \, \opket{\ubasisop{mn}} \,, ~~
\end{align}
where the coefficients are given by
\begin{align}
    \label{eq:A coefficients untiary basis}
    A^{\vpd}_{m,n} \, &= \, \opinprod{\ubasisop{mn}}{A} \, = \, \frac{1}{\LocDim} \, \tr{ \, \ubasisopdag{m,n} \, A \,} \,, ~~
\end{align}
via completeness. From \eqref{eq:SingleSiteOpUnitaryBasis}, we conclude that
\begin{equation}
    \label{eq:ZProjector}
    \BKop{m}{m} \, = \, \frac{1}{\LocDim} \, \sum\limits_{k=0}^{\LocDim-1} \, \omega^{-mk} \, \weight{k} \, ,~~
\end{equation}
where $\ket{m}$ satisfies $\weight{\,}\ket{m} = \omega^{m} \ket{m}$. By convention, the $\weight{\,}$ eigenbasis is the default basis for single-qudit states. 

Owing to self duality of the clock operators, we can express projectors onto eigenstates of $\shift{\,}$ as
\begin{equation}
    \label{eq:XProjector}
    \BKop{\widetilde{m}}{\widetilde{m}} \, = \, \frac{1}{\LocDim} \, \sum\limits_{\widetilde{k}=0}^{\LocDim-1} \, \omega^{-\widetilde{m} \widetilde{k}} \, \shift{\widetilde{k}} \, ,~~
\end{equation}
from which it is straightforward to verify that
\begin{align}
    \label{eq:ShiftEigenstateProj}
    \shift{\,} \BKop{\widetilde{m}}{\widetilde{m}} \, = \, \omega^{\widetilde{m}} \BKop{\widetilde{m}}{\widetilde{m}}  \, , ~~
\end{align}
meaning $\ket{\widetilde{m}}$ is an eigenstate of $\shift{\,}$ with eigenvalue $\omega^{\widetilde{m}}$, as expected. Specifically, 
\begin{align}
    \label{eq:ShiftEigenstate}
    \ket{\widetilde{m}} \, &\equiv \, \frac{1}{\sqrt{\LocDim}} \, \sum\limits_{k=0}^{\LocDim-1} \, \omega^{-\widetilde{m} \, k} \, \ket{k} \, , ~~
\end{align}
where $\ket{k}$ is the eigenstate of $\shift{}$ with eigenvalue $\omega^k$.

Generic \emph{many-body} operators can be written as
\begin{equation}
    \label{eq:ManyBodyOpUnitaryBasis}
    A \, = \, \frac{1}{\HilDim} \, \sum\limits_{\bvec{m},\bvec{n}} \tr{ \ubasisopdag{\bvec{m},\bvec{n}} \, A} \, \ubasisop{\bvec{m},\bvec{n}} \, , ~~
\end{equation}
where we have implicitly defined the shorthand
\begin{align}
    \ubasisop{\bvec{m},\bvec{n}} \, &\equiv \bigotimes_{i \in \Hilbert^{\vpp}_{\rm dil}} \, \Shift{m^{\,}_{i}}{i} \Weight{n^{\,}_{i}}{i}\, ,~~
    \label{eq:VectorBasisOp}
\end{align}
where the integer-valued vectors $\bvec{m}$ and $\bvec{n}$ reflect the powers of $\shift{\,}$ and $\weight{\,}$ that act on each degree of freedom in the dilated Hilbert space, as in \eqref{eq:MB Weyl basis op}.

\subsection{Projectors and other useful relations}
\label{app:ProjectorBasis}

Using the na\"ive operator basis \eqref{eq:NaiveBasisOp} with Weyl  $\weight{}$-basis states, we define the orthonormal set of \emph{diagonal} operators (or normalized $\weight{}$-state projectors) according to
\begin{align}
    \nProj{j}{(n)} \, &\equiv \, \nbasisop{n,n} \, = \, \sqrt{q} \, \BKop{n}{n}^{\vpp}_j \, = \,  \sum\limits_{k=0}^{\LocDim-1} \, \frac{\omega^{-nk}}{\sqrt{\LocDim}}\,\Weight{k}{j} \,,~\label{eq:Normalized projectors}
\end{align}
where $\weight{} \ket{n} \, = \omega^n \ket{n}$. We  similarly define the set of projectors onto Weyl $\shift{}$ states of site $j$ via
\begin{align}
    \tnProj{j}{(n)} \, &\equiv \, \nbasisop{\widetilde{n},\widetilde{n}} \, = \, \sqrt{q} \, \BKop{\widetilde{n}}{\widetilde{n}}^{\vpp}_j \, = \,   \sum\limits_{k=0}^{\LocDim-1} \,\frac{ \omega^{-nk}}{\sqrt{\LocDim}}\,\Shift{k}{j} \,,~\label{eq:Normalized X projectors}
\end{align}
where $\shift{} \ket{\widetilde{n}} \, = \omega^{\widetilde{n}} \ket{\widetilde{n}}$ \eqref{eq:ShiftEigenstate}. The projectors \eqref{eq:Normalized projectors} and \eqref{eq:Normalized X projectors} are orthonormal \eqref{eq:BasisOpInner},
\begin{equation}
    \label{eq:nproj orthonormality}
    \opinprod{\nProj{j}{(n)}}{\nProj{j}{(m)}} \, = \,\opinprod{\tnProj{j}{(n)}}{\tnProj{j}{(m)}} \, = \, \kron{m,n} \, , ~~
\end{equation}
as can be verified from \eqref{eq:OpInner} and the definition \eqref{eq:Normalized projectors}. These operators therefore comprise a complete orthonormal basis for the \emph{diagonal} operators acting on site $j$ in either the Weyl $\weight{}$ ($\nProj{}{(n)}$) or $\shift{}$ ($\tnProj{}{(n)}$) eigenbases.

For reference, we also state the following relations:
\begin{subequations}
\label{eq:naive to Weyl}
\begin{align}
    \nbasisop{ab} \, &= \, \sqrt{\LocDim} \, \BKop{a}{b} \, = \, \sum\limits_{n=0}^{\LocDim-1} \, \frac{\omega^{-n \, b}}{\sqrt{\LocDim}} \, \shift{a-b} \, \weight{n} \, \label{eq:naive Z to Weyl} \\
    \nbasisop{\widetilde{i} \widetilde{j}} \, &= \, \sqrt{\LocDim} \, \BKop{\widetilde{i}}{\widetilde{j}} \, = \, \sum\limits_{m=0}^{\LocDim-1} \, \frac{\omega^{-m \, \widetilde{i} }}{\sqrt{\LocDim}} \, \shift{m} \, \weight{\widetilde{j}-\widetilde{i}} \, \label{eq:naive X to Weyl} \, ,~~
\end{align}
\end{subequations}
with inner products given by
\begin{subequations}
\label{eq:naive Weyl inner}
\begin{align}
    \opinprod{\nbasisop{ab}}{\ubasisop{mn}} \, &= \, \LocDim^{-1/2} \, \omega^{-b \, n} \, \kron{a,m+b} \label{eq:naive Z Weyl inner} \\
    \opinprod{\nbasisop{\widetilde{i} \widetilde{j}}}{\ubasisop{mn}} \, &= \, \LocDim^{-1/2} \, \omega^{-\widetilde{i} \, m} \, \kron{\widetilde{j},n+\widetilde{i}} \label{eq:naive X Weyl inner} \, , ~~
\end{align}
\end{subequations}
and it is also useful to define
\begin{subequations}
\label{eq:naive change of basis}
\begin{align}
    \opket{ \nbasisop{ \widetilde{m},\widetilde{n} } } \, &= \, \frac{1}{ \sqrt{\LocDim} } \,\sum\limits_{i,j=0}^{\LocDim-1} \, \omega^{j \, \widetilde{n}-i\, \widetilde{m} } \, \BKop{i}{j} \notag \\
    &= \, \frac{1}{\LocDim} \, \sum\limits_{i,j=0}^{\LocDim-1} \, \omega^{ j \, \widetilde{n}-i \, \widetilde{m} } \, \opket{ \nbasisop{ij} } \label{eq:naive change of basis ket} \\
    \opbra{\nbasisop{\widetilde{m},\widetilde{n}}} \, &= \, \frac{1}{\sqrt{\LocDim}} \,\sum\limits_{k,l=0}^{\LocDim-1} \, \omega^{k \widetilde{m}-l \widetilde{n}} \, \BKop{l}{k} \notag \\
    &= \, \frac{1}{\LocDim} \, \sum\limits_{k,l=0}^{\LocDim-1} \, \omega^{k \widetilde{m}-l \widetilde{n} } \, \opbra{\nbasisop{kl}} \label{eq:naive change of basis bra} \, ,~~
\end{align}
\end{subequations}
where we used the relation \eqref{eq:ShiftEigenstate}.

\section{The onefold Haar channel}
\label{app:Haar Channel}
We now introduce the Haar-averaging procedure in the context of generic, Haar-averaged time evolution (i.e., without conservation laws or  constraints). In the Schr\"odinger picture, one considers the time evolution of density matrices (or similar objects like $\evo \mobserv \DensMat \evo^{\dagger}$), while operators remain constant in time; in the Heisenberg picture, one considers the density matrices (and states) to be independent of time, and instead evolves operators according to $\evo^{\dagger} \observ \evo$. Note that our unitary Stinespring formalism  of Sec.~\ref{sec:Stinespring} allows for the evolution of operators $\observ$ in the presence of projective measurements.

In this work, we primarily consider expressions involving one copy of a given random unitary $\Haar$ (and its conjugate $\Haar^{\dagger}$) over the unitary group $\U{\LocDim^{\ell}}$ with uniform (Haar) measure. We find that $\overline{\Haar^{\dagger} \, \observ \,\Haar} = \overline{\Haar\,\observ \,\Haar^{\dagger}}$; in other words, Haar-averaged time-evolution updates of \emph{either} density matrices in the Schr\"odinger picture or operators in the Heisenberg picture are both captured by the \emph{same} one-fold Haar channel \cite{YoshidaCBD},
\begin{align}
    \Phi \left[ \, \observ \, \right] \, &\equiv \, \overline{ \, \Haar^{\dagger} \, \observ \, \Haar \, }  \, = \, \frac{1}{\HilDim} \, \ident \, \tr{ \, \observ \, } \label{eq:1FoldHaarAvg} \, , ~~
\end{align}
where $\HilDim$ is the dimension of the underlying Hilbert space (i.e., the unitaries $\Haar$ are averaged over $\U{\HilDim}$ with uniform measure). The equivalence of the Haar-averaged Heisenberg- and Schr\"odinger-picture update channels can be verified 
by noting that $\Phi \left[ \observ^{\dagger} \right] = \Phi \left[ \observ^{\ast} \right]$.

In the absence of symmetries or constraints (i.e., each $\ell$-site unitary gate is a $\LocDim^{\ell} \times \LocDim^{\ell}$ Haar-random unitary without block structure), the Haar-averaged Heisenberg update rule for the time-evolution layer labeled $\lambda$ is given by
\begin{align}
    \overline{ \, \observ \left( t + \delta t \right) \,} \, &= \, \overline{ \, \evo^{\dagger}_{t,\lambda} \, \overline{\, \observ \left( t \right)\,} \, \evo^{\vpd}_{t,\lambda} \, } \, , ~~\label{eq:SingleLayerTevoUpdateAvgd}
\end{align}
where the ``upper'' average applies only to the unitaries in layer $t,\lambda$, and the ``lower'' average on the right-hand side includes averages for all prior Heisenberg time steps. 

Using  \eqref{eq:ManyBodyOpUnitaryBasis}, any physical observable $\observ$ can be decomposed according to
\begin{align}
    \label{eq:Observ for 1fold}
    \overline{\observ(t)} \, &= \,  \sum\limits_{\vec{m},\vec{n}} \,\overline{C_{\vec{m},\vec{n}}^{(t)} } \,\bigotimes\limits_{j=1}^{\Nsite} \,   \Shift{m^{\vpp}_j}{j}\, \Weight{n^{\vpp}_j}{j} \,   ,~~
\end{align}
where $\vec{m},\vec{n}$ are restricted to the 
physical slice $\tau=0$, and the coefficients $C$ are given by
\begin{align}
    \label{eq:Observ for 1fold Coeff}
    \overline{C_{\vec{m},\vec{n}}^{(t)} } \, &\equiv \,\frac{1}{\LocDim^{\Nsite}} \,  \underset{\rm dil}{\trace} \, \left[ \,  \overline{\observ (t)} ~ \bigotimes\limits_{j=1}^{\Nsite} \,  \Weight{-n^{\vpp}_j}{j} \, \Shift{-m^{\vpp}_j}{j}\, \right]  \, , ~~
\end{align}
which may be averaged over previous time-evolution gates. 

Assuming that $\evo$ is a circuit of the form \eqref{eq:TevoGate}, the Haar-averaged time-evolution channel corresponding to layer $\lambda$ of time step $t$ acts on $\observ$ \eqref{eq:Observ for 1fold} according to
\begin{align}
    \overline{\, \observ \left( t + \delta t \right) \,} \, &= \,  \sum\limits_{\vec{m},\vec{n}} \overline{ C^{(t)}_{\vec{m},\vec{n}} } \, \overline{ \evo^{\vpd}_{t,\lambda} \,\ubasisop{\vec{m},\vec{n}} \, \evo^{\dagger}_{t,\lambda} } \notag \\
    &= \, \sum\limits_{\vec{m},\vec{n}} \, \overline{ C^{(t)}_{\vec{m},\vec{n}} }\, \bigotimes\limits_{r \in \lambda}\, \overline{\, \gate^{\vpd}_{t,\lambda,r} \, \ubasisoppow{\vec{m},\vec{n}}{(r)} \, \gate^{\dagger}_{t,\lambda,r} \,} \notag \\
    &=\,  \sum\limits_{\vec{m},\vec{n}} \, \overline{ C^{(t)}_{\vec{m},\vec{n}} } \, \bigotimes\limits_{r \in \lambda}\, \frac{1}{\LocDim^{\ell}}\, \ident^{\,}_{r} \tr{\, \ubasisoppow{\vec{m},\vec{n}}{(r)} \,} \notag \\
    &= \, \sum\limits_{\vec{m},\vec{n}} \, \overline{ C^{(t)}_{\vec{m},\vec{n}} } \,\ident \,  \prod\limits_{j=1}^{\Nsite}\, \kron{m^{\vpp}_{j},0} \, \kron{n^{\vpp}_{j},0} \notag \\
    &= \, \overline{ C^{(t)}_{\vec{0},\vec{0}} } \, \ident \, = \, \frac{1}{\LocDim^{\Nsite}} \, \tr{ \, \overline{\observ(t)} \,} \, \ident 
    \,, ~~\label{eq:Gen Haar Update}
\end{align}
which means generic operators are trivialized by a single layer of featureless Haar-random gates that tile all sites of the system. Note that the same update applied to the density matrix in the Schr\"odinger picture gives
\begin{equation}
    \label{eq:Gen Haar Update Dens}
    \overline{\, \DensMat \left(t + \delta t \right)\,} \, = \,  \overline{ C^{(t)}_{\vec{0},\vec{0}} } \, \ident  \, = \,  \frac{1}{\LocDim^{\Nsite}} \, \tr{\, \overline{\DensMat (t)} \,} \, \ident \, = \, \DensMat^{\vpp}_{\infty} \, ,~~
\end{equation}
the infinite-temperature density matrix $\DensMat^{\vpp}_{\infty}= \LocDim^{-\Nsite} \, \ident$ (a featureless, maximally mixed state), for \emph{any} $\overline{\DensMat(t)}$.

\section{Transition-matrix formulation}
\label{app:Observ T Mat}
There are two approaches to evaluating the quantities of interest \eqref{eq:n-point-expval} in Sec.~\ref{sec:OneCopy}: The first is to evolve the dilated density matrix $\DensMatSS (t)$ \eqref{eq:DilatedInitialState} in the Schr\"odinger picture; the second is to evolve the probe observables $\observ^{\,}_j$ in \eqref{eq:n-fold Heisenberg} in the Heisenberg picture. 
In the former scenario, one includes $n$ additional Stinespring qubits to store the outcomes of measuring each of the $n$ probe observables $\observ^{\,}_i$.

\subsection{Evolution in operator space}
\label{subsubsec:Intro TMat}
In either case, we consider hybrid evolution under a dilated unitary circuit $\evo \in \operatorname{Aut}(\Hilbert^{\,}_{\rm dil})$ \eqref{eq:Tevo to t} comprising layers of local gates corresponding to (\emph{i}) maximally chaotic time evolution, possibly with block structure (to encode arbitrary combinations of Abelian symmetries and/or kinetic constraints) and (\emph{ii}) projective measurements of the circuit observables $\mobserv^{\,}_{r,\tau}$. As noted in Sec.~\ref{sec:Stinespring}, the initial state of all Stinespring registers is $\ket{0}$ by default, and we trace over all Stinespring qudits in evaluating \eqref{eq:n-point-expval} to capture repeated experimental trials\footnote{As argued in Sec.~\ref{subsec:NoTrajectories}, there is no value in considering individual measurement trajectories: The combination of projective measurements and unitary dynamics can produce \emph{any} mixed state, so there are no useful predictions to be made.}.

The ensemble- and outcome-averaged evolution of either the density matrix $\DensMatSS (t)$ or the probe observables $\observ^{\,}_i$ is efficiently captured using a \emph{transition matrix}. Much like the hybrid evolution unitary $\evo (t,t')$ (for $t>t'$) is a linear operator that maps the state $\ket{\psi ( t')}$ at time $t'$ to the state $\ket{\psi (t)}$ at time $t$, the transition matrix $\tmat (t,t')$ is a linear \emph{superoperator} that maps the density matrix $\DensMat (t')$ at time $t'$ to the density matrix $\DensMat (t) = \evo (t,t') \, \DensMat (t') \, \evo^{\dagger} (t,t')$ at time $t$ (in the Schr\"odinger picture). Additionally, the superoperator $\tmat^{\dagger} (t,t')$ maps the operator $\observ (t')$ at time $t'$ to $\observ (t) = \evo^{\dagger} (t,t') \, \observ (t') \, \evo (t,t')$ at time $t$ (in the Heisenberg picture). Importantly, in the absence of outcome-dependent feedback, all dilated unitary updates can be restricted to the physical Hilbert space. In other words,
\begin{subequations}
\label{eq:Tmat evo def}
\begin{align}
    \tmat^\dagger (t,t') \circ \observ  &\equiv  \trace_{\rm ss}  \left[  \evo^{\dagger} (t,t') \, \observ  \, \evo^{\vpd} (t,t') \, \BKop{\bvec{0}}{\bvec{0}}^{\vpp}_{\rm ss} \, \right] \label{eq:Tmat evo Heis def} \\
    \tmat (t,t') \circ \DensMat  &\equiv  \trace_{\rm ss}  \left[ \evo (t,t')  \DensMat \otimes \BKop{\bvec{0}}{\bvec{0}}^{\vpp}_{\rm ss}  \evo^{\dagger} (t,t')  \right]  \label{eq:Tmat evo Schro def} \, .
\end{align}
\end{subequations}
Suppose that the operator $\evo (t,t')$ takes the form
\begin{align}
    \label{eq:example evo full}
    \evo (t,t')  &=  \TimeOrder  \prod\limits_{s=t'}^t \, \prod\limits_{\sigma} \bigotimes\limits_{r \in \sigma}^{\MeasRounds^{\,}_s}  \measunitary^{\vpd}_{s,\sigma,r} \, \prod\limits_{\lambda=1}^{\ell}  \bigotimes\limits_{r' \in \lambda}  \gate^{\vpd}_{s,\lambda,r'} \, ,~
\end{align}
where $\TimeOrder$ is a time-ordering operator, $s$ runs over time steps $t'$ through $t$, $\sigma$ runs over the $\MeasRounds^{\,}_s$ measurement layers in round $s$ with individual gates $\measunitary$ acting on clusters $r$, and $\lambda$ runs over time-evolution layers with individual gates $\gate$ acting on $\ell$-site clusters $r'$. Then the  transition matrix $\tmat (t,t')$ corresponding to $\evo (t,t')$ \eqref{eq:example evo full} is given by
\begin{align}
    \tmat (t,t')  &\equiv  \TimeOrder  \prod\limits_{s=t'}^t \, \prod\limits_{\sigma=1}^{\MeasRounds^{\,}_s}  \bigotimes\limits_{r \in \sigma}  \tgate^{\rm (meas)}_{s,\sigma,r} \, \prod\limits_{\lambda=1}^{\ell}  \bigotimes\limits_{r' \in \lambda}  \tgate^{\rm (evo)}_{s,\lambda,r'} \label{eq:example tmat full} \, , ~
\end{align}
where $\tgate$ is a Hermitian transition-matrix \emph{gate}. The above realizes Schr\"odinger evolution of density matrices; the conjugate $\tmat^\dagger (t,t')$ for Heisenberg evolution is given by
\begin{align}
    \tmat^{\dagger} (t,t')  &\equiv  \AntiTimeOrder  \prod\limits_{s=t}^{t'} \, \prod\limits_{\lambda=\ell}^{1} \bigotimes\limits_{r' \in \lambda}  \tgate^{\rm (evo)}_{s,\lambda,r'} \, \prod\limits_{\sigma=\MeasRounds^{\,}_s}^1  \bigotimes\limits_{r \in \sigma}  \tgate^{\rm (meas)}_{s,\sigma,r} \, \label{eq:example tmat dag full} , ~
\end{align}
where $\AntiTimeOrder$ is the \emph{anti}-time-ordering operator (which arranges the transition-matrix gates in reverse chronological order compared to the time-ordering operator $\TimeOrder$). 

Importantly, the transition matrix $\tmat$ is a \emph{circuit} of transition-matrix gates (corresponding to both time evolution and measurements) with precisely the same circuit structure as the corresponding unitary operator $\evo$. Likewise, $\tmat^\dagger$ has the same circuit structure as $\evo^\dagger$. 

By restricting to hybrid quantum circuits without adaptive feedback (i.e., in which the results of measurements are not \emph{utilized}, we need only consider the evolution of observables (and density matrices) acting on the \emph{physical} degrees of freedom. It will prove convenient to regard operators as kets that live in the operator Hilbert space,
\begin{align}
    \opket{\observ} \, \in \, \operatorname{End} \left( \Hilbert^{\vpp}_{\rm ph} \right) \, ,~~ \label{eq:op ket gen}
\end{align}
where $\operatorname{End} ( \Hilbert^{\vpp}_{\rm ph} )$ is the space of operators acting on the physical Hilbert space $\Hilbert^{\vpp}_{\rm ph}$, which is detailed in App.~\ref{app:OpSpace}. 

The operator space is imbued with the inner product
\begin{align}
    \label{eq:op inner prod}
    \opinprod{A}{B} \, \equiv \, \frac{1}{\HilDim} \, \tr{ \, A^\dagger \, B \, } \, ,~~
\end{align}
where $\HilDim$ is the dimension of the underlying Hilbert space. Note that $\opinprod{B}{A} = \opinprod{A}{B}^*$ is skew symmetric, meaning that the operator space is a Hilbert space.

The elements of the operator space span all observables (and density matrices) acting on $\Hilbert^{\,}_{\rm ph}$. As with generic inner-product spaces, the elements of $\operatorname{End} ( \Hilbert^{\vpp}_{\rm ph} )$ can be decomposed onto an orthonormal basis (see App.~\ref{app:OperatorGym}). Of particular interest is the Weyl basis
\begin{align}
    \ubasisop{\vec{m},\vec{n}} \, &\equiv \, \bigotimes\limits_{j} \, \Shift{m^{\,}_j}{j} \, \Weight{n^{\,}_j}{j} \, ,~~ \label{eq:MB Weyl basis op} 
\end{align}
where $\shift{}$ is the Weyl ``shift'' operator \eqref{eq:ShiftOpDef} and $\weight{}$ is the Weyl ``weight'' operator \eqref{eq:WeightOpDef}, which provide a unitary extension of the Pauli $\PX{}$ and $\PZ{}$ operators to $\LocDim>2$. Importantly, this basis is orthonormal, i.e.,
\begin{align}
    \opinprod{\ubasisop{\vec{m}',\vec{n}'}}{\ubasisop{\vec{m},\vec{n}}} \, 
    &= \, \kron{\vec{m},\vec{m}'} \, \kron{\vec{n},\vec{n}'} \label{eq:Weyl basis ortho} \, ,~~
\end{align}
and also complete, i.e.,
\begin{align}
    \superident \, &= \, \sum\limits_{\vec{m},\vec{n}} \, \opBKop{\ubasisop{\vec{m},\vec{n}}}{\ubasisop{\vec{m},\vec{n}}} \, , ~~\label{eq:Weyl basis complete}
\end{align}
where the superidentity $\superident$  maps any operator to itself. 

The foregoing relations imply the decomposition
\begin{align}
    \opket{\observ (t)} \, &\equiv   \,  \sum\limits_{\vec{m},\vec{n}} \, \opket{ \ubasisop{ \vec{m},\vec{n} } } \hspace{-0.3mm} \opinprod{ \ubasisop{ \vec{m},\vec{n} } }{ \observ (t) } \notag \\
    &= \,  \sum\limits_{\vec{m},\vec{n}} \, O^{(t)}_{\vec{m},\vec{n}}  \,  \opket{\ubasisop{\vec{m},\vec{n}}} \, , ~~ \label{eq:OpSS ket}
\end{align}
and relating the two lines above gives
\begin{align}
    \label{eq:Op t-dep coeff}
    O^{(t)}_{\vec{m},\vec{n}} \, &= \, \opinprod{ \ubasisop{\vec{m},\vec{n}} }{ \observ (t) } \, = \, \frac{1}{\LocDim^{\Nsite}} \,  \tr{ \, \ubasisopdag{\vec{m},\vec{n}} \, \observ(t) \,}\, , ~~
\end{align}
for the time-dependent coefficients. Analogously, we decompose density matrices according to
\begin{align}
    \opket{\DensMatSS(t)} \, &\equiv \, \sum\limits_{\vec{m},\vec{n}} \, \opket{ \ubasisop{ \vec{m},\vec{n} } } \hspace{-0.3mm} \opinprod{ \ubasisop{ \vec{m},\vec{n} } }{ \DensMat (t) } \notag \\
    &= \, \sum\limits_{\vec{m},\vec{n}} \, C^{(t)}_{ \vec{m},\vec{n} }  \, \opket{ \ubasisop{\vec{m},\vec{n}} } \, ,~\label{eq:DensMatSS ket}
\end{align}
where the coefficeint $C$ for the density matrix is
\begin{align}
    \label{eq:C of t coeff}
    C^{(t)}_{\vec{m},\vec{n}} \, &= \, \opinprod{\ubasisop{\vec{m},\vec{n}}}{\DensMat(t)} \, = \, \frac{1}{\LocDim^{\Nsite}} \,  \tr{ \, \ubasisopdag{\vec{m},\vec{n}} \, \DensMat(t) \,}\, . ~
\end{align}
The transition matrix $\tmat(t,t')$ satisfies the relations \eqref{eq:Tmat evo def}
\begin{subequations}
\label{eq:tmat op space}
    \begin{align}
        \opket{\observ (t) } \, &= \, \tmat^{\dagger} (t,t') \, \opket{\observ (t')} \label{eq:tmat op space observ} \\
        \opket{\DensMat (t) } \, &= \, \tmat (t,t') \, \opket{\DensMat (t') } \label{eq:tmat op space rho} \, ,~~
    \end{align}
\end{subequations}
where $\tmat (t,t')$ [$\tmat^\dagger (t,t')$] is a circuit with the same structure as the underlying unitary circuit $\evo (t,t')$ [$\evo^{\dagger} (t,t')$]. 

The identification of $\tmat (t,t')$ as a \emph{matrix} is most transparent when one considers the updates to the coefficients of observables \eqref{eq:Op t-dep coeff} and density matrices \eqref{eq:C of t coeff} under hybrid time evolution. Taking the operator inner product of \eqref{eq:tmat op space} with Weyl basis operators \eqref{eq:MB Weyl basis op} to the following expressions for a single update $t \to t+\delta t$ corresponding to layer $t,\lambda$ of the hybrid circuit.
\begin{subequations}
\label{eq:tmat op coeffs}
\begin{align}
    \overline{ O^{(t+\delta t)}_{ \vec{m},\vec{n} } } \, &= \, \sum\limits_{ \vec{m}',\vec{n}' }\, \Tmat{ \vec{m},\vec{n}; \vec{m}',\vec{n}'}{ (t,\lambda) } \, \overline{ O^{(t)}_{ \vec{m}',\vec{n}' } } 
    \label{eq:Tmat Heis on coeffs} \\
    \overline{ C^{ (t+\delta t) }_{ \vec{m},\vec{n} } } \, &= \, \sum\limits_{ \vec{m}', \vec{n}' }\, \Tmat{ \vec{m}, \vec{n}; \vec{m}',\vec{n}' }{ (t,\lambda) } \, \overline{ C^{(t)}_{ \vec{m}',\vec{n}' } }\, , ~~\label{eq:Tmat Schro on coeffs}
\end{align}
\end{subequations}
where the superscripts denote which layer of transition-matrix gates acts, and may correspond to time evolution or measurement. The gates $\tgate$ are all Hermitian, and each layer $t,\lambda$ is the same for either the Heisenberg or Schr\"odinger picture---only the ordering of the layers changes, as indicated in \eqref{eq:example tmat full} and \eqref{eq:example tmat dag full}.

We now work out the elements of the transition matrices corresponding to layers of time-evolution and measurement gates in the Weyl basis $\vec{m},\vec{n}$ \eqref{eq:MB Weyl basis op}.

\subsection{Time-evolution transition matrices}
\label{subsubsec:Tevo TMat}
We first consider the time-evolution transition-matrix gates $\tgate^{\rm (evo)}_{t,\lambda,r}$ corresponding to local unitary gates $\gate^{\,}_{t,\lambda,r}$ \eqref{eq:GenProjGate} that  act on operators supported on the cluster $r$ in layer $\lambda$ of time step $t$. The superoperator $\tgate^{\rm (evo)}_{t,\lambda,r}$ encodes the averaging of the unitary gate $\gate$ \eqref{eq:GenProjGate} over a chaotic ensemble (possibly within various blocks  corresponding to Abelian symmetries and/or kinetic constraints) with Haar (i.e., uniform) measure \cite{YoshidaCBD,NahumRUC1, NahumOperator, RUCNCTibor, CDLC1, RUCconTibor, RUCconVedika, U1FRUC, ConstrainedRUC}. 

In the absence of block structure we find
\begin{equation}
    \tgate^{\rm (evo)}_{t,\lambda,r} \, = \, \opBKop{\ident}{\ident}^{\,}_r \implies \tgate^{\rm (evo)}_{t,\lambda} \, = \, \opBKop{\ident}{\ident}\, , ~~\label{eq:Tevo Tmat no blocc}
\end{equation}
where we assume that the gates in layer $\lambda$ tile the entire system. In other words, a \emph{single} layer of featureless, maximally chaotic time evolution annihilates \emph{any} nontrivial observable $\observ$, and replaces \emph{any} density matrix $\DensMat$ with the infinite-temperature Gibbs state $\DensMat^{\,}_{\infty} \propto \ident$. The matrix elements of $\tgate^{\rm (evo)}_{t,\lambda}$ \eqref{eq:Tevo Tmat no blocc} are given by
\begin{align}
    \label{eq:Tgate Mat El No Sym}
    \Tgate{\vec{m},\vec{n};\vec{m}',\vec{n}'}{(t,\lambda)} \, = \, \kron{\vec{m},\vec{m}'} \, \kron{\vec{n},\vec{n}'} \, \kron{\vec{m}',\vec{0}} \, \kron{\vec{n}',\vec{0}} \, ,~~
\end{align}
for any layer $t,\lambda$ that tiles the system.

For block-structured time-evolution gates $\gate^{\,}_{t,\lambda,r}$ of the form \eqref{eq:GenProjGate}---corresponding to generic combinations of Abelian symmetries and/or constraints---where the blocks $\alpha$ are defined in the Weyl $\weight{}$ basis \eqref{eq:WeightDag} and all realize chaotic dynamics, the matrix elements of the corresponding transition-matrix gate are given by 
\begin{align}
    \tgate^{(t,\lambda,r)}_{\vec{m},\vec{n};\vec{m'},\vec{n}'} &=  \kron{\vec{m},\vec{m}'} \, \kron{\vec{m}',\vec{0}} \, \sum\limits_{\alpha}  \sum\limits_{\vec{a},\vec{a}' \in \alpha}  \frac{\omega^{\vec{a}'\cdot \vec{n}' - \vec{a}\cdot \vec{n}}}{n^{\,}_{\alpha} \, \LocDim^{\ell}} \, , ~\label{eq:Tgate nice form Z} 
\end{align}
and the basis-independent form of the gate is
\begin{align}
    \label{eq:Tgate result}
    {\tgate}^{\vpp}_{t,\lambda,r}  &=  \sum\limits_{\alpha} \frac{1}{n^{\,}_{\alpha}} \sum\limits_{\vec{a},\vec{a}'\in \alpha} \,\opBKop{\nProj{r}{(\vec{a})}}{\nProj{r}{(\vec{a}')}} \, ,~~
\end{align}
where $\nProj{r}{(\vec{a})}$ is the normalized projector \eqref{eq:Normalized projectors} onto the $\weight{}$-basis configuration $\vec{a}$ of cluster $r$ and $n^{\,}_{\alpha}$ is the number of configurations $\vec{a}$ in block $\alpha$. Though we default to the $\weight{}$ basis \eqref{eq:WeightOpDef}, more generally, normalized projectors can be defined for arbitrary choices of orthonormal basis.

\subsection{Measurement transition matrices}
\label{subsubsec:Meas TMat}
We next consider the measurement gates $\tgate^{\rm (meas)}_{t,\sigma,r}$ that act on operators in the cluster $r$ in measurement round $\sigma$ of time step $t$. This gate captures the update due to projective measurement of the circuit observable $\mobserv^{\,}_{t,\sigma,r}$ with spectral decomposition \eqref{eq:ASpectralDecomp}. We primarily consider $\mobserv^{\,}_{t,\sigma,r}$ that are diagonal in the $\weight{}$ basis \eqref{eq:WeightOpDef} (i.e., the eigenstates of $\mobserv$ are $\weight{}$-basis states); however, we also consider  $\mobserv^{\,}_{t,\sigma,r}$ that are diagonal in the $\shift{}$ basis \eqref{eq:ShiftOpDef}\footnote{In general, arbitrary bases can realized by including a unitary change of basis before and after the measurement gate.}.

Here we consider protocols in which the outcome of a given projective measurement is not utilized in evaluation of quantities of the form \eqref{eq:n-point-expval}. The outcomes are not used adaptively (i.e., to determine subsequent gates) nor are they used for post processing (i.e., the assignment of outcome-dependent weights to different trajectories). Thus, we average each individual measurement gate over all outcomes, weighted by their probabilities. This corresponds to taking the trace over the Stinespring degree of freedom; accordingly, the transition matrices act on the physical operators alone, as in \eqref{eq:Tmat evo def}. 

In either the Heisenberg or Schr\"odinger picture, the transition-matrix gate $\tgate^{\rm (meas)}_{t,\sigma,r}$---corresponding to the measurement of $\mobserv^{\,}_{t,\sigma,r}$ \eqref{eq:ASpectralDecomp}---is defined via
\begin{align}
    \tgate^{ ( t,\sigma,r,{\rm Heis} ) }_{\vec{m},\vec{n};\vec{m}',\vec{n}'}  &= \frac{1}{\LocDim^{\ell}} \sum\limits_{\mu=0}^{\Noutcome-1}  \tr{  \ubasisopdag{ \vec{m},\vec{n} }  \Proj{r}{(\mu)} \ubasisop{\vec{m}',\vec{n}'} \Proj{r}{(\mu)} } \, , ~\label{eq:Tmat meas elements nonadapt}
\end{align}
where $\mu$ labels the $\Noutcome$ distinct eigenvalues of $\mobserv^{\,}_{t,\sigma,r}$ \eqref{eq:ASpectralDecomp}, which acts on the $\ell$-site cluster $r$.

If $\mobserv^{\,}_{t,\sigma,r}$ is diagonal in the $\weight{}$ basis \eqref{eq:WeightOpDef}, then the elements of the corresponding transition-matrix gate are
\begin{align}
    \Tgate{\vec{m},\vec{n};\vec{m}',\vec{n}'}{(t,\sigma,r,\weight{})} \, &= \, \kron{\vec{m},\vec{m}'}\kron{\vec{n},\vec{n}'} \kron{\vec{m}',\vec{0}} \notag \\
    ~+ \kron{\vec{m},\vec{m}'} &\, \sum\limits_{B} \sum\limits_{\vec{b} \neq \vec{b}' \in B } \frac{\omega^{\vec{b} \cdot \left( \vec{n}'-\vec{n} \right)}}{\LocDim^{\ell}} \, \kron{\vec{m}',\vec{b}'-\vec{b}}   \,, ~ \label{eq:Tmat meas elements nonadapt degen Z} 
\end{align}
and, alternatively, if $\mobserv^{\,}_{t,\sigma,r}$ is diagonal in the $\shift{}$ basis \eqref{eq:ShiftOpDef}, the elements 
are instead given by
\begin{align}
    \Tgate{\vec{m},\vec{n};\vec{m}',\vec{n}'}{(t,\sigma,r,\shift{})} \, &= \, \kron{\vec{m},\vec{m}'}\kron{\vec{n},\vec{n}'} \kron{\vec{n}',\vec{0}} \notag \\
    ~+ \kron{\vec{n},\vec{n}'} &\, \sum\limits_{B} \sum\limits_{\vec{b} \neq \vec{b}' \in B } \frac{\omega^{\vec{b} \cdot \left( \vec{m}'-\vec{m} \right)}}{\LocDim^{\ell}} \, \kron{\vec{n}',\vec{b}'-\vec{b}}   \,, ~ \label{eq:Tmat meas elements nonadapt degen X} 
\end{align}
where, in both expressions above, $B$ labels blocks of degenerate eigenstates of $\mobserv^{\,}_{t,\sigma,r}$ \eqref{eq:ASpectralDecomp} and $\vec{b} \neq \vec{b}'$ label distinct eigenstates within that block. The basis-independent form of the transition-matrix gate is simply
\begin{align}
    \tgate^{\rm (meas)}_{t,\sigma,r} \, &= \, \sum\limits_{\mu=0}^{\Noutcome-1} \sum\limits_{\vec{a},\vec{a}' \in \mu} \, \opBKop{\nbasisop{\vec{a},\vec{a}'}}{\nbasisop{\vec{a},\vec{a}'}} \, ,~~\label{eq:Tgate meas basis-indep form}
\end{align}
where $\mu$ runs over all eigenvalues of $\mobserv^{\,}_{t,\sigma,r}$ \eqref{eq:ASpectralDecomp} and $\nbasisop{\vec{a},\vec{a}'} = \LocDim^{\ell/2} \, \BKop{\vec{a}}{\vec{a}'}^{\,}_r$ is the  na\"ive basis operator \eqref{eq:NaiveBasisOp} defined in the eigenbasis of the circuit observable $\mobserv^{\,}_{t,\sigma,r}$.

If $\mobserv^{\,}_{t,\sigma,r}$ has a nondegenerate (ND) spectrum, the second term in both \eqref{eq:Tmat meas elements nonadapt degen Z} and \eqref{eq:Tmat meas elements nonadapt degen X} vanishes, leaving
\begin{equation}
    \label{eq:Tmat meas nondegen nonadapt}
    \tgate^{\rm ND}_{t,\sigma,r} \, = \, \sum\limits_{\vec{k}} \,\opBKop{\nProj{r}{(\vec{k})}}{\nProj{r}{(\vec{k})}}  \, , ~~
\end{equation}
where the normalized projectors \eqref{eq:Normalized projectors} may realize \emph{any} basis for the $\LocDim^{\ell}$ states of cluster $r$, corresponding to the (nondegenerate) eigenbasis of $\mobserv^{\,}_{t,\sigma,r}$. 

We now derive the  \emph{measurement} transition-matrix gates $\tgate^{\,}_{t,\sigma,r}$ corresponding to conjugation of observables or density matrices by the unitary measurement gates $\measunitary^{\,}_{t,\sigma,r}$ \eqref{eq:MeasUnitaryGate} without adaptive feedback. The gates $\measunitary^{\,}_{t,\sigma,r}$ \eqref{eq:MeasUnitaryGate} capture the measurement of some observable $\mobserv^{\,}_{t,\sigma,r}$ with $\Noutcome^{\,}_{t,\sigma,r}$ outcomes (i.e., eigenvalues); the observed outcome is stored in the Stinespring register $r,\tau(t,\sigma)$ \eqref{eq:Combined SS label} of the spacetime lattice defined in Sec.~\ref{subsec:SpacetimeLattice}. Thus, the gate $\measunitary^{\,}_{s,\sigma,r}$ \eqref{eq:MeasUnitaryGate} acts only on the physical qudits in cluster $r$ and the $\Noutcome^{\,}_{t,\sigma,r}$-state Stinespring qudit $r,\tau (t,\sigma)$. 

In the absence of outcome-dependent feedback, we can restrict the transition-matrix gate $\tgate^{\,}_{t,\sigma,r}$ to the physical operator space by defining, e.g.,
\begin{subequations}
\label{eq:nonadapt meas tgate def}
    \begin{align}
        \tgate^{ ( t,\sigma,r,{\rm Heis} ) }_{\vec{m},\vec{n};\vec{m}',\vec{n}'}  &\equiv \frac{1}{ \LocDim^{\ell} }  \trace\limits_{\rm dil}  \left[  \ubasisopdag{ \vec{m},\vec{n} }  \BKop{0}{0}^{\vpp}_{\rm ss}  \measunitary^{\dagger}_{t,\sigma,r} \ubasisop{ \vec{m}',\vec{n}' }  \ident^{\vpp}_{\rm ss} \measunitary^{\vpd}_{t,\sigma,r} \right] \notag \\
        = \frac{1}{ \LocDim^{\ell} } \sum\limits_{\mu,\nu=0}^{\Noutcome-1} &\trace\limits_{\rm ph} \left[ \ubasisopdag{ \vec{m},\vec{n} }  \Proj{r}{(\mu)} \ubasisop{ \vec{m}',\vec{n}' }  \Proj{r}{(\nu)} \right] \,  \matel{0}{\shift{\nu-\mu}}{0}^{\vpp}_{\rm ss} \notag \\
        = \frac{1}{\LocDim^{\ell}} &\sum\limits_{\mu=0}^{\Noutcome-1} \, \tr{  \ubasisopdag{ \vec{m},\vec{n} }  \Proj{r}{(\mu)} \ubasisop{\vec{m}',\vec{n}'} \Proj{r}{(\mu)} } \, , ~
        \label{eq:nonadapt meas tgate Heis def} \\
        \tgate^{ ( t,\sigma,r,{\rm Schr} ) }_{\vec{m},\vec{n};\vec{m}',\vec{n}'}  &\equiv \frac{1}{ \LocDim^{\ell} }  \trace\limits_{\rm dil}  \left[  \ubasisopdag{ \vec{m},\vec{n} }   \measunitary^{\vpd}_{t,\sigma,r} \ubasisop{ \vec{m}',\vec{n}' }   \BKop{0}{0}^{\vpp}_{\rm ss} \measunitary^{\dagger}_{t,\sigma,r} \right] \notag \\
        = \frac{1}{ \LocDim^{\ell} } \sum\limits_{\mu,\nu=0}^{\Noutcome-1} &\trace\limits_{\rm ph} \left[ \ubasisopdag{ \vec{m},\vec{n} }  \Proj{r}{(\mu)} \ubasisop{ \vec{m}',\vec{n}' }  \Proj{r}{(\nu)} \right] \, \trace\limits_{\rm ss} \left[ \shift{\mu} \BKop{0}{0} \shift{-\nu} \right] \notag \\
        = \frac{1}{\LocDim^{\ell}} &\sum\limits_{\mu=0}^{\Noutcome-1} \, \tr{  \ubasisopdag{ \vec{m},\vec{n} }  \Proj{r}{(\mu)} \ubasisop{\vec{m}',\vec{n}'} \Proj{r}{(\mu)} } \, , ~\label{eq:nonadapt meas tgate Schro def}
    \end{align}
\end{subequations}
where $\Noutcome$ is a shorthand for $\Noutcome^{\,}_{t,\sigma,r}$, and we see that the transition-matrix gate $\tgate^{\,}_{t,\sigma,r}$ is the same in both the Heisenberg and Schr\"odinger pictures,
\begin{align*}
    \tgate^{ ( t,\sigma,r) }_{\vec{m},\vec{n};\vec{m}',\vec{n}'}  \equiv \frac{1}{\LocDim^{\ell}} \sum\limits_{\mu=0}^{\Noutcome-1}  \tr{  \ubasisopdag{ \vec{m},\vec{n} }  \Proj{r}{(\mu)} \ubasisop{\vec{m}',\vec{n}'} \Proj{r}{(\mu)} } \, , \, \tag{\ref{eq:Tmat meas elements nonadapt}}
\end{align*}
and further progress can be made from \eqref{eq:Tmat meas elements nonadapt} by splitting the sum into two terms. The first term captures the contributions from (\emph{i}) all blocks $\mu$ with a single configuration and (\emph{ii}) realizing the same state in every multi-configuration block $\mu$. The second term captures the contributions from blocks $\mu$ with multiple configurations, where the two projectors in \eqref{eq:Tmat meas elements nonadapt} realize distinct states in the block. We now rewrite \eqref{eq:Tmat meas elements nonadapt} as
\begin{align}
    &= \, \frac{1}{\LocDim^{\ell}} \sum\limits_{\vec{k}} \tr{ \, \ubasisopdag{\vec{m},\vec{n}} \, \Proj{r}{(\vec{k})} \, \ubasisop{\vec{m}',\vec{n}'}\, \Proj{r}{(\vec{k})}  \,} \notag \\
    &~~+\frac{1}{\LocDim^{\ell}} \sum\limits_{B} \, \sum\limits_{\substack{\vec{b},\vec{b}' \in B \\ \vec{b} \neq \vec{b}'}} \tr{ \, \ubasisopdag{\vec{m},\vec{n}} \, \Proj{r}{(\vec{b}')} \, \ubasisop{\vec{m}',\vec{n}'}\, \Proj{r}{(\vec{b})}  \,} \, , ~~\label{eq:Tmat meas elements nonadapt 2}
\end{align}
where $\vec{k}$ runs over \emph{all} $\LocDim^{\ell}$ eigenstates for $\mobserv^{\,}_{t,\sigma,r}$, $B$ labels degenerate ``blocks'' containing multiple eigenstates of $\mobserv^{\,}_{t,\sigma,r}$ with the same eigenvalue, and $\vec{b}$ and $\vec{b}'$ label the distinct, degenerate eigenstates of $\mobserv^{\,}_{t,\sigma,r}$ in the block $B$.

First, consider the special scenario in which the measured operator $\mobserv^{\,}_{t,\sigma,r}$  has a nondegenerate (ND) spectrum \eqref{eq:ASpectralDecomp}. Correspondingly, there are no degenerate blocks $B$ in the second term in \eqref{eq:Tmat meas elements nonadapt 2}, which vanishes, leaving 
\begin{align}
    \tgate^{ ( t,\sigma,r,{\rm ND}) }_{\vec{m},\vec{n};\vec{m}',\vec{n}'}  &\equiv \opmatel{\ubasisop{\vec{m},\vec{n}}}{\tgate^{\rm ND}_{t,\sigma\,r}}{\ubasisop{\vec{m}',\vec{n}'}} \notag \\
    = \frac{1}{\LocDim^{\ell}}  &\sum\limits_{\vec{k}} \tr{ \, \ubasisopdag{\vec{m},\vec{n}} \, \Proj{r}{(\vec{k})} \, \ubasisop{\vec{m}',\vec{n}'}\, \Proj{r}{(\vec{k})}  \,} \notag \\
    = \frac{1}{\LocDim^{\ell}}  &\sum\limits_{\vec{k}} \tr{ \, \ubasisopdag{\vec{m},\vec{n}} \, \Proj{r}{(\vec{k})} \,} \, \tr{ \, \Proj{r}{(\vec{k})}  \,  \ubasisop{\vec{m}',\vec{n}'} \,} \notag \\
    =\LocDim^{\ell} \,  &\sum\limits_{\vec{k}} \, \opinprod{\ubasisop{\vec{m},\vec{n}}}{\Proj{r}{(\vec{k})}} \hspace{-0.3mm} \opinprod{\Proj{r}{(\vec{k})}}{\ubasisop{\vec{m}',\vec{n}'}} \, , ~\label{eq:Tmat meas nondegen nonadapt elements}
\end{align}
and thus, the measurement of a nondegenerate observable $\mobserv^{\,}_{t,\sigma,r}$ on cluster $r$ is captured by the gate
\begin{equation}
    \tag{\ref{eq:Tmat meas nondegen nonadapt}}
    \tgate^{\rm ND}_{t,\sigma,r} \, = \, \sum\limits_{\vec{k}} \,\opBKop{\nProj{r}{(\vec{k})}}{\nProj{r}{(\vec{k})}}  \, , ~~
\end{equation}
where we have used the normalized projector \eqref{eq:Normalized projectors} onto the $k$th eigenstate $\ket{\vec{k}}^{\,}_r$ of $\mobserv^{\,}_{t,\sigma,r}$, 
\begin{equation}
    \nProj{r}{(\vec{k})} \, = \, \LocDim^{\ell/2} \, \Proj{r}{(\vec{k})} \, = \, \LocDim^{\ell/2} \, \BKop{\vec{k}}{\vec{k}}^{\,}_r \, , ~ \label{eq:norm proj nondegen}
\end{equation}
as defined in App.~\ref{app:ProjectorBasis}. For example, if the measured observable $\mobserv$ is diagonal in the $\weight{}$ basis, \eqref{eq:norm proj nondegen} reduces to \eqref{eq:Normalized projectors}; if $\mobserv$ is diagonal in the $\shift{}$ basis, \eqref{eq:norm proj nondegen} reduces to \eqref{eq:Normalized X projectors}. It will prove sufficient to consider only these two scenarios: More general, intermediate scenarios can always be represented by applying unitary channels before and after the measurement (which result in additional transition-matrix gates to effect the change of basis).

However, for generic observables with degenerate eigenvalues, we must consider the second term in \eqref{eq:Tmat meas elements nonadapt 2}. We consider two limiting cases, corresponding to observables that are diagonalized by the $\weight{}$ \eqref{eq:WeightOpDef} and $\shift{}$ \eqref{eq:ShiftOpDef} Weyl bases. Of course, generic observables $\mobserv$ will not commute with either $\weight{}$ or $\shift{}$, and the corresponding eigenbasis will correspond to some rotation of, e.g., the weight basis. This can be captured by inserting appropriate unitary gates (acting on the physical qudits) on either side of the measurement gate $\measunitary$. However, in the main text, we consider (\emph{i}) featureless time evolution, in which case we take the measurement basis to be $\weight{}$; (\emph{ii}) block-diagonal evolution (corresponding to symmetries and/or constraints) with \emph{compatible} measurements, in which case we take the charge (or computational) and measurement bases to be $\weight{}$; and (\emph{iii}) measurements that are \emph{incompatible} with block-diagonal evolution, where we consider the limiting case in which the blocks are defined in the $\weight{}$ basis while the measurements are in the $\shift{}$ basis.

We first consider the case of a $\weight{}$-basis observable $\mobserv$ measured on the $\ell$-site cluster $r$ (with $\Noutcome < \LocDim^{\ell}$ unique eigenvalues). Using \eqref{eq:ZProjector}  to expand the projectors onto configurations $\vec{b} \neq \vec{b}' \in B$ in each \emph{degenerate} block $B$, the second term in \eqref{eq:Tmat meas elements nonadapt 2} becomes
\begin{align}
    &= \frac{1}{\LocDim^{\ell}} \sum\limits_{B} \, \sum\limits_{\vec{b} \neq \vec{b}' \in B} \tr{ \, \ubasisopdag{\vec{m},\vec{n}} \, \Proj{r}{(\vec{b}')} \, \ubasisop{\vec{m}',\vec{n}'}\, \Proj{r}{(\vec{b})}  \,}  \notag \\
    &= \frac{1}{\LocDim^{3 \, \ell}} \sum\limits_{B} \, \sum\limits_{\vec{b} \neq \vec{b}' \in B}  \sum\limits_{\vec{k},\vec{k}'} \, \omega^{\vec{k} \cdot \vec{b}-\vec{k}' \cdot \vec{b}'} \, \times \,\notag \\
    ~&~ \quad \quad \tr{ \, \weight{-\vec{n}} \shift{-\vec{m}} \, \weight{\vec{k}'}\,  \shift{\vec{m}'} \weight{\vec{n}'} \, \weight{-\vec{k}}\,} \, , ~\notag
\end{align}
and we use the multiplication rule \eqref{eq:ClockMultRule} and cyclic invariance to reorder the operators in the trace, finding
\begin{align}
    &= \frac{1}{\LocDim^{3 \, \ell}} \sum\limits_{B} \, \sum\limits_{\vec{b} \neq \vec{b}' \in B }  \sum\limits_{\vec{k},\vec{k}'} \, \omega^{\vec{k} \cdot \vec{b}-\vec{k}' \cdot \vec{b}' + \vec{k}' \cdot \vec{m}'} \, \times \,  \notag \\
    &~ \quad \quad \tr{ \,  \shift{\vec{m}'-\vec{m}} \, \weight{\vec{n}'-\vec{n}+\vec{k}'-\vec{k}}\, } \, , ~\notag
\end{align}
and evaluating the trace contributes a factor of $\LocDim^{\ell}$ along with Kronecker deltas to ensure that the operators inside the trace above realize the identity,
\begin{align}
    &= \, \kron{\vec{m},\vec{m}'} \, \frac{1}{\LocDim^{2 \, \ell}} \sum\limits_{B} \, \sum\limits_{\vec{b} \neq \vec{b}' \in B }  \sum\limits_{\vec{k},\vec{k}'} \,\omega^{\vec{k} \cdot \vec{b}-\vec{k}' \cdot \vec{b}' + \vec{k}' \cdot \vec{m}'} \, \kron{\vec{k}',\vec{k}+\vec{n}-\vec{n}'}  \notag \\
    &= \, \kron{\vec{m},\vec{m}'}  \sum\limits_{B} \, \sum\limits_{\vec{b} \neq \vec{b}' \in B } \,\frac{\omega^{\left( \vec{m}'-\vec{b}' \right) \cdot \left( \vec{n}-\vec{n}' \right)}}{\LocDim^{\ell}} \, \sum\limits_{\vec{k}} \,\frac{\omega^{\vec{k} \cdot \left( \vec{b}- \vec{b}' + \vec{m}' \right) }}{\LocDim^{\ell}}  \,  ,~\notag 
\end{align}
and the rightmost sum (over all $\vec{k}$) simplifies to
\begin{align}
    &= \, \kron{\vec{m},\vec{m}'}  \sum\limits_{B} \, \sum\limits_{\vec{b} \neq \vec{b}' \in B } \frac{\omega^{\vec{b} \cdot \left( \vec{n}'-\vec{n} \right)}}{\LocDim^{\ell}} \, \kron{\vec{m}',\vec{b}'-\vec{b}}  \, , ~
    \label{eq:Tmat meas nonadapt degen elements}
\end{align}
which is zero unless $\vec{m}'$ is compatible with the difference between the configurations $\vec{b}$ and $\vec{b}'$ in block $B$. Further simplifications to \eqref{eq:Tmat meas nonadapt degen elements} are not possible without considering a particular choice of $\mobserv$.

Summarizing the above, the measurement transition matrix gate \eqref{eq:Tmat meas elements nonadapt 2} for a generic, possibly degenerate observable $\mobserv$ acting on the $\ell$-site cluster $r$ in measurement layer $\sigma$ of time step $t$, whose $\Noutcome$ unique eigenvalues correspond to  $\weight{}$-basis states \eqref{eq:WeightOpDef} has elements
\begin{align}
    \Tgate{\vec{m},\vec{n};\vec{m}',\vec{n}'}{(t,\sigma,r,\weight{})} \, &= \, \kron{\vec{m},\vec{m}'}\kron{\vec{n},\vec{n}'} \kron{\vec{m}',\vec{0}} \notag \\
    ~+ \kron{\vec{m},\vec{m}'} &\, \sum\limits_{B} \sum\limits_{\vec{b} \neq \vec{b}' \in B } \frac{\omega^{\vec{b} \cdot \left( \vec{n}'-\vec{n} \right)}}{\LocDim^{\ell}} \, \kron{\vec{m}',\vec{b}'-\vec{b}}   \,, ~ \tag{\ref{eq:Tmat meas elements nonadapt degen Z}}
\end{align}
where $B$ labels \emph{degenerate} blocks, and $\vec{b},\vec{b}'$ are distinct $\ell$-site, $\weight{}$-basis configurations of cluster $r$, which must be compatible with $\vec{m}'$ (the powers of $\shift{}$ present in a given term in the observable or density matrix being evolved). The second term vanishes for nondegenerate observables.

We now repeat the derivation of \eqref{eq:Tmat meas elements nonadapt degen Z} for degenerate $\ell$-site observables that are diagonal in the Weyl $\shift{}$ basis \eqref{eq:ShiftOpDef}. The second term in \eqref{eq:Tmat meas elements nonadapt 2} becomes
\begin{align}
    &= \frac{1}{\LocDim^{\ell}} \sum\limits_{B} \, \sum\limits_{\vec{b} \neq \vec{b}' \in B} \tr{ \, \ubasisopdag{\vec{m},\vec{n}} \, \Proj{r}{(\vec{b}')} \, \ubasisop{\vec{m}',\vec{n}'}\, \Proj{r}{(\vec{b})}  \,}  \notag \\
    &= \frac{1}{\LocDim^{3 \, \ell}} \sum\limits_{B} \, \sum\limits_{\vec{b} \neq \vec{b}' \in B}  \sum\limits_{\vec{k},\vec{k}'} \, \omega^{\vec{k} \cdot \vec{b}-\vec{k}' \cdot \vec{b}'} \, \times \,\notag \\
    ~&~ \quad \quad \tr{ \, \weight{-\vec{n}} \shift{-\vec{m}} \, \shift{\vec{k}'}\,  \shift{\vec{m}'} \weight{\vec{n}'} \, \shift{-\vec{k}}\,} \, , ~\notag
\end{align}
and we again use \eqref{eq:ClockMultRule} and cyclic invariance of the trace to rewrite the above as
\begin{align}
    &= \frac{1}{\LocDim^{3 \, \ell}} \sum\limits_{B} \, \sum\limits_{\vec{b} \neq \vec{b}' \in B }  \sum\limits_{\vec{k},\vec{k}'} \, \omega^{\vec{k} \cdot \vec{b}-\vec{k}' \cdot \vec{b}' - \vec{k} \cdot \vec{n}'} \, \times \,  \notag \\
    &~ \quad \quad \tr{ \,  \shift{\vec{m}'-\vec{m}+\vec{k}'-\vec{k}} \, \weight{\vec{n}'-\vec{n}}\, } \, , ~\notag
\end{align}
and evaluating the trace leads to
\begin{align}
    &= \, \kron{\vec{n},\vec{n}'} \, \frac{1}{\LocDim^{2 \, \ell}} \sum\limits_{B} \, \sum\limits_{\vec{b} \neq \vec{b}' \in B }  \sum\limits_{\vec{k},\vec{k}'} \,\omega^{\vec{k} \cdot \vec{b}-\vec{k}' \cdot \vec{b}' - \vec{k} \cdot \vec{n}'} \, \kron{\vec{k}',\vec{k}+\vec{m}-\vec{m}'}  \notag \\
    &= \, \kron{\vec{n},\vec{n}'}  \sum\limits_{B} \, \sum\limits_{\vec{b} \neq \vec{b}' \in B } \,\frac{\omega^{\vec{b}'\cdot \left( \vec{m}'-\vec{m} \right)}}{\LocDim^{\ell}} \, \sum\limits_{\vec{k}} \,\frac{\omega^{\vec{k} \cdot \left( \vec{b}- \vec{b}' - \vec{n}' \right) }}{\LocDim^{\ell}}  \,  ,~\notag 
\end{align}
and performing the rightmost sum (over all $\vec{k}$) and also relabelling $\vec{b} \leftrightarrow \vec{b}'$ for convenience gives
\begin{align}
    &= \, \kron{\vec{n},\vec{n}'}  \sum\limits_{B} \, \sum\limits_{\vec{b} \neq \vec{b}' \in B } \frac{\omega^{\vec{b} \cdot \left( \vec{m}'-\vec{m} \right)}}{\LocDim^{\ell}} \, \kron{\vec{n}',\vec{b}'-\vec{b}}  \, , ~
    \label{eq:Tmat meas nonadapt degen X elements}
\end{align}
which, as with \eqref{eq:Tmat meas nonadapt degen elements} for the $\weight{}$-basis case, is zero unless $\vec{n}'$ is compatible with the difference between the configurations $\vec{b}$ and $\vec{b}'$ in block $B$. Again, further simplifications to \eqref{eq:Tmat meas nonadapt degen X elements} are not possible in generality.

Summarizing the above, the measurement transition matrix gate \eqref{eq:Tmat meas elements nonadapt 2} for a generic, possibly degenerate observable $\mobserv$ acting on the $\ell$-site cluster $r$ in measurement layer $\sigma$ of time step $t$, whose $\Noutcome$ unique eigenvalues correspond to  $\shift{}$-basis states \eqref{eq:ShiftOpDef} has elements
\begin{align}
    \Tgate{\vec{m},\vec{n};\vec{m}',\vec{n}'}{(t,\sigma,r,\shift{})} \, &= \, \kron{\vec{m},\vec{m}'}\kron{\vec{n},\vec{n}'} \kron{\vec{n}',\vec{0}} \notag \\
    ~+ \kron{\vec{n},\vec{n}'} &\, \sum\limits_{B} \sum\limits_{\vec{b} \neq \vec{b}' \in B } \frac{\omega^{\vec{b} \cdot \left( \vec{m}'-\vec{m} \right)}}{\LocDim^{\ell}} \, \kron{\vec{n}',\vec{b}'-\vec{b}}   \,, ~ \tag{\ref{eq:Tmat meas elements nonadapt degen X}}
\end{align}
where, as before, $B$ labels \emph{degenerate} blocks, whereas now, $\vec{b},\vec{b}'$ are distinct $\ell$-site, $\shift{}$-basis configurations of cluster $r$, which must be compatible with $\vec{n}'$ (the powers of $\weight{}$ present in a given term in the observable or density matrix being evolved). The second term vanishes for nondegenerate observables.

Most generally, starting from \eqref{eq:Tmat meas elements nonadapt}, we find
\begin{align*}
    \tgate^{ ( t,\sigma,r) }_{\vec{m},\vec{n};\vec{m}',\vec{n}'}  &\equiv \frac{1}{\LocDim^{\ell}} \sum\limits_{\mu=0}^{\Noutcome-1}  \tr{  \ubasisopdag{ \vec{m},\vec{n} }  \Proj{r}{(\mu)} \ubasisop{\vec{m}',\vec{n}'} \Proj{r}{(\mu)} } \tag{\ref{eq:Tmat meas elements nonadapt}} \\
    &= \frac{1}{\LocDim^{\ell}} \sum\limits_{\mu=0}^{\Noutcome-1} \sum\limits_{\vec{a},\vec{a}' \in \mu} \, \matel{ \vec{a}' }{ \ubasisopdag{ \vec{m},\vec{n} } }{ \vec{a} } \, \matel{ \vec{a} }{ \ubasisop{ \vec{m}',\vec{n}' } }{ \vec{a}' } \, , ~ 
\end{align*}
where $\vec{a}$ and $\vec{a}'$ label degenerate eigenstates of $\mobserv$ that correspond to eigenvalue $\eig{\mu}$; we rewrite the above as
\begin{align*}
    = \frac{1}{\LocDim^{\ell}} &\sum\limits_{\mu=0}^{\Noutcome-1} \sum\limits_{ \vec{a}, \vec{a}' \in \mu} \, \tr{ \, \ubasisopdag{ \vec{m},\vec{n} } \, \BKop{ \vec{a} }{ \vec{a}' } \, } \, \tr{ \,  \BKop{ \vec{a}' }{ \vec{a} } \,\ubasisop{ \vec{m}',\vec{n}' } \, }   \\
    &= \sum\limits_{\mu=0}^{\Noutcome-1} \sum\limits_{\vec{a},\vec{a}' \in \mu} \, \opinprod{\ubasisop{ \vec{m},\vec{n} } }{\nbasisop{\vec{a},\vec{a}'}} \, \opinprod{\nbasisop{\vec{a},\vec{a}'}}{\ubasisop{ \vec{m}',\vec{n}' } } \, , ~
\end{align*}
from which we conclude that
\begin{align*}
    \tgate^{\rm (meas)}_{t,\sigma,r} \, &= \, \sum\limits_{\mu=0}^{\Noutcome-1} \sum\limits_{\vec{a},\vec{a}' \in \mu} \, \opBKop{\nbasisop{\vec{a},\vec{a}'}}{\nbasisop{\vec{a},\vec{a}'}} \, ,~~\tag{\ref{eq:Tgate meas basis-indep form}}
\end{align*}
where $\nbasisop{\vec{a},\vec{a}'} = \LocDim^{\ell/2} \, \BKop{\vec{a}}{\vec{a}'}^{\,}_r$ is the normalized, na\"ive basis operator \eqref{eq:NaiveBasisOp} defined in the eigenbasis of the measured observable $\mobserv^{\,}_{t,\sigma,r}$ \eqref{eq:ASpectralDecomp}.

\let\oldaddcontentsline\addcontentsline%
\renewcommand{\addcontentsline}[3]{}%
\bibliography{circuits}
\let\addcontentsline\oldaddcontentsline%
\end{document}

%% file: preamble.tex
\usepackage{amsmath,amssymb,amsfonts,mathtools,dsfont,relsize}
\usepackage{microtype} 
\numberwithin{equation}{section}

\usepackage{suffix} 

\renewcommand{\thesection}{\arabic{section}}
\renewcommand{\thesubsection}{\thesection.\arabic{subsection}}
\renewcommand{\thesubsubsection}{\thesubsection.\arabic{subsubsection}}
\renewcommand{\theequation}{\thesection.\arabic{equation}}
\makeatletter
\renewcommand{\p@subsection}{}
\def\l@subsubsection#1#2{}
\makeatother

\usepackage[colorlinks=true, urlcolor=cyan, linkcolor=blue, citecolor=red, hyperindex=true, linktocpage=true]{hyperref}

\usepackage{xpatch}
\makeatletter
\patchcmd{\@ssect@ltx}
    {\addcontentsline{toc}{#1}{\protect\numberline{}#8}}
    {}
    {}
    {}
\makeatother

\newcommand{\vpd}[0]{\vphantom{\dagger}}
\newcommand{\vps}[0]{\vphantom{*}}
\newcommand{\vpp}[0]{\vphantom{\prime}}

\makeatletter
\g@addto@macro\bfseries{\boldmath}
\makeatother

\DeclarePairedDelimiter{\set}{\lbrace}{\rbrace}

\DeclarePairedDelimiter{\floor}{\lfloor}{\rfloor}
\DeclarePairedDelimiter{\norm}{\lVert}{\rVert}
\DeclarePairedDelimiter{\abs}{\lvert}{\rvert}
\DeclarePairedDelimiter{\expval}{\langle}{\rangle}

\DeclarePairedDelimiter{\ket}{\lvert}{\rangle}
\DeclarePairedDelimiter{\bra}{\langle}{\rvert}

\newcommand{\ii}[0]{\mathrm{i}}
\newcommand{\bvec}[1]{\boldsymbol{#1}}

\newcommand{\Order}[1]{\mathsf{O} \left( #1 \right)}
\newcommand{\order}[1]{\mathsf{o} \left( #1 \right)}

\newcommand{\Emean}[1]{\mathbb{E} \left[ #1 \right] }

\newcommand{\kron}[1]{\delta^{\,}_{#1}}

\newcommand{\ident}[0]{\mathds{1}}

\newcommand{\Comps}{\mathbb{C}}
\newcommand{\Ints}{\mathbb{Z}}

\newcommand{\U}[1]{\mathsf{U} \hspace{-0.5mm} \left( #1 \right)}

\newcommand{\BKop}[2]{\ket{#1} \hspace{-0.3mm} \bra{#2}}

\newcommand{\inprod}[2]{ \left\langle #1 \middle| #2 \right\rangle}
\newcommand{\matel}[3]{\left\langle #1 \middle| #2 \middle| #3 \right\rangle}
\WithSuffix\newcommand\matel*[3]{\langle #1 | #2 | #3 \rangle}

\newcommand{\tr}[1]{{\rm tr} \left[  #1  \right]}
\DeclareMathOperator*{\trace}{tr}

\newcommand{\com}[2]{\left[ #1, \, #2 \right]}

\newcommand{\Hilbert}{\mathcal{H}}

\newcommand{\LocDim}{q}
\newcommand{\HilDim}{\mathcal{D}}
\newcommand{\SpaceDim}{D} 
\newcommand{\Nsite}{N}

\newcommand{\size}{L}

\newcommand{\Ham}{H}

\newcommand{\Haar}{U}
\newcommand{\gate}{\mathcal{U}}
\newcommand{\evo}{\mathcal{W}}

\newcommand{\Floq}{\mathcal{F}}

\newcommand{\observ}{\mathcal{O}}
\newcommand{\mobserv}{A}

\newcommand{\eig}[1]{a^{\,}_{#1}}
\newcommand{\Eig}[1]{a^{\vpp}_{#1}}

\newcommand{\proj}[1]{\mathds{P}^{\,}_{#1}}
\newcommand{\Proj}[2]{\mathds{P}^{#2}_{#1}}

\newcommand{\nProj}[2]{\mathds{\pi}^{#2}_{#1}}

\newcommand{\tnProj}[2]{\widetilde{\mathds{\pi}}^{#2}_{#1}}

\newcommand{\Pauli}[2]{\sigma^{#1}_{#2}}

\newcommand{\PX}[1]{X^{\vps}_{#1}}
\newcommand{\PY}[1]{Y^{\vps}_{#1}}
\newcommand{\PZ}[1]{Z^{\vps}_{#1}}

\newcommand{\SSid}[1]{\ident^{\vps}_{#1}}
\newcommand{\SSProj}[2]{\mathcal{P}^{#2}_{#1}}

\newcommand{\superident}{\stackrel\frown{\ident}}
\newcommand{\opbra}[1]{\left( #1 \right|}
\newcommand{\opket}[1]{\left| #1 \right)}
\newcommand{\opinprod}[2]{\left( #1 \middle| #2 \right)}
\newcommand{\opmatel}[3]{\left( #1 \middle| #2 \middle| #3 \right)}
\newcommand{\opBKop}[2]{\left| #1 \middle) \hspace{-0.4mm} \middle( #2 \right|}
\newcommand{\opBKbkop}[4]{\left| #1 \middle) \hspace{-0.4mm}  \middle( #2 \middle| #3 \middle) \hspace{-0.4mm} \middle( #4 \right|}

\newcommand{\Nmeas}{\mathcal{N}}
\newcommand{\Noutcome}{M}
\newcommand{\MSites}{\Omega} 
\newcommand{\MeasRounds}{\mathcal{S}}

\newcommand{\measrate}{\gamma}

\newcommand{\isometry}{\mathds{V}}
\newcommand{\measunitary}[0]{\mathcal{V}}

\newcommand{\Umeas}[1]{\measunitary^{\vpd}_{\left[ #1 \right]}}
\newcommand{\Umeasdag}[1]{\measunitary^{\dagger}_{\left[ #1 \right]}}

\newcommand{\adapt}[0]{\mathcal{R}}
\newcommand{\Adapt}[1]{\adapt^{\vpd}_{#1}}
\newcommand{\AdaptDag}[1]{\adapt^{\dagger}_{#1}}

\newcommand{\rotgate}[0]{\mathsf{R}}
\newcommand{\RotGate}[1]{\rotgate^{\vpd}_{#1}}
\newcommand{\RotGateDag}[1]{\rotgate^{\dagger}_{#1}}

\newcommand{\DensMat}{\rho}
\newcommand{\DensMatSS}{\varrho}

\newcommand{\tmat}[0]{\mathcal{T}}
\newcommand{\Tmat}[2]{\mathcal{T}^{#2}_{#1}}
\newcommand{\tgate}[0]{T}
\newcommand{\Tgate}[2]{T^{#2}_{#1}}

\newcommand{\TimeOrder}{\mathcal{T}}
\newcommand{\AntiTimeOrder}{\widetilde{\mathcal{T}}}
\newcommand{\Tshift}{\mathds{T}}

\newcommand{\tfin}{T} 

\newcommand{\nbasisop}[1]{\observ^{\vpd}_{#1}}

\newcommand{\nbasisopdag}[1]{\observ^{\dagger}_{#1}}

\newcommand{\ubasisop}[1]{\Gamma^{\vpd}_{#1}}

\newcommand{\ubasisopdag}[1]{\Gamma^{\dagger}_{#1}}
\newcommand{\ubasisoppow}[2]{\Gamma^{#2}_{#1}}

\newcommand{\weight}[1]{Z^{#1}_{\,}}
\newcommand{\Weight}[2]{Z^{#1}_{#2}}
\newcommand{\shift}[1]{X^{#1}_{\,}}
\newcommand{\Shift}[2]{X^{#1}_{#2}}

\newcommand{\sweight}[1]{Z^{#1}_{\,}}
\newcommand{\sshift}[1]{X^{#1}_{\,}}
\newcommand{\SShift}[2]{X^{#1}_{#2}}

\newcommand{\charge}{\mathfrak{q}}
\newcommand{\totcharge}{Q}

